\documentclass{aa}  
\usepackage{txfonts}
\usepackage[]{hyperref}
\usepackage{graphicx}	
\usepackage{amsmath}	
\usepackage{float}
\usepackage{color}
\usepackage{soul}
\usepackage{caption}
\usepackage{subfig}
\usepackage{comment}

\newcommand{\G}{{\it Gaia}}
\newcommand{\Msun}{M$_{\odot}$}
\newcommand{\BPRP}{$G_{\rm BP}-G_{\rm RP}$\ }
\newcommand{\mycomment}[1]{}
\usepackage{multicol}

\begin{document}

   \title{A Random Forest spectral classification of the \emph{Gaia} 500-pc white dwarf population}

   \subtitle{}

   \author{Enrique Miguel Garc\'{\i}a-Zamora
          \inst{1}
          \and
          Santiago Torres\inst{1,2}\fnmsep\thanks{Email;
    santiago.torres@upc.edu}
          \and
          Alberto Rebassa-Mansergas\inst{1,2}
           \and
          Aina Ferrer-Burjachs\inst{1}
                  }

\institute{Departament de F\'\i sica, 
           Universitat Polit\`ecnica de Catalunya, 
           c/Esteve Terrades 5, 
           08860 Castelldefels, 
           Spain
           \and
           Institut d'Estudis Espacials de Catalunya, Esteve Terradas, 1, Edifici RDIT, Campus PMT-UPC, 08860 Castelldefels, Spain}

 \date{\today}
\titlerunning{Random Forest spectral classification of \emph{Gaia} white dwarfs}
\authorrunning{Garc\'{\i}a-Zamora et al.}

\offprints{S. Torres}
 
  \abstract
   {The third {\G} Data Release has provided the astronomical community with astrometric data of more than 1.8 billion sources, and low resolution spectra for 220 million. Such a large amount of data is difficult to handle by means of visual inspection. In the last years, artificial intelligence and machine learning algorithms have started to be applied in astronomy for data analysis and automatic classification, with excellent results.}
   {In this work, we present a spectral analysis of the {\G} white dwarf population up to 500\,pc from the Sun based on artificial intelligence algorithms to classify the sample into their main spectral types and subtypes.
   }
   {In order to classify the sample, which consists of 78\,920 white dwarfs with available {\G} spectra, we have applied a Random Forest algorithm to the {\G} spectral coefficients. We used the Montreal White Dwarf Database of already labeled objects as our training sample. The classified sample is compared with other already published catalogs and with our own higher resolution Gran Telescopio Canarias (GTC) spectra, enabling the construction of a golden sample of well-classified objects.} 
   {The Random Forest spectral classification of the 500-pc white dwarf population achieves an excellent global accuracy of 0.91 and an F1-score of 0.88 for the DA versus non-DA classification. In addition, we obtain a very high accuracy of 0.76 and a global F1-score of 0.62 for the non-DA subtype classification. In particular, our classification shows an excellent recall for DAs, DBs and DCs (>90\%), and a very good precision ($\geq$80\%) for DQs, DZs and DOs. Unfortunately, our algorithm does not perform well in correctly classifying subtypes given the low resolution of the \G\,  spectra.} 
{The use of machine learning techniques, particularly the Random Forest algorithm, has enabled us to spectrally classify 78,920 white dwarfs -- an increase of 543.6\% over those previously labeled -- with reasonable accuracy. Having an estimate of the spectral type for the vast majority of white dwarfs up to 500 pc provides the possibility of making better estimates of cooling ages, star formation rates, and stellar evolution processes, among other fundamental aspects for the study of the white dwarf population.}
   \keywords{(Stars:) white dwarfs, atmospheres, catalogs }

   \maketitle
%

\section{Introduction}
\label{intro}

White dwarfs are the most common stellar remnants in the Galaxy. They are the end products of stars with initial masses M$\la$8-10 M$_{\odot}$ (see, e.g., \citealt{Althaus2010}). They are composed of a degenerate core with a typical mass of M$\approx$0.6 M$_{\odot}$, generally surrounded by a thin, partially degenerate atmospheric hydrogen layer. Nuclear reactions in their cores have practically ceased, the energy source in their deep interiors being primarily derived from gravothermal energy instead.

Spectroscopic observations of white dwarfs enable detection of atomic and molecular spectral lines and bands. This, in turn, makes it possible to create a white dwarf spectral classification, attending to the presence of certain spectral lines \citep{Sion1983}. A first distinction can be established between white dwarfs that show Balmer spectral lines, named DAs or hydrogen-rich white dwarfs, and those which do not, generally grouped under the generic non-DA class. Under this term, we may find DB white dwarfs, which show \ion{He}{I} spectral lines; DOs, which exhibit \ion{He}{II} lines; DQs, with prominent carbon features, either atomic or molecular, in their spectra; DZs, which display metallic spectral lines, such as \ion{Ca}{II}; or DCs, which display no spectral lines and a featureless continuum. 

The previously described types are referred to as the primary spectral type of the white dwarf (see Table 2 in \citealt{Sion1983}). Secondary spectral features, such as spectral lines arising from different atomic lines than the dominant or the presence of a magnetic field, are frequently denoted by adding the relevant characteristic as a spectral subtype. In this way, for instance, a DA white dwarf with weaker metallic features will be labelled as a DAZ, or a magnetic DB white dwarf will be labelled as DBH.

Not only is spectral classification of capital importance for accurate stellar parameter derivation \citep{Bergeron2019}, \citep{Tremblay2019b}, but also for understanding the process of spectral evolution in white dwarfs \citep[see][and references therein]{Bedard2024}. Processes such as convective mixing or convective dilution \citep[e.g][]{Blouin2019,Cunningham2020}, the presence of carbon in hydrogen-deficient atmospheres as a possible explanation of the {\it Gaia} color-magnitude bifurcation \citep{Camisassa2023,Blouin2023}, the high ratio of DQ white dwarfs in the so-named Q branch \citep{Tremblay2019} or the origin of accreted material in white dwarfs \citep[e.g][]{Zuckerman2007,Farihi2010}, require a correct identification of the spectral type in order to be understood.

Spectroscopic follow-up of white dwarfs, however, is a time-costly process. A completed survey of white dwarfs within 40-pc has been performed by \citet{Tremblay2020, McCleery2020, Obrien2023} and a full 100-pc spectroscopic classification will eventually be achieved thanks to current and forthcoming surveys such as WEAVE \citep{WEAVE2022}, DESI \citep{DESI2019} and 4MOST \citep{4MOST2019}. However, it has to be emphasised that a considerably fastest way of providing a spectral type for a large number of white dwarfs is to make use of already available spectroscopic samples such as the one provided by {\G}.

The third \G\,  Data Release \citep{GaiaDR32023} has delivered the astronomical community with astrometric data of 1.8 billion objects. More importantly in the context of this paper, \G\, has also provided low resolution (30$\lesssim$R$\lesssim$100, see \citealt{Carrasco2021}) spectra for 220 million sources, including more than 100\,000 white dwarfs. This is the largest spectroscopic sample to date, however, the vast quantity of data impedes its analysis through human visual inspection alone, making artificial intelligence algorithms and  machine learning approaches a necessity. These approaches are hardly recent and have proved their reliability for the analysis of large astronomical databases. For instance, see the pioneering use of self-organizing maps for the identification of halo stars in \citet{Torres1998}, the first efforts and the more recent ones that use the Random Forest algorithm to perform white dwarf identification in Galactic component \citep{Torres2019} or spectral identification \citealt{Echeverry2022, Montegriffo2023}, as well as the deep learning techniques used in \citealt{Kong2018} or \citealt{Olivier2023} for white dwarf type classification. Additionally, statistical classification methods have been performed, particularly regarding white dwarfs. Namely, the use of the Virtual Observatory Spectral energy distribution Analyzer tool \citep{VOSA2008} in \citet{Jimenez2023} and \citet{Torres2023}, in which the spectral energy distributions (SEDs) of the 100-pc and 500-pc white dwarf samples respectively, were fitted automatedly to different model atmospheres. In this way, white dwarfs were classified into DA and non-DA with an accuracy over 90\%.

In this work, we apply a Random Forest algorithm to classify the 500-pc \G\, white dwarf sample into their spectral types and subtypes, as well as an analysis dedicated to detect possible binary objects such as white dwarf-main sequence binaries or unresolved double-degenerates. This study updates our recent work performed in \citet{Garcia2023}, where we used the same technique to classify the 100-pc white dwarf sample, and complements the work by \cite{Vincent2024}, who also classified white dwarfs in a 500-pc radius using gradient boosting classifiers. The 500-pc distance limit imposed is capital to our research, since it allows to obtain a nearly-complete classification of all known white dwarfs with available \G\, spectra. This in turn allows us to derive more accurate percentages of the different spectral types.

The 500-pc white dwarf sample used in this work was selected following the same conditions established in \citet{Torres2023}, except for the $-0.5 \leq $\BPRP$ \leq 0.86$ color restriction. Unlike \citet{Torres2023}, we do not use atmospheric models for our spectral classification. Additionally, we select only objects lying below the 0.45~$M_{\odot}$ white dwarf cooling track, further reducing potential contamination by non-white dwarfs to an almost negligible level. Of the 93,439 objects with \G\ spectra, almost all (93,323) also appear in the recent classification by \citet{Vincent2024}.

In Section \ref{s:method}, the applied methodology is detailed. In Section \ref{s:valid}, the algorithm validation tests performed on the classified white dwarf sample, both for spectral and binarity classification, are shown. The classification of the unclassified sample is presented in Section \ref{s:classi}. The results of this work are analyzed in Section \ref{s:Analysis} and compared in Section \ref{s:Comp} with those obtained by various recent automated classifications. Finally, we outline our conclusions in Section \ref{s:conc}.

\section{The method: Random Forest classification of \emph{Gaia} spectral coefficients}
\label{s:method}

Random Forest \citep{Breiman2001} is a widely used machine learning algorithm. A set of labeled data, the training set, is used to create an ensemble of decision trees, the random forest. This process is known as training the algorithm; once it has taken place, new data can be classified into the training dataset categories. 

As we are using the same methodology described in \citealt{Garcia2023}, we will only outline the most important details in this section. For a more in-depth description, we refer to \citealt{Garcia2023} and \citealt{Torres2019}.

The \G\, spectra cover the wavelength range 3\,300-10\,500 {\AA} (the Blue Photometer covers the 3\,300-6\,800 {\AA} wavelength range; while the Red Photometer covers the 6\,400-10\,500 {\AA} range; see \citealt{Carrasco2021}) at a resolution of $\lambda / \Delta\lambda \approx100$. Additionally, they are not provided in the classical flux versus wavelength representation, but rather as the coefficients of a linear combination of base functions (more concretely, Hermite functions that act as the basis for spectral representation; see \citealt{Carrasco2021}) internally calibrated in a pseudo-pixel scale. Each spectrum, both BP and BP, is provided through 55 coefficients, resulting in a total of 110 coefficients for the mean XP spectrum.

Following the same approach as \cite{Garcia2023} and \cite{Vincent2024}, these 110 coefficients are used as input data for our algorithm. The rationale behind this decision is that the use of the \G\, spectral coefficients provides better performance than other inputs \citep{Montegriffo2023}. This treatment is appropriate, as all the relevant spectral information (e.g., the continuum shape or the spectral lines that define the different spectral types) is contained within the 110 coefficients (see, for example, \citealt{Weiler2023} for a mathematical description of the spectral coefficients applied to hydrogen lines). No external calibration to wavelength-flux form was applied, as one consequence of the external calibration is the introduction of oscillatory behaviour in the spectra, known as ``wiggles'', which are introduced by the mathematical process used for external calibration \citep{DeAngeli2022}. While these wiggles affect the whole wavelength range, they are more prominent in its extremes. 

Furthermore, our analysis relies on a sample of externally classified white dwarfs (see Section\,\ref{s:valid}), the temperature range of which spans from the hotter objects to the cooler end. This poses an advantage over the classification approaches undertaken in \cite{Jimenez2023} and \cite{Torres2023}, since they rely on fitting SEDs to white dwarf theoretical atmosphere models. These models are known to harbor substantial uncertainties at temperatures below $T\approx5\,500 K$, and therefore the obtained classification might be inaccurate for the colder objects. Our approach, on the contrary, forgoes this model fitting and therefore more accurate classifications for colder objects are obtained.

In order to create the Random Forests and obtain the confusion matrices and classification metrics, as well as creating all unclassified white dwarf classifications, the Python package \texttt{scikit-learn} \citep{Pedregosa2011} was used.

\section{Algorithm training and validation}
\label{s:valid}

Before the sample classification can be performed, the Random Forest algorithm must first be trained and then validated using already classified white dwarfs. For this purpose, we resorted to the Montreal White Dwarf Database\footnote{\url{https://www.montrealwhitedwarfdatabase.org/}} (MWDD), which contains astrometric and photometric data, as well as spectral and binarity classification, for tens of thousands of white dwarfs sampling the entire effective temperature range \citep{Dufour2017}.

The white dwarfs in the MWDD with an assigned spectral type and a binary classification in a 500-pc radius around the Sun were collected. From these, those which possessed a \G\, spectra formed the training set for the spectral classification and subclassification validation tests. 

For the validation tests, we applied the cross-validation method called Stratified $k$-Fold. It consists in dividing the whole training set into $k$ subsets, all of which keep a very similar category ratio (i.e. the proportion of objects belonging to each class in the sample, as close as possible to the whole set category ratio). In this work, following the method outlined in \citet{Garcia2023}, the value $k=10$ was chosen and as many iterations as subsets exist were completed. In each iteration, $k-1$ folds were used for training the algorithm, while the remaining fold was used for testing. The chosen test fold was different for each iteration. Thus, once the cross-validation was completed, the whole sample has been used for training and testing the algorithm.

Two different sets of validation tests were carried out. The first set validated the spectral type classification algorithm; the second set tested the binarity classification algorithm. In all cases, the 110 \G\, spectral coefficients, as well as the spectral and binarity information contained in the MWDD, were used as input data for our algorithm.

\subsection{Spectral classification validation tests}
\label{ssec:val_types}

In the MWDD, 51\,319 objects have been classified into their spectral types, 34\,250 of them in a 500-pc radius around the Sun. From this subset, we derive a training set comprised of 14\,519 white dwarfs with \G\, spectra (10\,916 DAs, 1\,189 DBs, 1\,596 DCs, 26 DOs, 372 DQs and 420 DZs). As input data for the validation tests, their spectral labels and their 110 spectral coefficients are used in order to train the algorithm.

Following the pipeline described for validation tests in \cite{Garcia2023}, three different spectral type classification tests were performed. The first test  classified the whole sample as DA or non-DA; the second test classified non-DAs into their spectral types (DB, DC, DO, DQ and DZ). Finally, the third validation test focused on spectral subtype classification for the previously listed spectral types.

\subsubsection{First validation test: DA vs non-DA}
\label{sssec:DAnonDA}

\begin{figure}[ht]
\includegraphics[width=1\columnwidth,trim=35 0 30 20, clip]{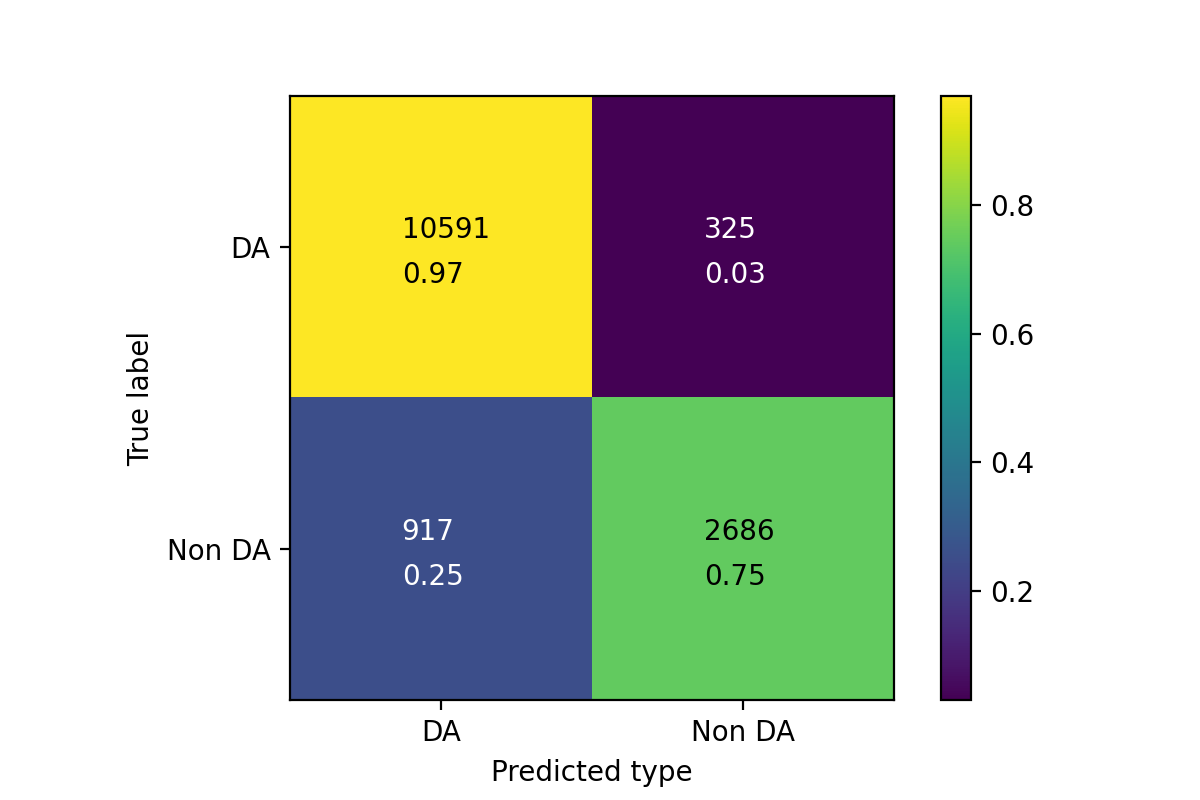}
\includegraphics[width=1\columnwidth,trim=30 0 30 20, clip]{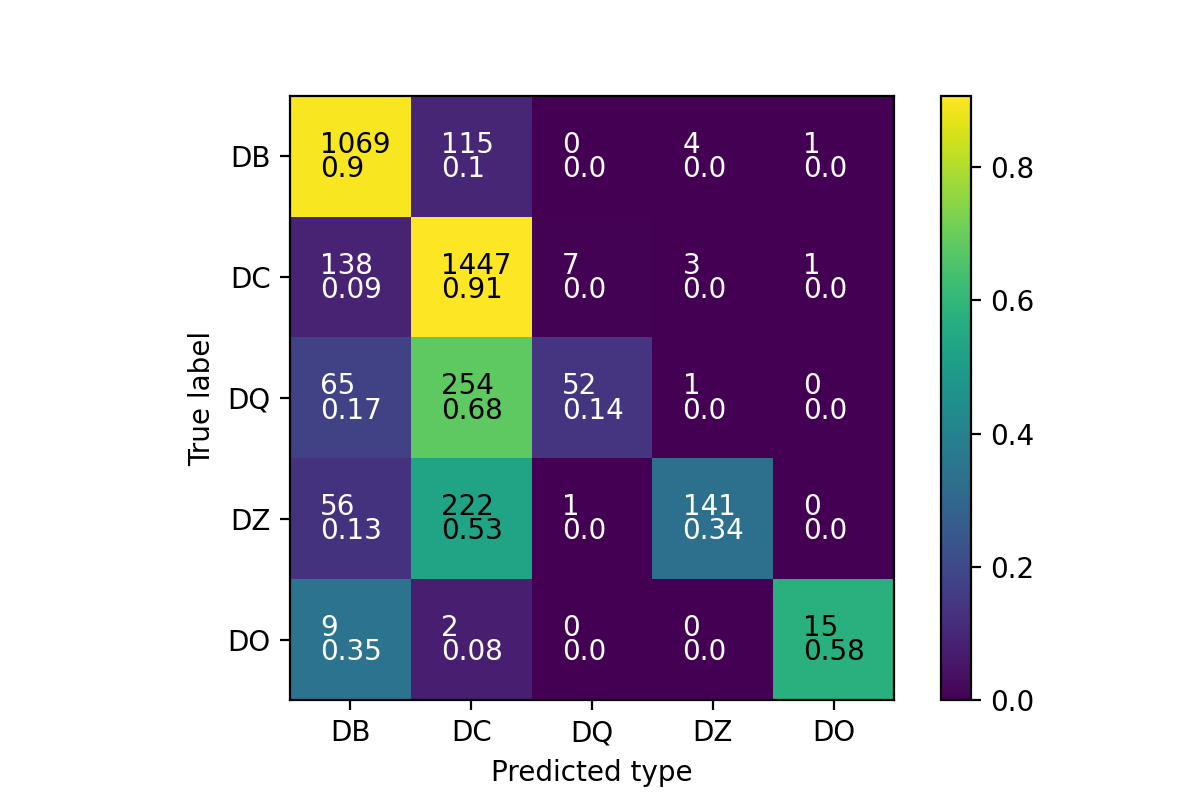}
    \caption{Confusion matrices for our validation tests: DA vs non-DA (top panel) and non-DA types (bottom panel). As true label (rows) we adopted the MWDD classification,  while the predicted label (columns) is the one resulting from our Random Forest algorithm.}
    \label{f:cmvaltests}
\end{figure}

The 14\,519 objects in our training sample were classified into DA (10\,916 objects, 75.18\%) and non-DAs (3\,603 objects, 24.82\%). The results are shown in the form of a confusion matrix in the upper panel of Figure \ref{f:cmvaltests}.

Confusion matrices show a visual representation of the classification model results. The rows contain the true values (i.e., the spectral types assigned to every object in the MWDD) and the columns contain the predicted values (namely, the spectral classification assigned by our algorithm). For every matrix element, the number of objects, as well as its ratio to the number of objects contained in each category in the MWDD, are given. A perfect classification would produce a diagonal matrix, with non-zero elements exclusively along the confusion matrix main diagonal. The classification metrics are also shown in Table \ref{tab:Metrics}.

\begin{table*}[h!]
    \caption{Validation tests classification metrics}
    \label{tab:Metrics}
\begin{center}
    \begin{tabular}{lccccccccc}
            \noalign{\smallskip}
            \hline\hline
            \noalign{\smallskip}
        Metric & Validation test & DA  & Non-DA & DB & DC & DO & DQ & DZ   \\
            \noalign{\smallskip}
            \hline
            \noalign{\smallskip}
        Recall & Test 1 &  0.97 & 0.75 & - & - & - & - & -  \\
        Recall & Test 2 & - & - & 0.90 & 0.91 & 0.58 & 0.14 & 0.34  \\
            \noalign{\smallskip}
            \hline
            \noalign{\smallskip}
        Precision & Test 1 & 0.92 & 0.89 & - & - & - & - & -  \\
        Precision & Test 2 & - & - & 0.80 & 0.71 & 0.88 & 0.87 & 0.95  \\
            \noalign{\smallskip}
            \hline
            \noalign{\smallskip}
        F1-score & Test 1 & 0.94& 0.81 & - & - & - & - & - \\
        F1-score & Test 2 & - & - & 0.85 & 0.80 & 0.70 & 0.24 & 0.50   \\
            \noalign{\smallskip}
            \hline
                   \end{tabular}
 \tablefoot{Classification metrics for the first validation test in which we classify white dwarfs of the MWDD into DA and non-DA classes (Test 1), and the second validation test (Test 2) in which non-DAs are classified into DB, DC, DO, DQ and DZ.}
\end{center}
\end{table*}

The confusion matrix (Figure \ref{f:cmvaltests}, top panel) shows an excellent recall for DAs (97\%), as well as a very good recall for non DAs (75\%).
We also took into account the global metrics. For this particular classification, the algorithm achieved an accuracy\footnote{Accuracy is defined as the proportion of correct predictions.} of 0.91, 
 and a mean F1-score{\footnote{F1-score is defined as ${2\times{\rm Recall}\times{\rm Precision}}/({{\rm Recall}+{\rm Precision}}).$} of 0.88. To take into account the class imbalance in the test sample, two other metrics were additionally considered: balanced accuracy\footnote{The balanced accuracy is defined as the average of the recalls for each class.} (0.86 value) and G-means\footnote{The G-mean score is defined as the geometric mean of the recalls for every class.} (0.85 value). In all cases, the results indicate that the Random Forest algorithm is able to classify DA and non-DA white dwarfs satisfactorily.

\subsubsection{Non-DAs validation test}

For this test, only the 3\,603 objects classified as non-DA in the MWDD (1\,189 DBs, 1\,596 DCs, 26 DOs, 372 DQs and 420 DZs) were considered. As it can be seen in the corresponding confusion matrix (see Figure \ref{f:cmvaltests}, bottom panel), our algorithm has an excellent recall for DBs and DCs ($\geq$90\%), as well as a relatively good recall for DOs ($\sim$60\%). While the recalls for DQs and DZs are improvable (14\% and 34\% respectively), it must be stressed that these categories, as well as DOs, also have a very good precision 
($\geq$80\%). This implies that, additionally to false positives being almost non-existent for these spectral types, our algorithm is highly useful for identifying members of these classes in the population, despite the fact that not all of them are found.

Globally, our algorithm achieves a mean F1-score of 0.62, an accuracy of 0.76, a balanced accuracy of 0.57 and a G-mean score  of 0.47. These values show the impact of the low resolution \G\, spectra in our non-DA spectral classification. 

\subsubsection{Spectral subtype validation test}
\label{ssec:val_subtypes}

Once the capability of our Random Forest algorithm to correctly classify white dwarfs into their primary spectral types has been demonstrated, its ability to identify subtypes inside a certain class is tested. Only DA, DB, DO, DQ and DZ subtypes are taken into account, as DCs, showing featureless spectra, are by definition not divided into subtypes. 

Separate validation tests were conducted, one for each primary spectral type. As the primary type validation sample contains, in some instances, spectral information that is useful for spectral type classification, but not useful for subtype classification (for example, spectral types being registered as DA/DAZ? or DB/DBAZ?), these elements were taken out of the validation sample. This process left us with a subtype validation sample comprised of 14\,319 objects. The excluded objects were not discarded but were moved to the subtype classification sample, keeping their MWDD primary spectral type.

The DA subtype training sample comprises the subtypes DA, DAB, DAH, DAO and DAZ (10\,417, 30, 259, 6 and 107 elements, respectively). No subtypes were correctly identified, while five DAs are misidentified as magnetic DAHs. No other types were correctly classified. Extreme numerical imbalance is believed to severely impact our classification.

For DB subtypes, the pure DB, DBA, DBH and DBZ subtypes (791, 291, 14 and 63 elements, respectively) were considered. Results show that the DBA subtype, despite reaching over 40\% precision, achieves only 5\% recall. The DBH and DBZ subtypes, on the other hand, show no cases of correct identification; moreover, one DBH and two DBZs are misclassified as DBAs.

The DO subtypes, DO, DOA, and DOZ (with 15, 3, and 8 elements, respectively), were considered. Although some DOZs are correctly classified (25 \% recall, albeit with a low precision, only 33\%), we opt for a more cautious interpretation of these results. Due to the small size of the whole DO set, which includes only 26 white dwarfs, model training is far from optimal.

DQs are divided into DQ, DQA, DQH, DQZ and DQpec (346, 11, 2, 7 and 6 elements, respectively). Only DQs are correctly classified and no other subtypes are found. Numerical imbalance is thought to be at the root of this result.

Finally, DZ subtypes are considered: DZ, DZA, and DZH (360, 50, and 10 elements, respectively). While two DZAs were correctly identified as such (4\% recall), three DZs were mislabeled as DZA (40\% precision). Therefore, our algorithm cannot correctly classify DZ subtypes.

Based on the above analysis, it seems that a reliable subspectral type classification based on \G\ white dwarf spectra is likely unfeasible. 
This result is not entirely surprising given the low spectral resolution of \G. Moreover, due to the fact that the numerical imbalance is now more extreme than in the 100-pc sample (see Section \ref{ss:Comparison}), which favors pure spectral types, the classification is biased towards them.

\subsection{Binarity validation test}
\label{ssec:val_bin}

The last test aimed at discerning the possibility of using the Random Forest algorithm to find white dwarf candidates that are members of binaries, either unresolved white dwarf-main sequence (WDMS) binaries  or unresolved double-degenerate (DD) objects. While we do not expect to find many such objects, especially WDMS candidates, since our considered objects lie in the white dwarf locus of the color-magnitude diagram, we nevertheless undertook this classification to fully assess the capabilities of our Random Forest algorithm.

For this test, we added to our training sample the white dwarfs in a 500-pc radius registered as WDMS or DD in the MWDD, as well as the WDMS from the Sloan Digital Sky Survey catalog of \cite{Rebassa2013}. The final training set comprises 14\,651 objects (14\,309 WDs, 20 DDs and 322 WDMS binaries, respectively). As already mentioned, all objects considered for the training set are inside the color-magnitude diagram white dwarf locus. Objects in the intermediate zone between white dwarfs and main sequence stars were not considered and will be analyzed elsewhere.

Out of the 322 WDMS considered, only three are correctly identified (a recall slightly lower than 1\%). Moreover, three white dwarfs are misidentified as WDMS objects (50\% precision). All DD are classified as single WDs (0\% recall), while no white dwarf or WDMS binary is wrongly classified as a DD. The parameters are summarized in Table\,\ref{tab:BinMetrics}.

\begin{table}
    \caption[]{Binarity validation test classification metrics}
    \label{tab:BinMetrics}
\begin{center}
    \begin{tabular}{lccc}
            \noalign{\smallskip}
            \hline\hline
            \noalign{\smallskip}
        Metric & WD & WDMS & DD \\
            \noalign{\smallskip}
            \hline
            \noalign{\smallskip}
        Recall & 1.0 & 0.01 & 0  \\
            \noalign{\smallskip}
            \hline
            \noalign{\smallskip}
        Precision & 0.98 & 0.5 & 0 \\
            \noalign{\smallskip}
            \hline
            \noalign{\smallskip}
        F1-score & 0.99 & 0.02 & 0 \\
            \noalign{\smallskip}
            \hline
    \end{tabular}
\end{center}
\end{table}

While disheartening, these results are not completely unexpected. Due to the location of these objects in the color-magnitude diagram, it is expected that the white dwarf component is dominant in the spectrum if the system is a WDMS. On visual inspection, most of them indeed show very little contribution from the main sequence, confirming our hypothesis. 

Regarding the identification of DDs, the task itself is extremely challenging even when using high-resolution spectra. As a consequence, classification of double white dwarfs is generally based on the detection of radial velocity shifts \citep{Breedt2017, Napiwotzki2020} rather than on direct analysis of the possible combined spectra.

We conclude this section confirming the robustness of our Random Forest algorithm in classifying DA and non-DA white dwarfs, its great utility in precisely assigning spectral types to non-DA white dwarfs (despite the fact that not all individual objects are expected to be identified) and the lack of use in differentiating between spectral subclasses and white dwarf binaries in the white dwarf locus. 

\section{Classification of the \emph{Gaia} 500-pc white dwarf population}
\label{s:classi}

Once the algorithm has been validated, it was applied to the unclassified 500-pc white dwarf sample, which comprises 78\,920 objects. We recall that our analysis is restricted to the white dwarf color-magnitude locus, i.e., the region below the 0.45\,\Msun\ white dwarf cooling track \citep[see Section 2 of][]{Torres2023}. As outlined in Section \ref{s:valid}, three different classifications are performed. The first spectroscopically classifies the whole sample into their primary spectral types. These are further classified into their subtypes in the second classification. Finally, the binarity classification algorithm is applied to the sample. Even if the results of both the spectral subtype and binarity validation tests do not give us much reason for optimism, we nevertheless wanted to test the algorithm on the unclassified sample.

\subsection{Primary spectral types}

Following the procedure described in subsection \ref{sssec:DAnonDA}, the first step classifies the white dwarfs into the DA and non-DA categories. The sample used for validation tests was adopted in order to train the classification algorithm. Of the 78\,920 objects, 64\,976 were classified as DA, while the remaining 13\,944 were categorized as non-DA. The objects classified as DAs are illustrated in the \emph{Gaia} color-magnitude diagram in the top panel of Figure \ref{f:DA_noDA_500pc}. They extend from the hottest region, \BPRP$\approx -0.5$, until \BPRP$\approx 1.2-1.3$. This corresponds to a white dwarf effective temperature of $\approx5\,000\,$K, where almost all the hydrogen atoms are in the ground state and therefore the spectra evolves into that of a featureless DC.

\begin{figure}[ht]
\centering
\includegraphics[width=1\columnwidth,trim=-20 0 0 0, clip]{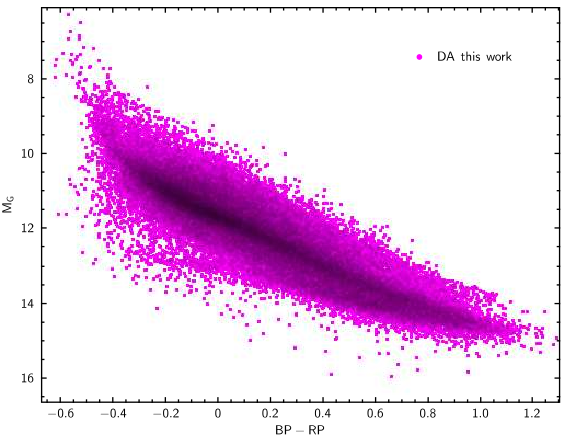}
\includegraphics[width=1\columnwidth,trim=-20 0 0 0, clip]{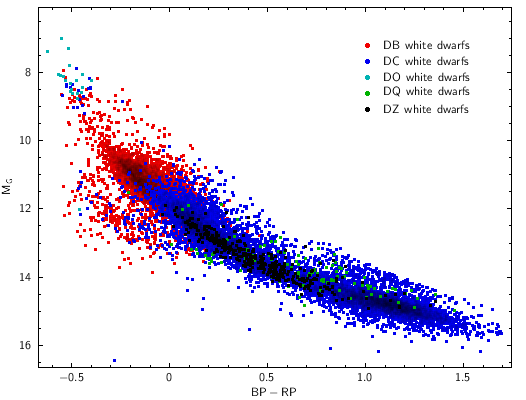}
   \caption{\emph{Gaia} color-magnitude diagram  for white dwarfs classified as DA (top panel), and  for those classified into various non-DA spectral types (bottom panel) by our algorithm.}
    \label{f:DA_noDA_500pc}
\end{figure}

The remainder 13\,944 non-DA objects were further classified into the DB, DC, DO, DQ and DZ primary spectral types, resulting in 4\,957 as DB, 8\,496 as DC, 21 as DO, 105 as DQ and 365 as DZ. These sources are shown in the \emph{Gaia} color-magnitude diagram in the bottom panel of Figure\,\ref{f:DA_noDA_500pc}. The white dwarfs of different spectral types appear in their expected locations across the diagram: DOs appear only at the brightest, hottest part (\BPRP$\approx-0.5$); DBs appear in the hotter region (\BPRP$\approx-0.5-0$, and DZs and DQs appear spread through a wider range of temperatures.

\subsection{Spectral subtypes}

We attempted further subclassification, despite the challenges indicated by the validation test, of the classification set created in the way described in subsection \ref{ssec:val_subtypes}. Among the 65\,073 DA white dwarfs, only six were identified as DAH, with no DAB, DAO or DAZ candidates. For the 4\,985 DB white dwarfs, only 67 were classified as DBA, while no DBH, DBO, DBQ, or DBZ were found. In the DO category, 12 of 21 objects remained DO, while nine were classified as DOZ. The small training sample limits its reliability, however. No subtypes were identified within the 105 DQ white dwarfs. Lastly, among 365 DZ white dwarfs, no DZA or DZH candidates were found, consistent with validation results.

\begin{table*}[h!]
    \caption[]{Random Forest algorithm classification of {\it Gaia} 500-pc white dwarfs}
    \label{tab:catalogue}
\begin{center}
    \begin{tabular}{rccccccc}
            \noalign{\smallskip}
            \hline\hline
            \noalign{\smallskip}
        \emph{Gaia} Source ID & RA & Dec & $M_G$ & BP-RP & $d$ & $S_{\rm type}$ & Binarity \\
            & (deg) & (deg) & (mag) & (mag) & (pc) & & \\
            \noalign{\smallskip}
            \hline
            \noalign{\smallskip}
        4282373867773042944 & 284.14 & 5.05 & 13.28 & 0.16 & 93.14 & DA & WD\\
        6257089948030509696 & 226.79 & -18.96 & 10.97 & 0.04 & 266.38 & DA & WD \\
                  $\ldots$  & $\ldots$ & $\ldots$ & $\ldots$ & $\ldots$ & $\ldots$ & $\ldots$ \\
        5396722000514439936 & 166.72 & -39.33 & 12.78 & 0.14 & 67.04 & DAH & WD\\
        3630648387747801088 & 204.92 & -7.22 & 12.57 & -0.01 & 56.33 & DAH & WD\\
                  $\ldots$  & $\ldots$ & $\ldots$ & $\ldots$ & $\ldots$ & $\ldots$ & $\ldots$ \\
        134554212918392704 & 39.53 & 33.03 & 11.34 & -0.12 & 139.45 & DB & WD\\
        3133168437589747712 & 101.61 & 6.91 & 11.19 & -0.19 & 172.33 & DB & WD\\
                  $\ldots$  & $\ldots$ & $\ldots$ & $\ldots$ & $\ldots$ & $\ldots$ & $\ldots$ \\
        2396881442417524480 & 344.05 & -21.59 & 11.59 & 0.06 & 216.49 & DBA & WD\\
        1216046326882605696 & 235.99 & 19.98 & 11.96 & 0.10 & 149.91 & DBA & WD\\
                  $\ldots$  & $\ldots$ & $\ldots$ & $\ldots$ & $\ldots$ & $\ldots$ & $\ldots$ \\
        1184536930671777408 & 221.73 & 12.58 & 11.89 & 0.02 & 242.37 & DC & WD\\
        6115057239274569728 & 206.14 & -39.11 & 12.97 & 0.28 & 128.40 & DC & WD\\
                  $\ldots$  & $\ldots$ & $\ldots$ & $\ldots$ & $\ldots$ & $\ldots$ & $\ldots$  \\
        3128765207057429504 & 105.52 & 5.24 & 7.39 & -0.62 & 325.62 & DO & WD\\
        6451742611825266176 & 318.89 & -61.98 & 8.07 & -0.57 & 359.89 & DO & WD\\
                  $\ldots$  & $\ldots$ & $\ldots$ & $\ldots$ & $\ldots$ & $\ldots$ & $\ldots$  \\
        3165340216537925120 & 114.03 & 14.75 & 8.09 & -0.56 & 151.07 & DOZ & WD\\
        6517863392828518272 & 336.51 & -49.15 & 8.10 & -0.54 & 288.44 & DOZ & WD\\
                  $\ldots$  & $\ldots$ & $\ldots$ & $\ldots$ & $\ldots$ & $\ldots$ & $\ldots$  \\
        484896928635700992 & 88.90 & 69.42 & 13.09 & 0.46 & 129.20 & DQ & WD\\
        4770661758988444672 & 77.65 & -54.31 & 13.87 & 0.73 & 131.93 & DQ & WD\\
                  $\ldots$  & $\ldots$ & $\ldots$ & $\ldots$ & $\ldots$ & $\ldots$ & $\ldots$ \\
        3921116197046844032 & 185.02 & 14.31 & 13.18 & 0.36 & 112.23 & DZ & WD\\
        5821373624752235264 & 243.32 & -67.20 & 14.13 & 0.69 & 106.07 & DZ & WD\\
                  $\ldots$  & $\ldots$ & $\ldots$ & $\ldots$ & $\ldots$ & $\ldots$ & $\ldots$ \\
            \noalign{\smallskip}
            \hline
    \end{tabular}
\end{center}
\end{table*}

\subsection{The {\it Gaia} 500-pc sample classification summary}
\label{s:500pcsummary}

We have spectroscopically classified 78\,920 objects into their primary spectral types. Of these, 64\,976 objects were classified as DA (six of which were subclassified as magnetic DAH), 4\,957 as DB (67 as DBA), 8\,496 as DC, 21 as DO (9 as DOZ), 105 as DQ and 365 as DZ. As a consequence, the number of DA, DB, DC, DO, DQ and DZ white dwarfs in a 500-pc radius around the Sun has increased in a 595\%, 417\%, 532\%, 80.8\%, 28.2\% and 86.9\%, respectively, with respect to the training sample. Whilst the primary spectral type classification can be considered as reliable, caution should be taken with the secondary spectral type assigned, mainly due to the low resolution of the \G\, spectra.

Additionally, 78\,791 objects have been classified into single and binary. Of them, only two have been classified as a WDMS candidate.

All data have been compiled into a single catalog, a representative excerpt of which is shown in Table \ref{tab:catalogue}. The complete table can be accessed at the CDS.

\section{Analysis of the classified 500-pc white dwarf population}
\label{s:Analysis}

In this section, we perform a detailed inspection and analysis of the white dwarf population within 500 pc in terms of their spectral type classification.

\begin{table}
    \caption{Spectral types percentages}
    \label{tab:Pops}
    \begin{small}
\begin{center}
    \begin{tabular}{lcccccc}
            \noalign{\smallskip}
            \hline\hline
            \noalign{\smallskip}
        Population & DA & DB & DC & DO & DQ & DZ\\
            \noalign{\smallskip}
            \hline
            \noalign{\smallskip}
        100 pc MWDD & 68.61 & 3.34 & 19.72 & - & 4.03 & 4.30\\
        100 pc classified & 50.15 & 0.80 & 46.97 & - & 0.66 & 1.40\\
        100 pc total & 54.49 & 1.40 & 40.56 & - & 1.45 & 2.08\\
        \noalign{\smallskip}
            \hline
            \noalign{\smallskip}
        500 pc MWDD & 75.18 & 8.19 & 10.99 & 0.18 & 2.56 & 2.89\\
        500 pc classified & 82.33 & 6.28 & 10.77 & 0.03 & 0.13 & 0.46\\
        500 pc total & 81.22 & 6.58 & 10.80 & 0.05 & 0.51 & 0.84\\
        
            \noalign{\smallskip}
            \hline
    \end{tabular}

\end{center}
    \tablefoot{Percentages of {\it Gaia} white dwarfs according to their spectral types within 100 and 500 pc for the training (MWDD) sample, the classified sample and the total sample, respectively.}
\end{small}
\end{table}

\subsection{Spectral content of the white dwarf population}

To begin with, we here provide and compare the percentages of white dwarfs with different spectral types within 100 pc (which was the sample of study in our previous work; \citealt{Garcia2023}) and 500 pc. We do this for the training sample (that is, the MWDD sample used), the classified sample and the complete sample. The results are shown in Table \ref{tab:Pops}, and can be visualized in Figure \ref{f:Fractions_distance}.

\begin{figure}[h!]
\includegraphics[width=1\columnwidth,trim=10 0 20 20, clip]{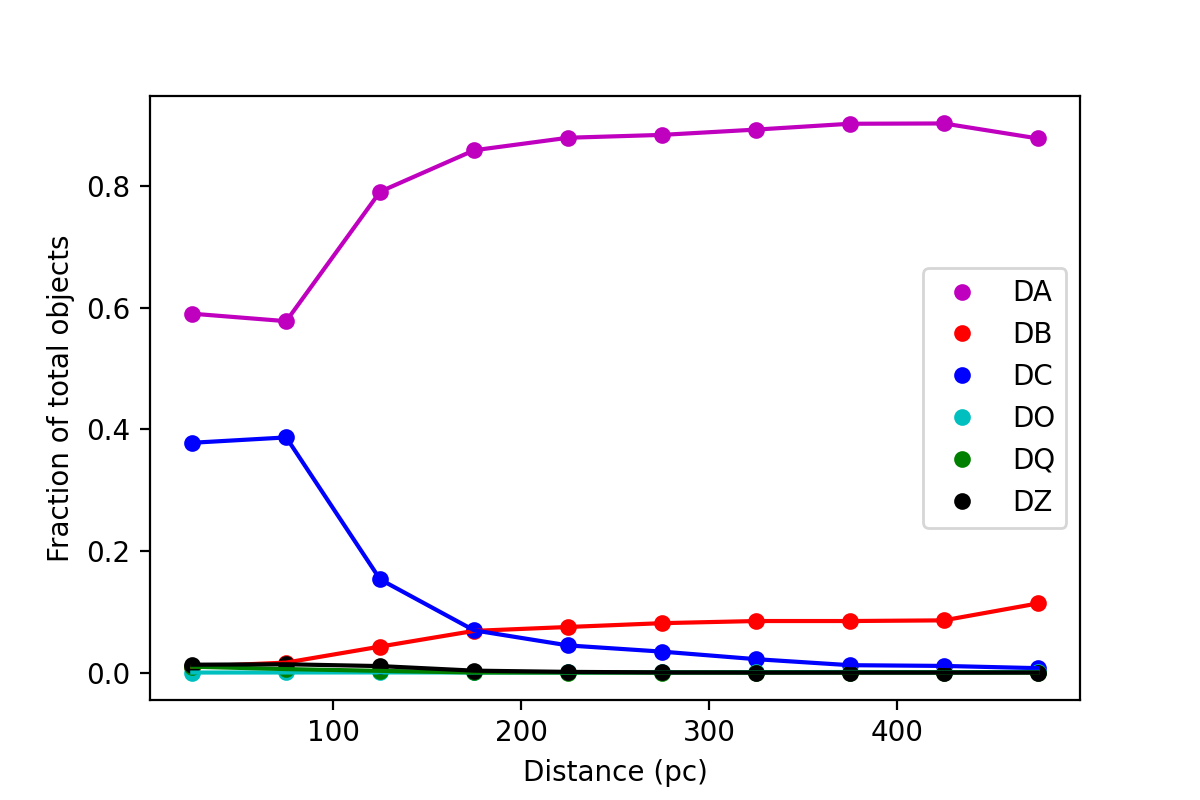}
\includegraphics[width=1\columnwidth,trim=0 0 20 20, clip]{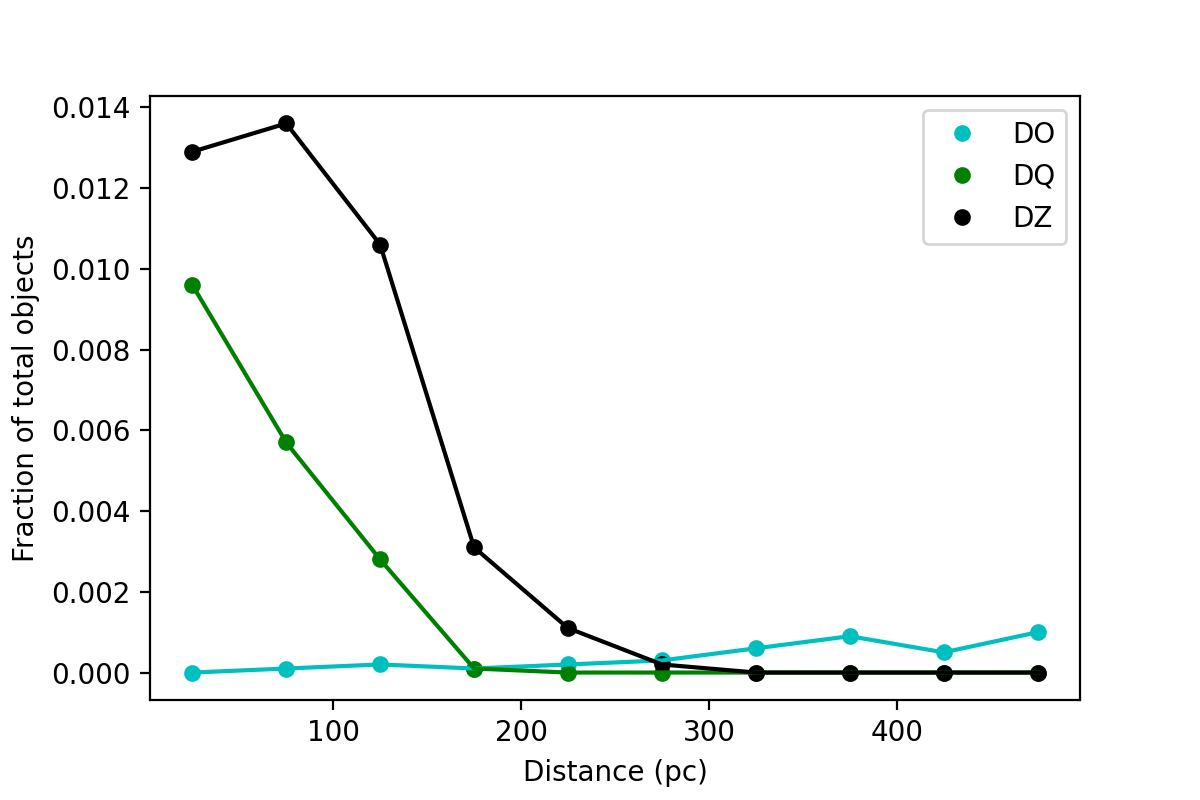}
    \caption{Top panel: Fraction of white dwarfs with different spectral type with respect to the total number of classified objects as a function of distance. Bottom panel: Same, but only for DOs, DQs and DZs.}
    \label{f:Fractions_distance}
\end{figure}

The most notable difference between the proportions in both populations is that, while in a 100 pc radius DAs comprise barely above half of all objects and only slightly more than 1\% are DBs, these percentages raise to over 80\% for DAs and 6\% for DBs in the 500 pc sample. This increase is a clear observational bias that can be understood as follows. As we move towards farther distances, the colder and fainter objects are more likely to fall below the observability threshold. This is clearly revealed by the dramatic drop of DC (intrinsically cool) white dwarfs from around 40\% at 100 pc to nearly 10\% at 500 pc. On the other hand, the observable white dwarfs will be increasingly the most luminous. These will be the hottest objects, which belong preferentially to the DA, DB and DO spectral types. It is also worth noting that the percentages of DQ and DZ white dwarfs also decrease considerably from 100 to 500 pc, indicating that these objects are generally cool and hence more affected by observational biases when observed at greater distances.

\subsection{Spectral types as a function of the \BPRP color}

In this section we analyze the percentages of white dwarfs with different spectral types of the classified 500-pc sample for five intervals of \BPRP color: \BPRP$\le 0$; $0\le$ \BPRP$\le 0.5$; $0.5\le $ \BPRP$\le 1$; $1\le$ \BPRP$\le 1.5$; and $1.5\le $ \BPRP. Given that the considered color is a good indicator of effective temperature, this exercise is a proxy of the spectral evolution of white dwarfs as they cool down. The proportion of objects for each interval is given in Table \ref{tab:Prop_col} and are illustrated in Figure \ref{f:Fractions_colour}.

\begin{table*}
    \caption{Spectral type percentages according to {\it Gaia} color}
    \label{tab:Prop_col}
\begin{center}
    \begin{tabular}{ccccccc}
            \noalign{\smallskip}
            \hline\hline
            \noalign{\smallskip}
        Interval & DA & DB & DC & DO & DQ & DZ\\
            \noalign{\smallskip}
            \hline
            \noalign{\smallskip}
     \BPRP$\le 0$ & 86.48 & 12.33 & 1.10 & 0.07 & 0.004 & 0.007\\
        $0\le$ \BPRP$\le 0.5$ & 85.39 & 4.34 & 9.58 & 0 & 0.08 & 0.62\\
        $0.5\le$ \BPRP$\le 1$ & 83.07 & 0.007 & 15.39 & 0 & 0.45 & 1.07\\
        $1\le$ \BPRP$\le 1.5$ & 11.51 & 0 & 87.61 & 0 & 0.56 & 0.31\\
        $1.5\le$ \BPRP & 0 & 0 & 100 & 0 & 0 & 0\\
        \noalign{\smallskip}
        \hline
    \end{tabular}
\end{center}
\end{table*}

As expected, the fraction of DAs is nearly constant and over 80\% until \BPRP$\sim1$. At this point, it starts to decrease to just over 11\% in the $1\le$ \BPRP$\le 1.5$ interval, and it becomes null for $1.5\le$ \BPRP. This mirrors the spectral evolution of DA white dwarfs. That is, at \BPRP$\sim1$, the temperatures are too cool for the white dwarfs to display Balmer lines in their spectra and they become DCs. Indeed, the dramatic drop of DAs at this limit is a direct consequence of the large increase of DCs, the fraction of which raises from just barely over 10\% to more than 80\% for \BPRP$>1$.

The same tendency observed for DAs is also true for DBs, which constitute over 12\% of the objects when \BPRP$\le 0$; just above 4\% when $0\le$ \BPRP$\le 0.5$. A similar percentage increase is observed among DCs, which clues us into this interval containing the DB-DC transition region. The few DOs, in turn, appear only at \BPRP$\le 0$.

The temperature distribution of DZs is also coherent, being over 0.3\% in the $0\le$ \BPRP$\le 1.5$, negligible outside it, and reaching a peak abundance over 1\% in the $0.5\le$ \BPRP$\le 1$ interval. This is consistent with the spectral behavior of metals: at higher temperatures, they enter higher ionization states, the spectral lines of which are located in the ultraviolet region.

Lastly, DQs only show a noticeable fraction in the $0.5\le$ \BPRP$\le 1.5$ interval (0.45\% in the hotter half and 0.56\% in the cooler half, respectively). It is also worth noting the continuous increase of DQ white dwarfs for redder \BPRP colors. This behavior is expected, since these white dwarfs are generally cool objects; and convection, which dredges carbon up into the white dwarf atmosphere, is more efficient at lower temperatures.

\begin{figure}[h!]
\includegraphics[width=1\columnwidth,trim=10 0 20 20, clip]{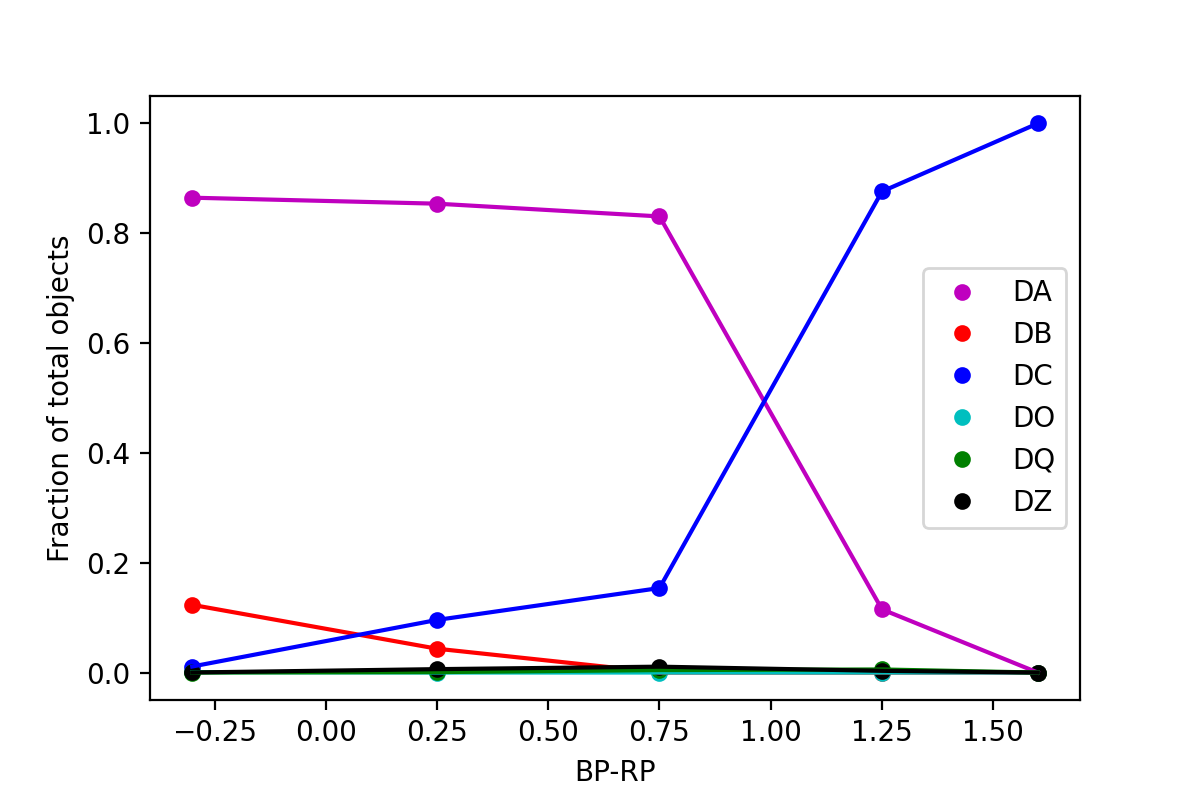}
\includegraphics[width=1\columnwidth,trim=0 0 20 20, clip]{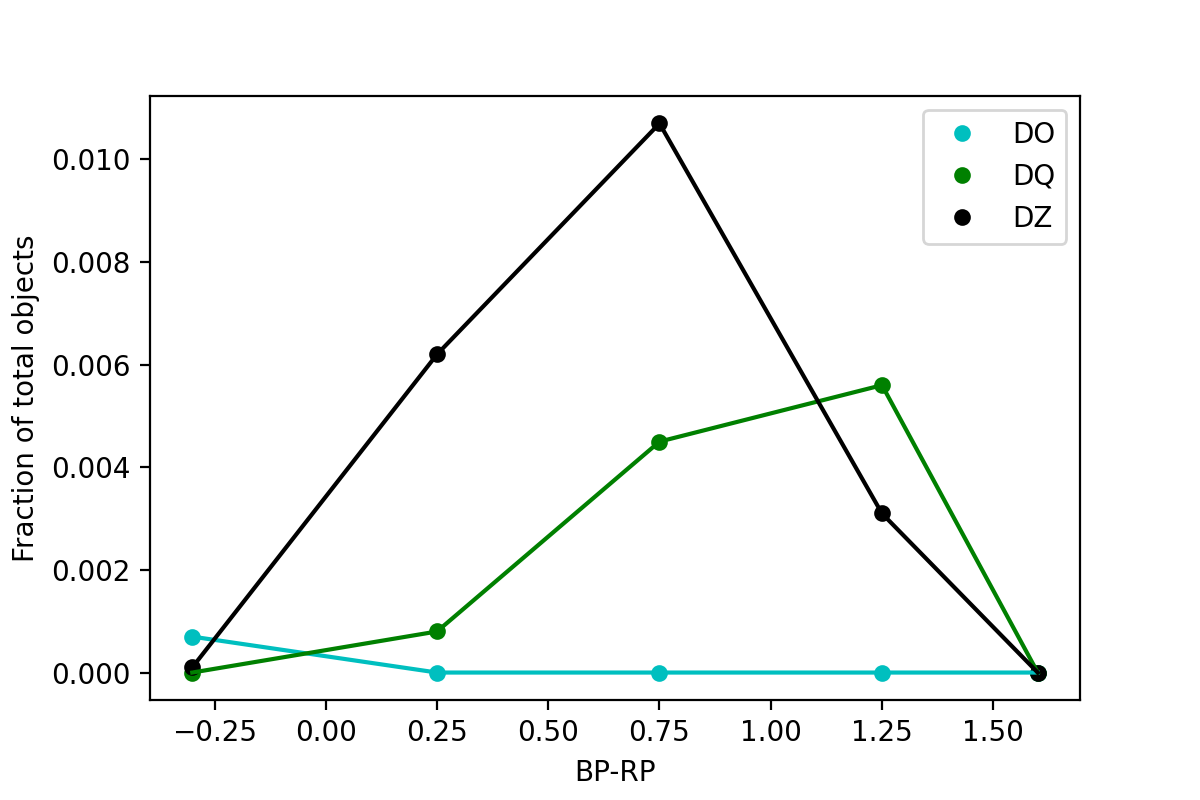}
    \caption{Top panel: The fraction of white dwarfs with different spectral type with respect to the total number of classified objects as a function of \BPRP color. Bottom panel: Same, but only for DOs, DQs and DZs.}
    \label{f:Fractions_colour}
\end{figure}

\subsection{The spectral content of the A, B and Q-branches}

One of the unexpected features revealed by the \emph{Gaia} space mission was the presence of remarkably distinct branches in the color-magnitude diagram of white dwarfs \citep[see Figure 13, ][]{Gaia2018}. While the so-called A branch follows the classical white dwarf cooling sequence for hydrogen-pure atmospheres, additional branches, such as the B and Q branches, deviate from it, indicating the influence of other physical processes, such as atmospheric composition changes, crystallization, and distillation effects, among others (see, for instance, \citealt{Camisassa2021}; \citealt{Camisassa2023}); \citealt{Tremblay2019}). We should recall that the branches are not composed of a single spectral type (i.e., the A-branch also includes non-DAs, and vice versa). Hence, a detailed and thorough characterization of the spectral types of these branches is of crucial importance for constraining theoretical models.

\subsubsection{A branch}  

We define the A branch as limited to the color region between $0.1\le$ \BPRP$\le0.5$ and bounded by the parallel lines $M_{\rm G}=3.5\cdot($\BPRP$)+11.9$ and  $M_{\rm G}=3.5\cdot($\BPRP$)+11.5$ (see red box in Figure \ref{f:A_B_Branches}).

In the classification using our Random Forest algorithm, we found that, for the 500 pc sample, 8\,904 out of the 78\,920 objects are classified as belonging to the A branch. Of them, 8\,088 (90.84\%) are DAs, 149 (1.67\%) are DBs, 626 (7.03)\% are DCs, 5 (0.06\%) are DQs, and  36 (0.4\%) are DZs. No DOs, as expected, appear in the A branch.

When these objects classified by our algorithm are added to the already labeled sample from MWDD, we find that 10\,599 objects belong to this branch: 9\,569 DAs (90.28\%), 162 DBs (1.53\%), 747 DCs (7.05\%), 45 DQs (0.42\%), and 76 DZs (0.72\%).

The results found here are in perfect agreement with those presented in the literature, which indicate that the A branch is predominantly composed of hydrogen-pure atmosphere white dwarfs \citep[e.g.][]{Jimenez2023}. Moreover, the DA versus non-DA proportion, both in the 100 pc and 500 pc samples, remains nearly identical, with the non-DA types constituting only a very small fraction of the total (less than 10\%). Among these few non-DAs, the majority are DBs and DCs, consistent with the expected spectral evolution of helium-atmosphere white dwarfs \citep[see Figure 13 in ][]{Bedard2024}.

\begin{figure}[h!]
\includegraphics[width=1\columnwidth,trim=-10 0 0 0, clip]{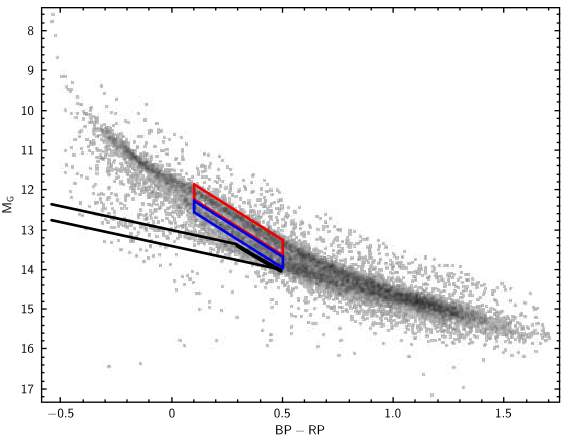}
\includegraphics[width=1\columnwidth,trim=-10 0 0 0, clip]{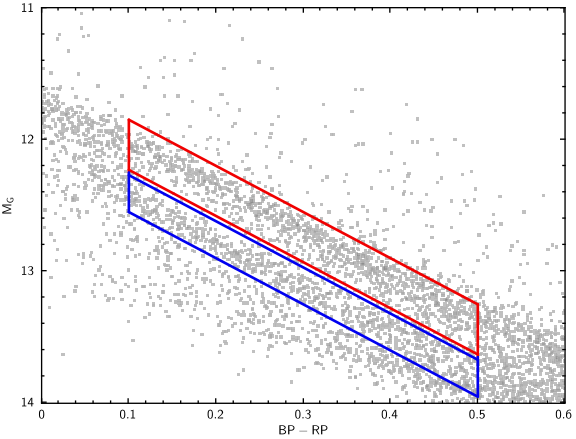}
    \caption{Top panel: the adopted definition for the A (red), B (blue) and Q (black) branches in this analysis. Bottom panel: Same, but zoomed in the A and B branches. For clarity, the regions have been represented over the 100-pc white dwarf sample of \citep{Garcia2023}.}
    \label{f:A_B_Branches}
\end{figure}

\subsubsection{B branch}

Similarly to the A branch, the B branch is defined by the color region between $0.1\le$ \BPRP$\le0.5$, but now limited by the lines $M_{\rm G}=3.5\cdot($\BPRP$)+11.9$ and  $M_{\rm G}=3.5\cdot($\BPRP$)+12.2$ (see blue box in Figure \ref{f:A_B_Branches}).

We found 4\,400 objects in the B branch, which were classified by algorithm as follows: 2\,877 DAs (65.39\%), 224 DBs (5.09\%), 1\,171 are DCs (26.61\%), 3 DQs (0.07\%), and 125 DZs (2.84\%).

When all classified objects in 500 pc are considered (those of our algorithm and those previously labeled)}, out of the 5\,319 objects in the B branch, 3\,190 (59.97\%) are DAs, 229 (4.21\%) are DBs, 1\,470 (27.64\%) are DCs, 160 are DQs (3.01\%), and 270 (5.08\%) are DZs.

It is important to note that in the 100 pc classification, however, the ratio of classified DAs to non-DAs in the B branch is close to the 35\%-65\% ratio, respectively, reported by \citet{Jimenez2023}. However, for the 500 pc sample, the classified DA population predominates, making up nearly 60\% of the B branch. It is possible that distance-related observational bias plays a role in this result.

\begin{figure}[h!]
\includegraphics[width=1\columnwidth,trim=-20 0 0 0, clip]{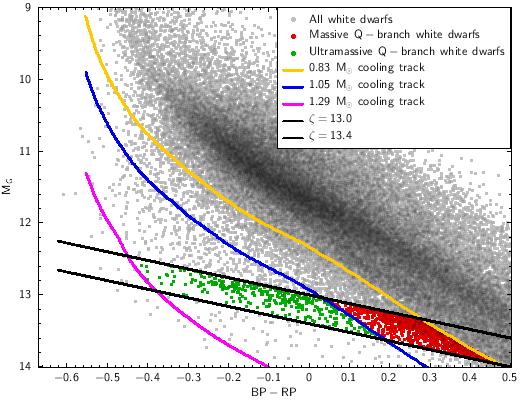}
   \caption{Massive and ultramassive Q-branch objects and selected DA cooling tracks used in \cite{Camisassa2019}. Objects found between the $\zeta=13.0$, $\zeta=13.4$ and the 0.83 and 1.05 $M_{\odot}$ cooling tracks are considered to belong to the massive Q branch; objects found between the $\zeta=13.0$, $\zeta=13.4$ and the 1.05 and 1.29 $M_{\odot}$ cooling tracks are considered to belong to the massive Q branch.}
    \label{f:MassiveUltramassive}
\end{figure}

\subsubsection{Q branch}
\label{q-branch}

In this section we analyze the fraction of white dwarfs with different spectral types in the Q-branch.  Following the analysis in \citet{Cheng2019} and \citet{Camisassa2021}, we first define the parameter $\zeta = M_{\rm G} - 1.2 \cdot (G_{\rm BP} - G_{\rm RP})$. The Q-branch is then defined as the region of the color-magnitude diagram between $13.0 \leq \zeta \leq 13.4$.

Furthermore, within the Q-branch, we defined two distinct regions to differentiate between massive and ultra-massive white dwarfs. The selection is based on the cooling tracks of \citet{Camisassa2019} for white dwarf masses: one region for massive white dwarfs with masses in the range from 0.83 to 1.05 $M_{\odot}$, and another for ultra-massive white dwarfs with masses between 1.05 and 1.29 $M_{\odot}$. A color-magnitude diagram displaying the cooling tracks, the constant $\zeta$ straight lines, and the identified massive and ultra-massive white dwarfs is shown in Figure \ref{f:MassiveUltramassive}.

The massive Q branch comprises a total of 965 objects, 813 (84.25\%) are DAs; 9 (0.93\%) are DBs; 125 (12.95\%) are DCs; 12 (1.24\%) are DQs, and 6 (0.62\%) are DZs. No DOs are found in the massive Q branch region.

Meanwhile, the ultramassive Q branch is formed by a total  338 objects, 259 (76.63\%) are DAs; 57 (16.86\%) are DBs; 21 (6.21\%) are DCs, and 1 (0.30\%) DQs.

Compared to the entire population distribution at 500 pc, we observe a slight decrease in the proportion of DAs, along with a 2.5-fold increase in the proportion of DBs for the ultramassive Q-branch (see subsection \ref{ss:Mass_DBs} for a more detailed analysis of this result). Similarly, when taking into account the whole (massive plus ultramassive) 500-pc Q branch, the proportion of DQs in it (2.2\%) is four times as large as the DQ ratio in the total 500-pc population (0.5\%).

Finally, while the fact that a higher percentage of DQs appears in the massive Q-branch rather than in the ultramassive Q-branch may suggest that DQs in the Q-branch are more commonly massive rather than ultramassive, we advise against drawing such a conclusion. The low number of objects found (12 and 1, respectively); as well as the low recall our Random Forest algorithm has shown for DQs, prevent us from reaching any conclusions regarding the mass distribution of DQs.

 Finally, we compared our results with those of \citet{Manser2024}, who provided spectral classifications for 288 objects in the Q branch (specifically corresponding to our definition of the ultramassive Q branch) using spectra from the DESI survey. We find a similar fraction of DAs ($\sim70\%$), but our analysis yields approximately half the number of DQs and nearly ten times the proportion of DBs reported in their study. A detailed discussion of this discrepancy can be found in Subsection \ref{ss:Mass_DBs}.

\subsection{Peculiar features of non-DA white dwarfs}

Finally, we discuss in this section two  peculiar results that arise from our classification concerning non-DA white dwarfs. First, we center our attention on the apparent deficit of DC white dwarfs at around $0.5\le $ \BPRP$\le 0.8$. Second, we discuss a seeming subpopulation ($N\approx$250) of massive objects classified as DB that are located at the top of the Q branch.

\begin{table*}[h!]
    \caption[]{Spectral types percentages of {\it Gaia} white dwarfs in the A, B and Q branches.}
    \label{tab:Prop_rama}
\begin{center}
    \begin{tabular}{lcccccc}
            \noalign{\smallskip}
            \hline\hline
            \noalign{\smallskip}
        Interval & DA & Non DA & DB & DC & DQ & DZ\\
            \noalign{\smallskip}
            \hline
            \noalign{\smallskip}
        \multicolumn{7}{c}{A branch} \\
        \noalign{\smallskip}
            \hline
            \noalign{\smallskip}
        100 pc classified & 96.49\% & 3.51\% & 0\% & 2.81\% & 0.35\% & 0.35\%\\
        100 pc total & 96.88\% & 3.12\% & 0\% & 2.30\% & 0.49\% & 0.33\%\\
        500 pc classified & 90.84\% & 9.16\% & 1.67\% & 7.03\% & 0.06\% & 0.4\%\\
        500 pc total& 90.28\% & 9.72\% & 1.53\% & 7.05\% & 0.42\% & 0.72\%\\
        \noalign{\smallskip}
        \hline
        \noalign{\smallskip}
        \multicolumn{7}{c}{B branch} \\
        \noalign{\smallskip}
            \hline
            \noalign{\smallskip}
        100 pc classified & 40.70\% & 59.30\% & 0.16\% & 49.76\% & 1.75\% & 7.63\%\\
        100 pc total & 37.80\% & 62.20\% & 0.23\% & 44.79\% & 6.76\% & 10.42\%\\
        500 pc classified & 65.39\% & 34.61\% & 5.09\% & 26.61\% & 0.07\% & 2.84\%\\
        500 pc total& 59.97\% & 40.03\% & 4.31\% & 27.64\% & 3.01\% & 5.08\%\\
        \noalign{\smallskip}
        \hline
        \noalign{\smallskip}
        \multicolumn{7}{c}{Massive Q branch} \\
        \noalign{\smallskip}
            \hline
            \noalign{\smallskip}
        100 pc classified & 73.37\% & 26.63\% & 0\% & 20.65\% & 4.89\% & 1.09\%\\
        100 pc total & 76.19\% & 23.81\% & 0\% & 18.65\% & 3.97\% & 1.19\%\\
        500 pc classified & 84.25\% & 15.75\% & 0.93\% & 12.95\% & 1.24\% & 0.62\%\\
        500 pc total& 82.54\% & 17.46\% & 0.79\% & 14.04\% & 1.32\% & 1.32\%\\
        \noalign{\smallskip}
        \hline
        \noalign{\smallskip}
        \multicolumn{7}{c}{Ultramassive Q branch} \\
        \noalign{\smallskip}
            \hline
            \noalign{\smallskip}
        100 pc classified & 72.73\% & 27.27\% & 5.45\% & 16.36\% & 5.45\% & 0\%\\
        100 pc total & 67.95\% & 32.05\% & 5.13\% & 15.38\% & 11.54\% & 0\%\\
        500 pc classified & 76.63\% & 23.37\% & 16.86\% & 6.21\% & 0.30\% & 0\%\\
        500 pc total& 74.33\% & 25.67\% & 14.67\% & 6.36\% & 4.65\% & 0\%\\
        \noalign{\smallskip}
        \hline
    \end{tabular}
\end{center}
\end{table*}

\subsubsection{A DC deficit}

An interesting feature found in this work is the emergence of an apparent deficit in the DC number in the $0.5\le $ \BPRP$\le0.85$ region (see bottom left panel of Figure \ref{f:500pc_HR} in Annex 1). This deficit, which corresponds to temperatures between 6500\,K and 5000\,K, approximately, has also been detected by \citet{Vincent2024}. Moreover, \citet{Blouin2019} also report an increase in the number of hydrogen-rich objects in this temperature range, in consonance with the DC deficit found. On the other hand, such a deficit is not observed in the 40-pc volume-limited sample of \citet{Obrien2024}, raising the possibility of this deficit being an observational bias rather than a real feature.

\subsubsection{Massive DBs}
\label{ss:Mass_DBs}

\begin{figure}
\includegraphics[width=1\columnwidth,trim=-20 0 0 0, clip]{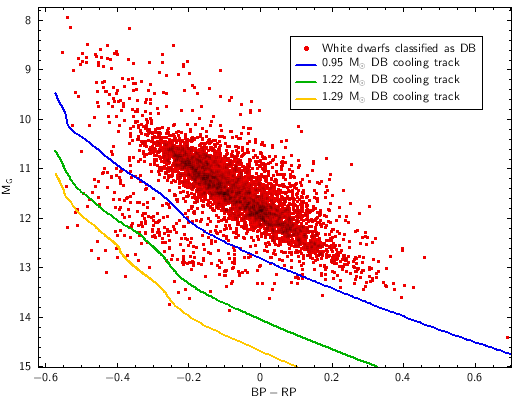}
   \caption{White dwarfs classified as DBs (red dots) and selected DB cooling tracks  from \cite{Camisassa2019}. Objects found below the 0.95 \Msun\ cooling track (blue line) are considered to belong to a massive DB subpopulation found by our algorithm.}
    \label{f:Massive_DBs}
\end{figure}

A second interesting finding from our Random Forest classification is the presence of a subsample ($N\approx250$) of massive white dwarfs classified as DBs. This subsample represents approximately 5\% of the total DB population and appears to be located over the Q branch, near its hottest tip; see Figure \ref{f:Massive_DBs}, along with selected DB cooling tracks from \citet{Camisassa2019}.

This apparently massive DB white dwarf subpopulation has not been found by any other spectroscopic works \citep[e.g.][]{Genest2019, Bergeron2011} nor by any \G\, volume-limited sample of white dwarfs \citep{Hollands2018, Garcia2023, Obrien2023}. However, the recent spectral classification by \cite{Vincent2024} of \G\, spectra also finds this subpopulation. Insight into this issue shows that our Random Forest algorithm assigns them a slightly higher probability of being magnetic DAH white dwarf as compared to non-massive DBs.  Visual inspection of the externally calibrated \G\,spectra, however, does not allow us to confirm or discard the possibility of these objects harboring a magnetic field, due to their low resolution.

Additionally, a search for these objects in the MWDD reveals that, among the few pre-classified ones, only $\sim$20\% of them are indeed DB white dwarfs, while the rest are mainly magnetic DAH white dwarfs. Other exotic objects, such as DAQ or DQA white dwarfs, are also included. In order to discern the true nature of this subpopulation, we are performing higher resolution follow-up spectroscopy of these objects.

\section{Comparison with other works}
\label{s:Comp}

In this section, we compare the classification results obtained in this work with those obtained through automatic classification algorithms, both supervised and unsupervised, in other studies. We also compare our Random Forest classification with higher resolution spectra obtained by our team.

The interest of comparing these classifications lies in the fact that they use the same input material (i.e., the 110 \G\,  spectral coefficients), but different classification algorithms, and training sets in the case of supervised machine learning algorithms. In this way, by comparing the results, including the coincidences and discrepancies, we can gain an understanding of how these classifications work, as well as their strengths and weaknesses.

\subsection{Comparison of 100-pc and 500-pc Classifications}
\label{ss:Comparison}

In this subsection we perform a comparison between the 100-pc classification we obtained in \citet{Garcia2023} and the assigned type to those same objects in the 500-pc classification. The rationale behind this comparison is that the assigned spectral type to common objects should coincide. However, since the training sets are different and contain not only different object proportions but also different number of objects in each category, a small fraction of white dwarfs with different classifications is expected.

It is important to emphasize that the 100-pc spectral classification by \cite{Garcia2023} considered as DA all white dwarfs with \BPRP$\leq0.86$ that had been classified as such in \cite{Jimenez2023}. Since the present work does not rely on any previous classification for such objects, a second 100-pc classification was derived for the purpose of this comparison, in the way as explained in Section \ref{s:classi}: a first DA vs. non-DA classification followed by a spectral type classification of the non-DAs.

\begin{figure}[ht]
\includegraphics[width=1\columnwidth,trim=35 0 30 20, clip]{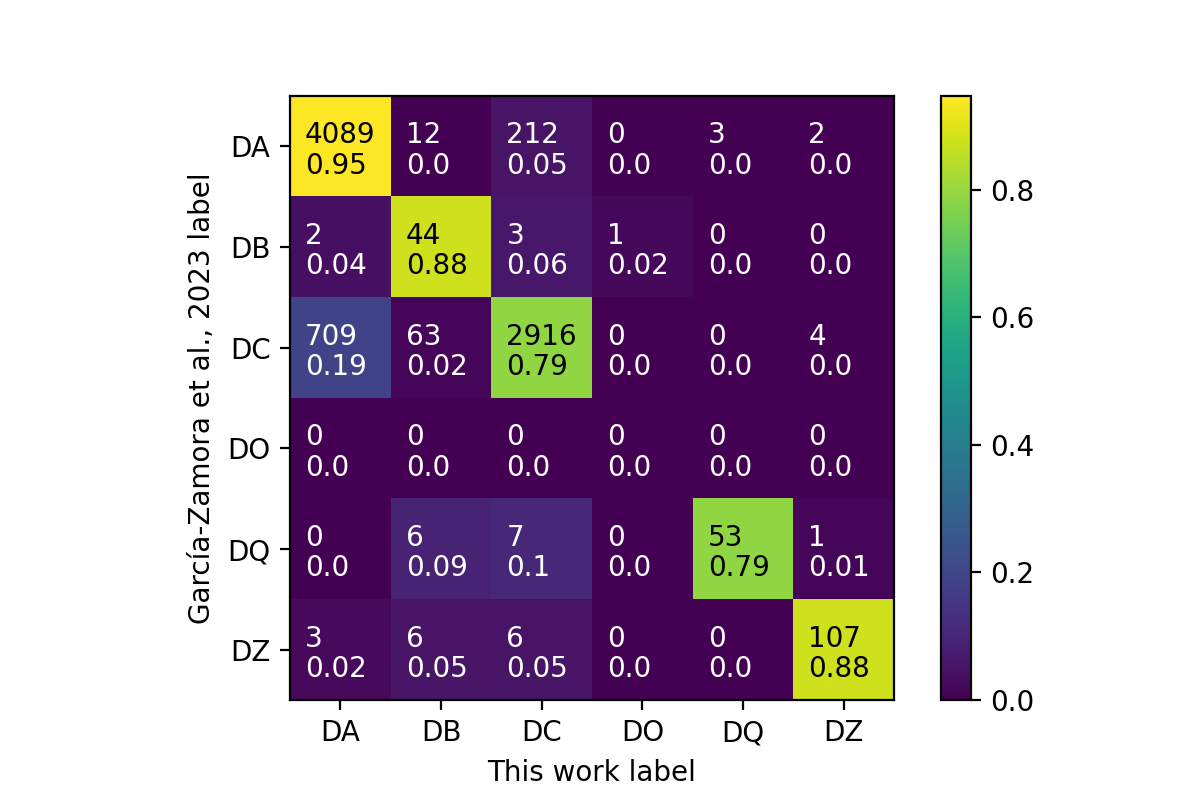}
    \caption{Confusion matrix comparing 100-pc versus 500-pc classifications. The rows represent the classification from \cite{Garcia2023}, while the columns show the classification assigned in this work.}
    \label{f:100vs500}
\end{figure}

The comparison result is shown in the confusion matrix in Figure\,\ref{f:100vs500}. 
It is clear that both classifications are largely equivalent, supported by the following metrics: an accuracy of 0.87, a balanced accuracy of 0.87, a mean F1-score of 0.80 and a G-means score of 0.85. The biggest discrepancy corresponds to the 709 objects classified as DC in the 100-pc classification, but as DA in the 500-pc classification. These are mostly cold objects, with \BPRP$>0.8$. On the other hand, the 212 objects classified as DA in the 100-pc classification, but as DC in the 500 pc classification, are mostly warm, with color $0.3< $\BPRP$<0.8$.  Numerical imbalance in the training set might be the cause of this discrepancies, as the DA-DC ratio in the training set is higher in the 500 pc sample than in the 100 pc sample.

\subsection{Comparison with \citet{Vincent2024}}
\label{ss:vin}

In this section, we compare the result of our Random Forest classification algorithm with the recent classification provided by \cite{Vincent2024}, who used a Gradient Boosted Decision Trees algorithm to spectroscopically classify 100\,886 white dwarfs with \G\, spectra. As a training set, the SDSS-\G\, catalog described in \cite{GentilleFusillo2021} was used.

\begin{figure}[ht]
\includegraphics[width=1\columnwidth,trim=35 0 30 20, clip]{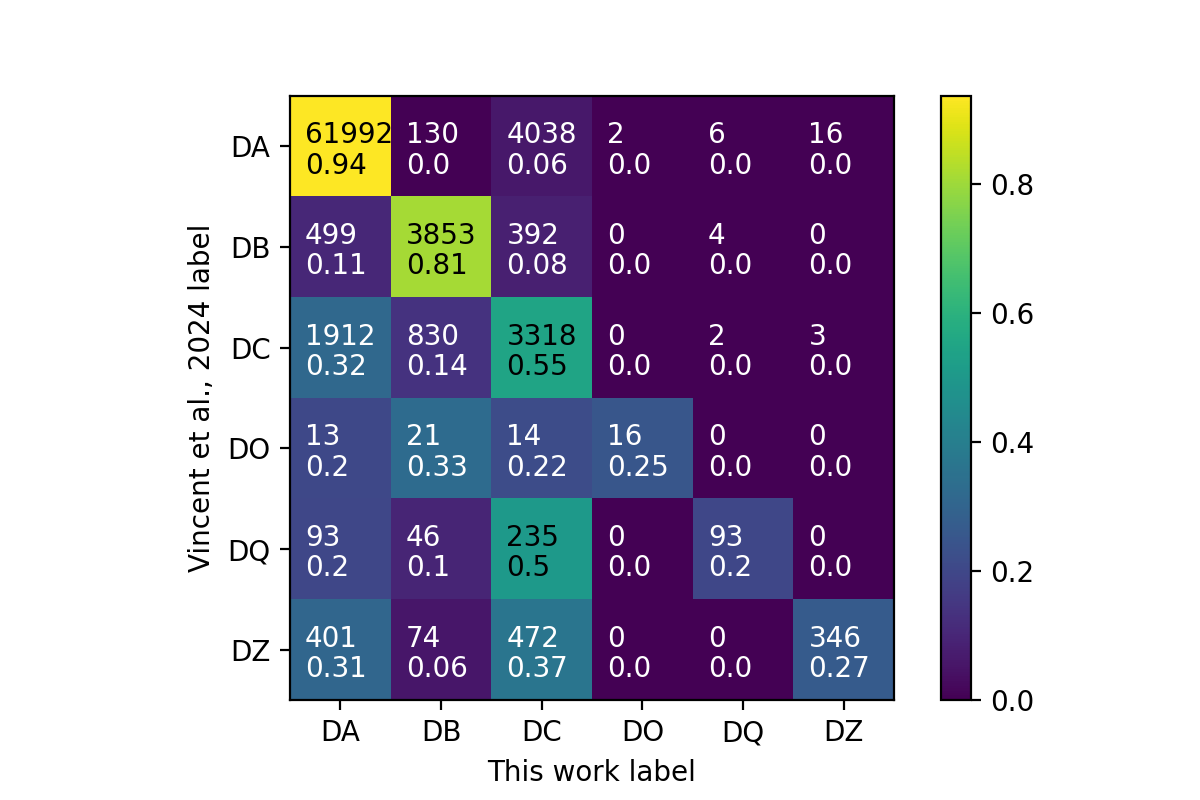}
    \caption{Confusion matrix for the comparison between this work and the classification in \cite{Vincent2024}. The rows represent the classification from \cite{Vincent2024}, while the columns show the classification assigned in this work.}
    \label{f:yovsvincent}
\end{figure}

A total of 69\,618 objects were found to be in both catalogs. The results of the comparison are collected in a confusion matrix, shown in Figure \ref{f:yovsvincent}.

From the confusion matrix, we find an excellent agreement for DAs, and a very good agreement for DBs (0.94 recall for DAs and 0.81 for DBs). On the other hand, the agreement is rather poor for DOs, DQs and DZs (recalls of 0.25, 0.2 and 0.27, respectively). Many white dwarfs classified as such in \cite{Vincent2024} are classified as DC by our algorithm. A possible explanation could be the difference in classification algorithms, as well as the different training sets. Even though both works use the 110 \G\, XP spectral coefficients as input for the automatic classifiers, the classification by \cite{Vincent2024} was trained on the spectral classification of white dwarfs from the SDSS-\G\, catalog described in \cite{GentilleFusillo2021}, whereas we used the spectral classification from the MWDD. Since the output of any automatic classifier depends on the dataset used as the training sample, different results are expected.  

It also should be noted that the low agreement for DOs, DQs and DZs is not indicative of our classification being less accurate. On the contrary, more conservative; as the objects assigned the spectral DQ and DZ types have a very high probability of belonging to them. This also implies a more realistic classification in some instances, such as the DA-DC classification at the cooler end of the spectrum.

The biggest discrepancy appears in the DA/DC classification, with more than 4\,000 objects classified as DA in \cite{Vincent2024} being classified as DC by our algorithm. This discrepancy arises mainly for the coldest white dwarfs, as from the 4\,038 discrepant objects, 2\,254 of them have \BPRP$>1$ and 3,050 have \BPRP$>0.86$. In this temperature range, hydrogen atoms are found mostly in their ground state, therefore the optical Balmer lines are not expected to be very prominent, if at all present. Therefore, we believe our DC classification to be more realistic, especially since the training sample used in \citealt{Vincent2024} does not contain many cold objects to train the algorithm.

On the other hand, the 1\,912 objects classified as DC by \cite{Vincent2024} and as DA in this work are found in the middle temperature region, with the approximate boundaries \BPRP$=0$ and \BPRP$=0.7$. A clear interpretation of this discrepancy is more difficult to find, though it is possible that, at least at the coldest end, the DC deficit is playing a role.

As for global metrics, this comparison attains an accuracy of 0.88, a balanced accuracy of 0.50; a mean F1- score of 0.56 and a G-means score of 0.42. These results show that both supervised classifications are compatible.

\subsection{Comparison with \citet{Kao2024}}
\label{ss:Kao}

Our next comparison is with the unsupervised classification presented in the recent work of \cite{Kao2024}, who applied the UMAP (Uniform Manifold Approximation and Projection for Dimension Reduction; see \citealt{McInnes2018}) algorithm to 96\,134 white dwarfs. Their 110 \G\, coefficients were reduced to two coordinates using the UMAP algorithm and later visualized in a graph. DA, DB, DO and DZ white dwarfs with spectral classification in the MWDD were used to identify clustered regions on the map. However, it is worth noting that no spectral classification was attempted for the unclassified white dwarfs. As such, the present comparison will focus on the location of our classified white dwarfs in their UMAP coordinates and the clustered regions, rather than on ascertaining how many white dwarfs share the same classification.

\begin{figure}[h!]
\includegraphics[width=1\columnwidth,trim=0 0 10 20, clip]{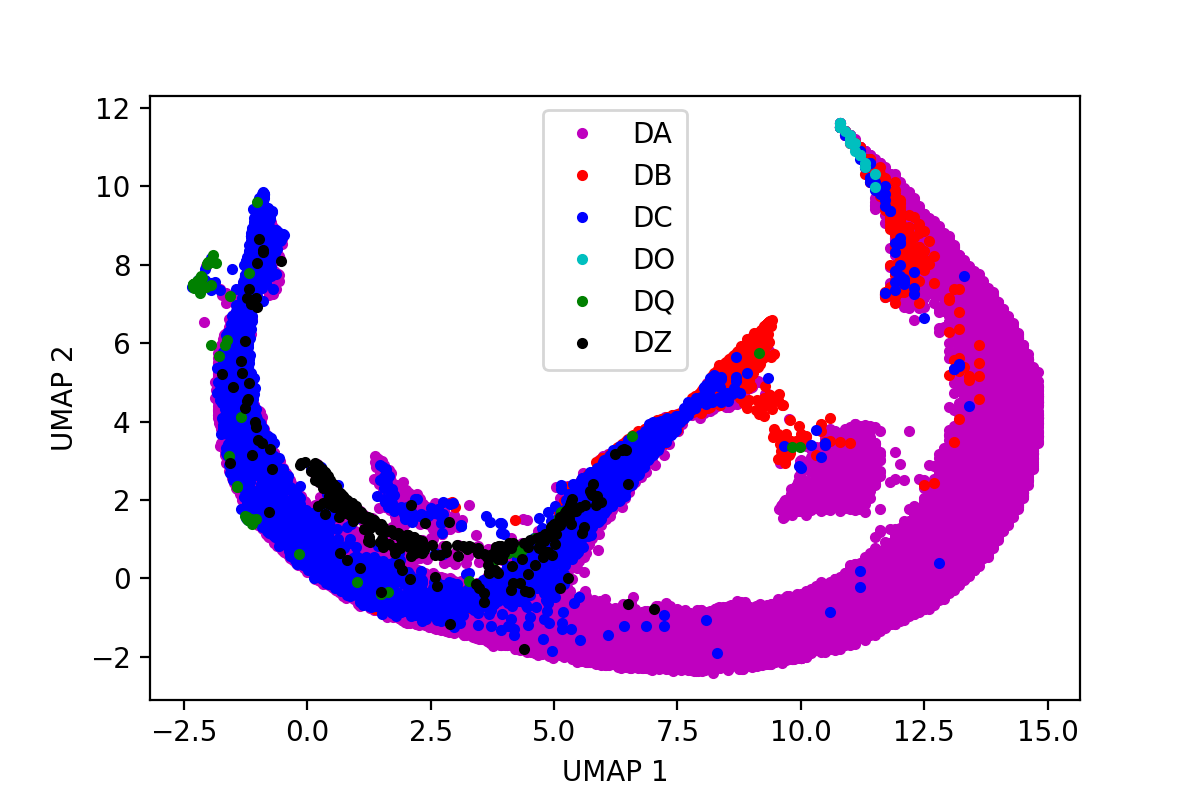}
   \caption{UMAP using the coordinates from \citet{Kao2024} and color-coded according to the spectral classification obtained in this work.}
    \label{f:Kaovsme}
\end{figure}

In order to make this comparison, we used the UMAP coordinates of \citet{Kao2024} of the 77\,417 shared white dwarfs in both classifications, and represented them using the spectral classification obtained in this work as a color code. The result can be seen in Figure\,\ref{f:Kaovsme}. Comparing this with Figure 3 of \cite{Kao2024}, we confirm that our classified white dwarfs can be found in the same regions where the MWDD classified objects by \citet{Kao2024} fall. As such, we also find the cool DZ bow in the $0\le{\rm UMAP1}\le3$ and $1\le{\rm UMAP2}\le3$ region; as well as some DZs in the lower half of the $3\le{\rm UMAP1}\le7.5$ and $1\le{\rm UMAP2}\le5$ region.

Moreover, we also find DB white dwarfs in the $7.5\le{\rm UMAP1}\le10$ and $4\le{\rm UMAP2}\le7$ region, as well as in the top right tip of the diagram, where \cite{Kao2024} finds the MWDD DB white dwarfs. DOs, both in the MWDD and found in this work, are also limited to this top right tip. In conclusion, the results of our primary spectral type classification are fully compatible with the spectral type distribution found in the UMAP of \citep{Kao2024}.

\subsection{Comparison with \citet{Perez-Couto2024}}

The last comparison contrasts this work with the unsupervised classification presented in \citet{Perez-Couto2024}, who used self-organizing maps \citep{Kohonen1982}, an unsupervised machine learning technique, to identify polluted objects in a white dwarf sample. 

\begin{figure}[h!]
\includegraphics[width=1\columnwidth,trim=35 0 30 20, clip]{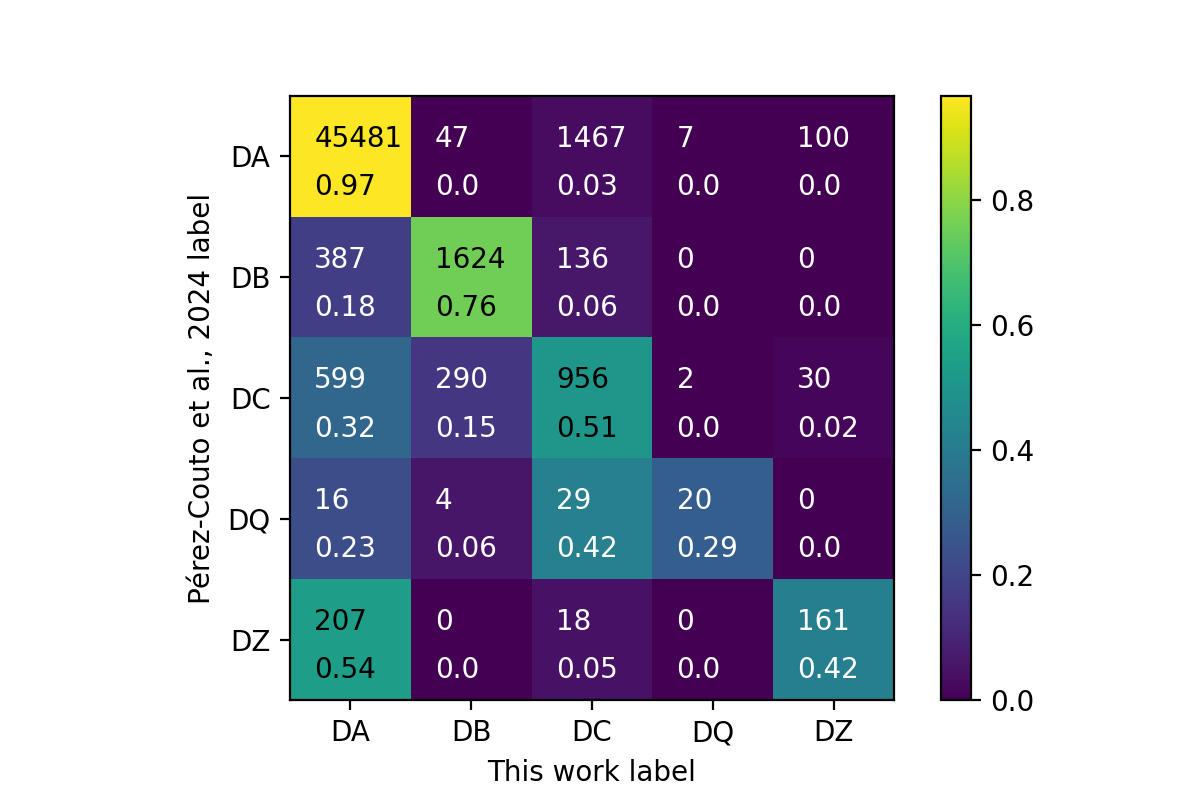}
   \caption{Confusion matrix for the comparison between this work and the classification in \cite{Perez-Couto2024}. The rows represent the classification from \cite{Perez-Couto2024}, while the columns show the classification assigned in this work.}
    \label{f:Coutovsme}
\end{figure}

In their work, they used self-organizing maps to identify 66\,337 white dwarf candidates, of which 61\,817 were then assigned primary spectral types. A cross-match with our catalog shows that 51\,581 objects have been classified in both works. To compare them, we present the corresponding confusion matrix in Figure \ref{f:Coutovsme}.

As the confusion matrix shows, the best agreements are found for DAs and DBs (97\% and 76\%, respectively), while the agreement is lower for DQs and DZs (29\% and 42\%, respectively). A possible cause is the low recall for both types in both works. While it is true that both works show high precision (>80\%) for them, this does not guarantee that the few identified objects will actually be the same in both classifications, and coincidences between them are unlikely.

This fact is supported by the global metrics (0.94 accuracy, 0.59 balanced accuracy, 0.61 mean F1-score, 0.54 G-means score). These, along with the confusion matrix, allow us to conclude the compatibility of both spectral classifications.

\subsection{A golden sample of classified white dwarfs}

From the different automatic classifications presented in this section, we can derive a golden sample; that is, a set of white dwarfs, whose primary spectral classification coincides among the different automatic classifications. The importance of such a set is that if different algorithms with different training sets assign an object the same primary spectral type, the classification can be consider as highly reliable.

The work from \citet{Kao2024} does not classify unclassified white dwarfs. While the authors of \citet{Perez-Couto2024} kindly provided a confusion matrix, no catalog with their spectral classification was published or accessible to us. Therefore, we are limited to deriving a golden sample from the work of \citet{Vincent2024}.

A golden sample is thus derived from our work and \citep{Vincent2024} using the 78\,821 common white dwarfs. The resulting populations were: 61\,992 DAs, 3\,853 DBs, 3\,318 DCs, 16 DOs, 93 DQs, and 346 DZs. The color-magnitude diagrams for the golden sample, by spectral type, can be found in Appendix \ref{s:ann2}.

It is also worth noting that certain features observed in our 500 pc classification, such as the DC deficit or the existence of the massive DB white dwarf subpopulation, also appear in the golden sample. This reinforces our conclusion that they are actual physical features, rather than artifacts introduced by the machine learning algorithms.

\subsection{Spectroscopic follow-up comparison}

Finally, in this section, we compare the predictions of the Random Forest algorithm with the spectra of selected targets for which higher-resolution spectroscopic follow-up has been conducted at the 10\,m GTC (Gran Telescopio Canarias), located in the island of La Palma. The OSIRIS (Optical System for Imaging and low-Intermediate-Resolution Integrated Spectroscopy; \citealt{Cepa2013}) instrument was used in the observations, together with the R1000B grism and the 0.6" slit width, which resulted in spectra covering the 3600-7800\AA\, wavelength range at a resolving power of $\simeq$1000. The complete sample has been observed between September 2024 and January 2025.

The observed spectra were reduced and calibrated using the pamela \citep{Marsh1989} and MOLLY packages\footnote{Developed by Tom Marsh and available at \url{https://cygnus.astro.warwick.ac.uk/phsaap/software/molly/html/INDEX.html}}, respectively, and visually inspected to assign a spectral classification. For a total of 65 objects (2 DAH, 7 DB, 7 DBA, 6 DO, 5 DQ and 38 DZ) a reliable spectral type could be assigned. The observational types of these 65 objects were then compared with the spectral type predicted by our Random Forest model, and the results are shown in the confusion matrix in Figure \ref{f:obs_vs_pred}. Finally, in Appendix \ref{a:Table}, we list the objects with the visual and the predicted spectral type.

\begin{figure}[h!]
\includegraphics[width=1\columnwidth,trim=35 0 30 20, clip]{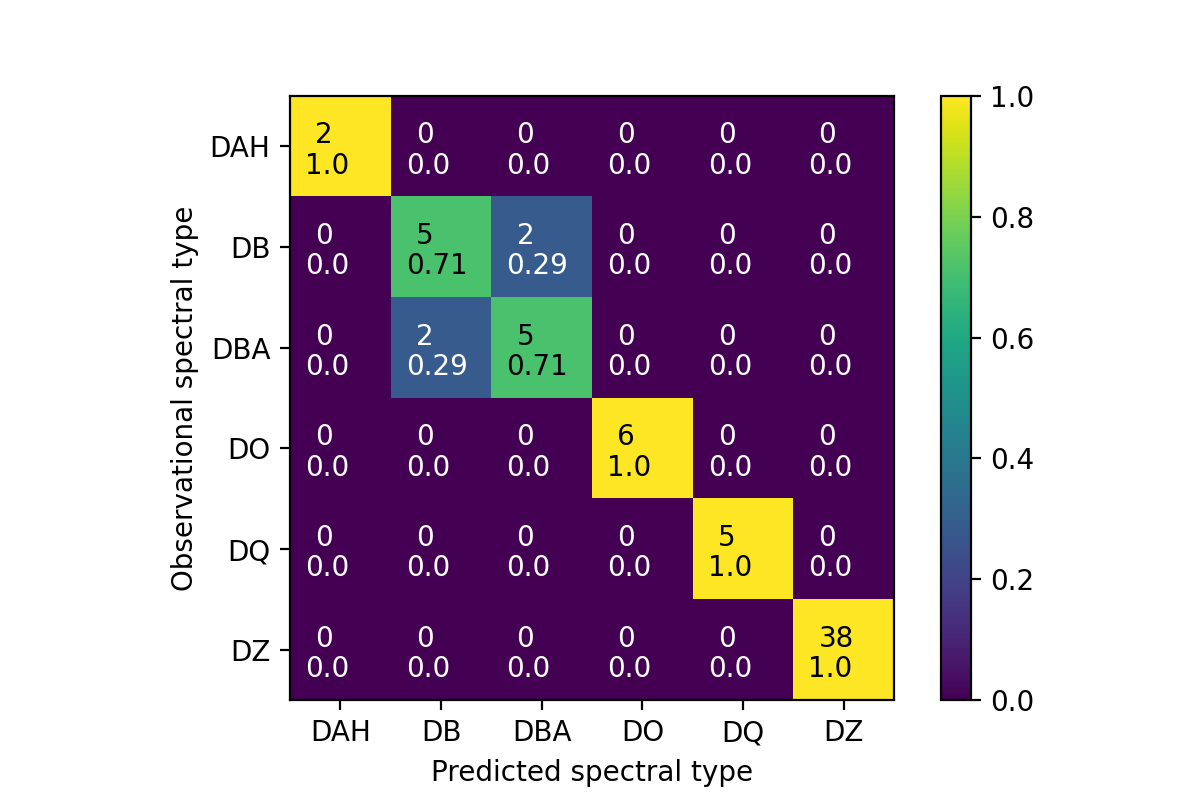}
   \caption{Confusion matrix for the comparison between the predicted spectral type of selected candidates and the observational types derived from medium resolution (R$\sim$1000) spectra obtained at GTC OSIRIS.}
    \label{f:obs_vs_pred}
\end{figure}

The results reflect the excellent precision found by our Random Forest algorithm with an accuracy of 0.94. There are minimal discrepancies between the predicted and observational types, with the only exceptions being objects classified as DB-DBA, which fall into the more ambiguous subtype classification. Additionally, some of the DO spectra present early features reminiscent of the PG 1159 type.

In Appendix \ref{a:gaiaGTC} we present examples of GTC spectra for correctly classified targets, along with the original \emph{Gaia} spectra used by our algorithm in the classification.

\section{Conclusions}
\label{s:conc}

The information contained in the coefficients of 78\,920 \G\, white dwarf spectra has been analyzed by means of an artificial intelligence, Random Forest, algorithm. In the validation tests, we have shown their usefulness for primary spectral type classification into DA, DB, DC, DO, DQ and DZ types. However, further categorization into subspectral types or binarity is found to be rather limited, due to the low \G\, spectral resolution in the subtype classification and the white dwarf component spectral dominance in the binarity case. Despite this, 6 DAH, 67 DBA and 9 DOZ (2.31\%, 23\% and 111\% increase with respect to the training set, respectively) candidates have been found.

A summary of our main finding is as follows:

\begin{enumerate}
\item A total of 78\,920 objects have been classified into their primary types. Of these, 64\,976 have been identified as DA (6 of which as DAH), 4\,957 as DB (66 of which as DBA), 8\,496 as DC, 21 as DO (9 of which as DOZ), 105 as DQ and 365 as DZ.
\item Our algorithm shows an excellent recall for DA, DB and DC WDs ($\geq$90\%), good recall for DOs ($\sim$60\%), and improvable recall for DQs and DZs (<35\%).
\item Despite their low recall, DQs and DZs show a very good precision ($\geq$80\%), which is also true for DOs.
\item The high precision achieved for DO, DQ and DZ types implies that, even though not all white dwarfs belonging to these spectral types are found, those which are will, with a very high probability, belong to those types.
\item With the possible exception of the DAH, DBA and DOZ subtypes, our algorithm does not seem to be able to recognize any spectral subtypes. This result is not entirely unexpected, as non-prominent spectral lines that define the spectral subtypes are not expected to be detectable in \G\, low resolution spectra.
\item Our algorithm does not seem to be able to recognize binary candidates in the white dwarf region of the color-magnitude diagram. Of the 78\,791 classified objects, only two are classified as WDMS candidates.
\item  Our Random Forest algorithm achieved an accuracy of 0.94 based on the comparison with 65 selected objects for which high-resolution GTC spectroscopic follow-up was performed, with minimal discrepancies mainly in the DB-DBA subtype classification.
\end{enumerate}

We also outline the physical implications of our analysis in the following points:

\begin{enumerate}
    \item We built a golden sample of 69\,618 objects, equally and independently classified by both our algorithm and the work of \citet{Vincent2024}, comprising 88.21\% of the classified set in the \G\ white dwarf catalog. Since two different automatic classifications assign the golden sample objects the same primary spectral type, we can conclude that this classification is highly reliable.
    
    \item DAs are found to constitute more than 80\% of all white dwarfs up to \BPRP$=1.0$ which  corresponds to an effective temperature of approximately 5\,000\,K; this proportion is higher than in volume-limited samples, and may result from the inherent observational bias of a magnitude-limited sample. For redder colors, DCs are the majority in line with the known spectral evolution of white dwarfs.

    \item In the A branch, DAs represent more than 90\% of the objects, in agreement with the findings of \cite{Jimenez2023}; whereas in the B branch they account for approximately $60\%$  of the objects, contrary to the 35\% ratio found in the 100-pc sample reported by \cite{Jimenez2023} The observational bias may also be the cause of the B-branch abundance discrepancies between volume-limited and magnitude-limited samples.

    \item The DA proportion in the Q branch is found to be $\sim$70\%, in line with the findings in \cite{Manser2024}. Additionally, the DQ ratio in the Q-branch is found to be approximately four times the DQ ratio in the total 500 pc population.
    
    \item We have identified a subpopulation ($N\sim250$) of massive objects classified as DB white dwarfs, 164 of which also appear in \cite{Vincent2024}. A search of these objects in the MWDD reveals that, among the few classified objects of this subpopulation, 20\% at most are classified as DB. As a means of probing into the true nature of these objects, we are performing follow-up observations of higher resolution, which are essential for a better understanding of this result, as well as comprehension of their true nature. 
    
    \item In both our 500 pc classification and the golden sample we find a decrease in the number of DCs in the $0.5\le$ \BPRP$\le 1$ region. 
\end{enumerate}

\begin{acknowledgements}
We acknowledge support from MINECO under the PID2023-148661NB-I00 grant and by the AGAUR/Generalitat de Catalunya grant SGR-386/2021. Enrique Miguel García Zamora also acknowledges financial support from Banco de Santander, under a Becas Santander Investigación/Ajuts de Formació de Professorat Universitari (2022\_FPU‐UPC\_16) grant. Based on observations made with the Gran Telescopio Canarias (programme GTC6-21B), installed in the Spanish Observatorio del Roque de los Muchachos of the Instituto de Astrofísica de Canarias, in the island of La Palma.
\end{acknowledgements}

\section*{Data Availability Statement}
The data underlying this article are available in the article.  Supplementary material will be shared on reasonable request to the corresponding author.

\bibliographystyle{aa}
\bibliography{RF500pc}

\begin{thebibliography}{55}
\expandafter\ifx\csname natexlab\endcsname\relax\def\natexlab#1{#1}\fi

\bibitem[{{Althaus} {et~al.}(2010){Althaus}, {C{\'o}rsico}, {Isern}, \& {Garc{\'{\i}}a-Berro}}]{Althaus2010}
{Althaus}, L.~G., {C{\'o}rsico}, A.~H., {Isern}, J., \& {Garc{\'{\i}}a-Berro}, E. 2010, \aapr, 18, 471

\bibitem[{{Bayo} {et~al.}(2008){Bayo}, {Rodrigo}, {Barrado Y Navascu{\'e}s}, {Solano}, {Guti{\'e}rrez}, {Morales-Calder{\'o}n}, \& {Allard}}]{VOSA2008}
{Bayo}, A., {Rodrigo}, C., {Barrado Y Navascu{\'e}s}, D., {et~al.} 2008, \aap, 492, 277

\bibitem[{{B{\'e}dard}(2024)}]{Bedard2024}
{B{\'e}dard}, A. 2024, \apss, 369, 43

\bibitem[{{Bergeron} {et~al.}(2019){Bergeron}, {Dufour}, {Fontaine}, {Coutu}, {Blouin}, {Genest-Beaulieu}, {B{\'e}dard}, \& {Rolland}}]{Bergeron2019}
{Bergeron}, P., {Dufour}, P., {Fontaine}, G., {et~al.} 2019, \apj, 876, 67

\bibitem[{{Bergeron} {et~al.}(2011){Bergeron}, {Wesemael}, {Dufour}, {Beauchamp}, {Hunter}, {Saffer}, {Gianninas}, {Ruiz}, {Limoges}, {Dufour}, {Fontaine}, \& {Liebert}}]{Bergeron2011}
{Bergeron}, P., {Wesemael}, F., {Dufour}, P., {et~al.} 2011, \apj, 737, 28

\bibitem[{{Blouin} {et~al.}(2023){Blouin}, {B{\'e}dard}, \& {Tremblay}}]{Blouin2023}
{Blouin}, S., {B{\'e}dard}, A., \& {Tremblay}, P.-E. 2023, \mnras, 523, 3363

\bibitem[{{Blouin} {et~al.}(2019){Blouin}, {Dufour}, {Thibeault}, \& {Allard}}]{Blouin2019}
{Blouin}, S., {Dufour}, P., {Thibeault}, C., \& {Allard}, N.~F. 2019, \apj, 878, 63

\bibitem[{{Breedt} {et~al.}(2017){Breedt}, {Steeghs}, {Marsh}, {Gentile Fusillo}, {Tremblay}, {Green}, {De Pasquale}, {Hermes}, {G{\"a}nsicke}, {Parsons}, {Bours}, {Longa-Pe{\~n}a}, \& {Rebassa-Mansergas}}]{Breedt2017}
{Breedt}, E., {Steeghs}, D., {Marsh}, T.~R., {et~al.} 2017, \mnras, 468, 2910

\bibitem[{{Breiman}(2001)}]{Breiman2001}
{Breiman}, L. 2001, Machine Learning, 45, 5

\bibitem[{{Camisassa} {et~al.}(2023){Camisassa}, {Torres}, {Hollands}, {Koester}, {Raddi}, {Althaus}, \& {Rebassa-Mansergas}}]{Camisassa2023}
{Camisassa}, M., {Torres}, S., {Hollands}, M., {et~al.} 2023, \aap, 674, A213

\bibitem[{{Camisassa} {et~al.}(2019){Camisassa}, {Althaus}, {C{\'o}rsico}, {De Ger{\'o}nimo}, {Miller Bertolami}, {Novarino}, {Rohrmann}, {Wachlin}, \& {Garc{\'\i}a-Berro}}]{Camisassa2019}
{Camisassa}, M.~E., {Althaus}, L.~G., {C{\'o}rsico}, A.~H., {et~al.} 2019, \aap, 625, A87

\bibitem[{{Camisassa} {et~al.}(2021){Camisassa}, {Althaus}, {Torres}, {C{\'o}rsico}, {Rebassa-Mansergas}, {Tremblay}, {Cheng}, \& {Raddi}}]{Camisassa2021}
{Camisassa}, M.~E., {Althaus}, L.~G., {Torres}, S., {et~al.} 2021, \aap, 649, L7

\bibitem[{{Carrasco} {et~al.}(2021){Carrasco}, {Weiler}, {Jordi}, {Fabricius}, {De Angeli}, {Evans}, {van Leeuwen}, {Riello}, \& {Montegriffo}}]{Carrasco2021}
{Carrasco}, J.~M., {Weiler}, M., {Jordi}, C., {et~al.} 2021, \aap, 652, A86

\bibitem[{{Cepa} {et~al.}(2013){Cepa}, {Bongiovanni}, {P{\'e}rez Garc{\'\i}a}, {Ederoclite}, {Gonz{\'a}lez-Serrano}, {Gonz{\'a}lez}, {S{\'a}nchez-Portal}, {Alfaro}, \& {Cabrera-Lavers}}]{Cepa2013}
{Cepa}, J., {Bongiovanni}, A., {P{\'e}rez Garc{\'\i}a}, A.~M., {et~al.} 2013, in Highlights of Spanish Astrophysics VII, ed. J.~C. {Guirado}, L.~M. {Lara}, V.~{Quilis}, \& J.~{Gorgas}, 868--873

\bibitem[{{Cheng} {et~al.}(2019){Cheng}, {Cummings}, \& {M{\'e}nard}}]{Cheng2019}
{Cheng}, S., {Cummings}, J.~D., \& {M{\'e}nard}, B. 2019, \apj, 886, 100

\bibitem[{Cunningham {et~al.}(2020)Cunningham, Tremblay, Gentile Fusillo, Hollands, \& Cukanovaite}]{Cunningham2020}
Cunningham, T., Tremblay, P.-E., Gentile Fusillo, N.~P., Hollands, M., \& Cukanovaite, E. 2020, Monthly Notices of the Royal Astronomical Society, 492, 3540

\bibitem[{{De Angeli} {et~al.}(2023){De Angeli}, {Weiler}, {Montegriffo}, {Evans}, {Riello}, {Andrae}, {Carrasco}, {Busso}, {Burgess}, {Cacciari}, {Davidson}, {Harrison}, {Hodgkin}, {Jordi}, {Osborne}, {Pancino}, {Altavilla}, {Barstow}, {Bailer-Jones}, {Bellazzini}, {Brown}, {Castellani}, {Cowell}, {Delchambre}, {De Luise}, {Diener}, {Fabricius}, {Fouesneau}, {Fr{\'e}mat}, {Gilmore}, {Giuffrida}, {Hambly}, {Hidalgo}, {Holland}, {Kostrzewa-Rutkowska}, {van Leeuwen}, {Lobel}, {Marinoni}, {Miller}, {Pagani}, {Palaversa}, {Piersimoni}, {Pulone}, {Ragaini}, {Rainer}, {Richards}, {Rixon}, {Ruz-Mieres}, {Sanna}, {Sarro}, {Rowell}, {Sordo}, {Walton}, \& {Yoldas}}]{DeAngeli2022}
{De Angeli}, F., {Weiler}, M., {Montegriffo}, P., {et~al.} 2023, \aap, 674, A2

\bibitem[{{de Jong} {et~al.}(2019){de Jong}, {Agertz}, {Berbel}, {Aird}, {Alexander}, {Amarsi}, {Anders}, {Andrae}, {Ansarinejad}, {Ansorge}, \& et~al.}]{4MOST2019}
{de Jong}, R.~S., {Agertz}, O., {Berbel}, A.~A., {et~al.} 2019, The Messenger, 175, 3

\bibitem[{{Dufour} {et~al.}(2017){Dufour}, {Blouin}, {Coutu}, {Fortin-Archambault}, {Thibeault}, {Bergeron}, \& {Fontaine}}]{Dufour2017}
{Dufour}, P., {Blouin}, S., {Coutu}, S., {et~al.} 2017, in Astronomical Society of the Pacific Conference Series, Vol. 509, 20th European White Dwarf Workshop, ed. P.~E. {Tremblay}, B.~{Gaensicke}, \& T.~{Marsh}, 3

\bibitem[{{Echeverry} {et~al.}(2022){Echeverry}, {Torres}, {Rebassa-Mansergas}, \& {Ferrer-Burjachs}}]{Echeverry2022}
{Echeverry}, D., {Torres}, S., {Rebassa-Mansergas}, A., \& {Ferrer-Burjachs}, A. 2022, \aap, 667, A144

\bibitem[{{Farihi} {et~al.}(2010){Farihi}, {Barstow}, {Redfield}, {Dufour}, \& {Hambly}}]{Farihi2010}
{Farihi}, J., {Barstow}, M.~A., {Redfield}, S., {Dufour}, P., \& {Hambly}, N.~C. 2010, \mnras, 404, 2123

\bibitem[{{Gaia Collaboration} {et~al.}(2018){Gaia Collaboration}, {Babusiaux}, {van Leeuwen}, {Barstow}, {Jordi}, {Vallenari}, {Bossini}, {Bressan}, {Cantat-Gaudin}, {van Leeuwen}, {Brown}, {Prusti}, {de Bruijne}, {Bailer-Jones}, {Biermann}, {Evans}, {Eyer}, {Jansen}, {Klioner}, {Lammers}, {Lindegren}, {Luri}, {Mignard}, {Panem}, {Pourbaix}, {Randich}, {Sartoretti}, {Siddiqui}, {Soubiran}, {Walton}, {Arenou}, {Bastian}, {Cropper}, {Drimmel}, {Katz}, {Lattanzi}, {Bakker}, {Cacciari}, {Casta{\~n}eda}, {Chaoul}, {Cheek}, {De Angeli}, {Fabricius}, {Guerra}, {Holl}, {Masana}, {Messineo}, {Mowlavi}, {Nienartowicz}, {Panuzzo}, {Portell}, {Riello}, {Seabroke}, {Tanga}, {Th{\'e}venin}, {Gracia-Abril}, {Comoretto}, {Garcia-Reinaldos}, {Teyssier}, {Altmann}, {Andrae}, {Audard}, {Bellas-Velidis}, {Benson}, {Berthier}, {Blomme}, {Burgess}, {Busso}, {Carry}, {Cellino}, {Clementini}, {Clotet}, {Creevey}, {Davidson}, {De Ridder}, {Delchambre}, {Dell'Oro}, {Ducourant}, {Fern{\'a}ndez-Hern{\'a}ndez}, {Fouesneau},
  {Fr{\'e}mat}, {Galluccio}, {Garc{\'\i}a-Torres}, {Gonz{\'a}lez-N{\'u}{\~n}ez}, {Gonz{\'a}lez-Vidal}, {Gosset}, {Guy}, {Halbwachs}, {Hambly}, {Harrison}, {Hern{\'a}ndez}, {Hestroffer}, {Hodgkin}, {Hutton}, {Jasniewicz}, {Jean-Antoine-Piccolo}, {Jordan}, {Korn}, {Krone-Martins}, {Lanzafame}, {Lebzelter}, {L{\"o}ffler}, {Manteiga}, {Marrese}, {Mart{\'\i}n-Fleitas}, {Moitinho}, {Mora}, {Muinonen}, {Osinde}, {Pancino}, {Pauwels}, {Petit}, {Recio-Blanco}, {Richards}, {Rimoldini}, {Robin}, {Sarro}, {Siopis}, {Smith}, {Sozzetti}, {S{\"u}veges}, {Torra}, {van Reeven}, {Abbas}, {Abreu Aramburu}, {Accart}, {Aerts}, {Altavilla}, {{\'A}lvarez}, {Alvarez}, {Alves}, {Anderson}, {Andrei}, {Anglada Varela}, {Antiche}, {Antoja}, {Arcay}, {Astraatmadja}, {Bach}, {Baker}, {Balaguer-N{\'u}{\~n}ez}, {Balm}, {Barache}, {Barata}, {Barbato}, {Barblan}, {Barklem}, {Barrado}, {Barros}, {Bartholom{\'e} Mu{\~n}oz}, {Bassilana}, {Becciani}, {Bellazzini}, {Berihuete}, {Bertone}, {Bianchi}, {Bienaym{\'e}}, {Blanco-Cuaresma}, {Boch},
  {Boeche}, {Bombrun}, {Borrachero}, {Bouquillon}, {Bourda}, {Bragaglia}, {Bramante}, {Breddels}, {Brouillet}, {Br{\"u}semeister}, {Brugaletta}, {Bucciarelli}, {Burlacu}, {Busonero}, {Butkevich}, {Buzzi}, {Caffau}, {Cancelliere}, {Cannizzaro}, {Carballo}, {Carlucci}, {Carrasco}, {Casamiquela}, {Castellani}, {Castro-Ginard}, {Charlot}, {Chemin}, {Chiavassa}, {Cocozza}, {Costigan}, {Cowell}, {Crifo}, {Crosta}, {Crowley}, {Cuypers}, {Dafonte}, {Damerdji}, {Dapergolas}, {David}, {David}, \& {de Laverny}}]{Gaia2018}
{Gaia Collaboration}, {Babusiaux}, C., {van Leeuwen}, F., {et~al.} 2018, \aap, 616, A10

\bibitem[{{Gaia Collaboration} {et~al.}(2023){Gaia Collaboration}, {Vallenari}, {Brown}, {Prusti}, {de Bruijne}, {Arenou}, {Babusiaux}, {Biermann}, {Creevey}, {Ducourant}, {Evans}, {Eyer}, {Guerra}, {Hutton}, {Jordi}, {Klioner}, {Lammers}, {Lindegren}, {Luri}, {Mignard}, {Panem}, {Pourbaix}, {Randich}, {Sartoretti}, {Soubiran}, {Tanga}, {Walton}, {Bailer-Jones}, {Bastian}, {Drimmel}, {Jansen}, {Katz}, {Lattanzi}, {van Leeuwen}, {Bakker}, {Cacciari}, {Casta{\~n}eda}, {De Angeli}, {Fabricius}, {Fouesneau}, {Fr{\'e}mat}, {Galluccio}, {Guerrier}, {Heiter}, {Masana}, {Messineo}, {Mowlavi}, {Nicolas}, {Nienartowicz}, {Pailler}, {Panuzzo}, {Riclet}, {Roux}, {Seabroke}, {Sordo}, {Th{\'e}venin}, {Gracia-Abril}, {Portell}, {Teyssier}, {Altmann}, {Andrae}, {Audard}, {Bellas-Velidis}, {Benson}, {Berthier}, {Blomme}, {Burgess}, {Busonero}, {Busso}, {C{\'a}novas}, {Carry}, {Cellino}, {Cheek}, {Clementini}, {Damerdji}, {Davidson}, {de Teodoro}, {Nu{\~n}ez Campos}, {Delchambre}, {Dell'Oro}, {Esquej},
  {Fern{\'a}ndez-Hern{\'a}ndez}, {Fraile}, {Garabato}, {Garc{\'\i}a-Lario}, {Gosset}, {Haigron}, {Halbwachs}, {Hambly}, {Harrison}, {Hern{\'a}ndez}, {Hestroffer}, {Hodgkin}, {Holl}, {Jan{\ss}en}, {Jevardat de Fombelle}, {Jordan}, {Krone-Martins}, {Lanzafame}, {L{\"o}ffler}, {Marchal}, {Marrese}, {Moitinho}, {Muinonen}, {Osborne}, {Pancino}, {Pauwels}, {Recio-Blanco}, {Reyl{\'e}}, {Riello}, {Rimoldini}, {Roegiers}, {Rybizki}, {Sarro}, {Siopis}, {Smith}, {Sozzetti}, {Utrilla}, {van Leeuwen}, {Abbas}, {{\'A}brah{\'a}m}, {Abreu Aramburu}, {Aerts}, {Aguado}, {Ajaj}, {Aldea-Montero}, {Altavilla}, {{\'A}lvarez}, {Alves}, {Anders}, {Anderson}, {Anglada Varela}, {Antoja}, {Baines}, {Baker}, {Balaguer-N{\'u}{\~n}ez}, {Balbinot}, {Balog}, {Barache}, {Barbato}, {Barros}, {Barstow}, {Bartolom{\'e}}, {Bassilana}, {Bauchet}, {Becciani}, {Bellazzini}, {Berihuete}, {Bernet}, {Bertone}, {Bianchi}, {Binnenfeld}, {Blanco-Cuaresma}, {Blazere}, {Boch}, {Bombrun}, {Bossini}, {Bouquillon}, {Bragaglia}, {Bramante}, {Breedt},
  {Bressan}, {Brouillet}, {Brugaletta}, {Bucciarelli}, {Burlacu}, {Butkevich}, {Buzzi}, {Caffau}, {Cancelliere}, {Cantat-Gaudin}, {Carballo}, {Carlucci}, {Carnerero}, {Carrasco}, {Casamiquela}, {Castellani}, {Castro-Ginard}, {Chaoul}, {Charlot}, {Chemin}, {Chiaramida}, {Chiavassa}, {Chornay}, {Comoretto}, {Contursi}, {Cooper}, {Cornez}, {Cowell}, {Crifo}, {Cropper}, {Crosta}, {Crowley}, {Dafonte}, {Dapergolas}, {David}, {David}, {de Laverny}, {De Luise}, {De March}, {De Ridder}, {de Souza}, {de Torres}, {del Peloso}, {del Pozo}, {Delbo}, {Delgado}, {Delisle}, {Demouchy}, {Dharmawardena}, {Di Matteo}, {Diakite}, {Diener}, {Distefano}, {Dolding}, {Edvardsson}, {Enke}, {Fabre}, {Fabrizio}, {Faigler}, {Fedorets}, {Fernique}, {Fienga}, {Figueras}, {Fournier}, {Fouron}, {Fragkoudi}, {Gai}, {Garcia-Gutierrez}, {Garcia-Reinaldos}, {Garc{\'\i}a-Torres}, {Garofalo}, {Gavel}, {Gavras}, {Gerlach}, {Geyer}, {Giacobbe}, {Gilmore}, {Girona}, {Giuffrida}, {Gomel}, {Gomez}, {Gonz{\'a}lez-N{\'u}{\~n}ez},
  {Gonz{\'a}lez-Santamar{\'\i}a}, {Gonz{\'a}lez-Vidal}, {Granvik}, {Guillout}, {Guiraud}, {Guti{\'e}rrez-S{\'a}nchez}, {Guy}, {Hatzidimitriou}, {Hauser}, {Haywood}, {Helmer}, {Helmi}, {Sarmiento}, {Hidalgo}, {Hilger}, {H{\l}adczuk}, {Hobbs}, {Holland}, {Huckle}, {Jardine}, {Jasniewicz}, {Jean-Antoine Piccolo}, {Jim{\'e}nez-Arranz}, {Jorissen}, {Juaristi Campillo}, {Julbe}, {Karbevska}, {Kervella}, {Khanna}, {Kontizas}, {Kordopatis}, {Korn}, {K{\'o}sp{\'a}l}, {Kostrzewa-Rutkowska}, {Kruszy{\'n}ska}, {Kun}, {Laizeau}, {Lambert}, {Lanza}, {Lasne}, {Le Campion}, {Lebreton}, {Lebzelter}, {Leccia}, {Leclerc}, {Lecoeur-Taibi}, {Liao}, {Licata}, {Lindstr{\o}m}, {Lister}, {Livanou}, {Lobel}, {Lorca}, {Loup}, {Madrero Pardo}, {Magdaleno Romeo}, {Managau}, {Mann}, {Manteiga}, {Marchant}, {Marconi}, {Marcos}, {Marcos Santos}, {Mar{\'\i}n Pina}, {Marinoni}, {Marocco}, {Marshall}, {Martin Polo}, {Mart{\'\i}n-Fleitas}, {Marton}, {Mary}, {Masip}, {Massari}, {Mastrobuono-Battisti}, {Mazeh}, {McMillan}, {Messina}, {Michalik},
  {Millar}, {Mints}, {Molina}, {Molinaro}, {Moln{\'a}r}, {Monari}, {Mongui{\'o}}, {Montegriffo}, {Montero}, {Mor}, {Mora}, {Morbidelli}, {Morel}, {Morris}, {Muraveva}, {Murphy}, {Musella}, {Nagy}, {Noval}, {Oca{\~n}a}, {Ogden}, {Ordenovic}, {Osinde}, {Pagani}, {Pagano}, {Palaversa}, {Palicio}, {Pallas-Quintela}, {Panahi}, {Payne-Wardenaar}, {Pe{\~n}alosa Esteller}, {Penttil{\"a}}, {Pichon}, {Piersimoni}, {Pineau}, {Plachy}, {Plum}, {Poggio}, {Pr{\v{s}}a}, {Pulone}, {Racero}, {Ragaini}, {Rainer}, {Raiteri}, {Rambaux}, {Ramos}, {Ramos-Lerate}, {Re Fiorentin}, {Regibo}, {Richards}, {Rios Diaz}, {Ripepi}, {Riva}, {Rix}, {Rixon}, {Robichon}, {Robin}, {Robin}, {Roelens}, {Rogues}, {Rohrbasser}, {Romero-G{\'o}mez}, {Rowell}, {Royer}, {Ruz Mieres}, {Rybicki}, {Sadowski}, {S{\'a}ez N{\'u}{\~n}ez}, {Sagrist{\`a} Sell{\'e}s}, {Sahlmann}, {Salguero}, {Samaras}, {Sanchez Gimenez}, {Sanna}, {Santove{\~n}a}, {Sarasso}, {Schultheis}, {Sciacca}, {Segol}, {Segovia}, {S{\'e}gransan}, {Semeux}, {Shahaf}, {Siddiqui}, {Siebert},
  {Siltala}, {Silvelo}, {Slezak}, {Slezak}, {Smart}, {Snaith}, {Solano}, {Solitro}, {Souami}, {Souchay}, {Spagna}, {Spina}, {Spoto}, {Steele}, {Steidelm{\"u}ller}, {Stephenson}, {S{\"u}veges}, {Surdej}, {Szabados}, {Szegedi-Elek}, {Taris}, {Taylor}, {Teixeira}, {Tolomei}, {Tonello}, {Torra}, {Torra}, {Torralba Elipe}, {Trabucchi}, {Tsounis}, {Turon}, {Ulla}, {Unger}, {Vaillant}, {van Dillen}, {van Reeven}, {Vanel}, {Vecchiato}, {Viala}, {Vicente}, {Voutsinas}, {Weiler}, {Wevers}, {Wyrzykowski}, {Yoldas}, {Yvard}, {Zhao}, {Zorec}, {Zucker}, \& {Zwitter}}]{GaiaDR32023}
{Gaia Collaboration}, {Vallenari}, A., {Brown}, A.~G.~A., {et~al.} 2023, \aap, 674, A1

\bibitem[{{Garc{\'\i}a-Zamora} {et~al.}(2023){Garc{\'\i}a-Zamora}, {Torres}, \& {Rebassa-Mansergas}}]{Garcia2023}
{Garc{\'\i}a-Zamora}, E.~M., {Torres}, S., \& {Rebassa-Mansergas}, A. 2023, \aap, 679, A127

\bibitem[{{Genest-Beaulieu} \& {Bergeron}(2019)}]{Genest2019}
{Genest-Beaulieu}, C. \& {Bergeron}, P. 2019, \apj, 871, 169

\bibitem[{{Gentile Fusillo} {et~al.}(2021){Gentile Fusillo}, {Tremblay}, {Cukanovaite}, {Vorontseva}, {Lallement}, {Hollands}, {G{\"a}nsicke}, {Burdge}, {McCleery}, \& {Jordan}}]{GentilleFusillo2021}
{Gentile Fusillo}, N.~P., {Tremblay}, P.~E., {Cukanovaite}, E., {et~al.} 2021, \mnras, 508, 3877

\bibitem[{{Hollands} {et~al.}(2018){Hollands}, {Tremblay}, {G{\"a}nsicke}, {Gentile-Fusillo}, \& {Toonen}}]{Hollands2018}
{Hollands}, M.~A., {Tremblay}, P.~E., {G{\"a}nsicke}, B.~T., {Gentile-Fusillo}, N.~P., \& {Toonen}, S. 2018, \mnras, 480, 3942

\bibitem[{{Jim{\'e}nez-Esteban} {et~al.}(2023){Jim{\'e}nez-Esteban}, {Torres}, {Rebassa-Mansergas}, {Cruz}, {Murillo-Ojeda}, {Solano}, {Rodrigo}, \& {Camisassa}}]{Jimenez2023}
{Jim{\'e}nez-Esteban}, F.~M., {Torres}, S., {Rebassa-Mansergas}, A., {et~al.} 2023, \mnras, 518, 5106

\bibitem[{{Jin} {et~al.}(2024){Jin}, {Trager}, {Dalton}, {Aguerri}, {Drew}, {Falc{\'o}n-Barroso}, {G{\"a}nsicke}, {Hill}, {Iovino}, {Pieri}, \& et~al.}]{WEAVE2022}
{Jin}, S., {Trager}, S.~C., {Dalton}, G.~B., {et~al.} 2024, \mnras, 530, 2688

\bibitem[{{Kao} {et~al.}(2024){Kao}, {Hawkins}, {Rogers}, {Bonsor}, {Dunlap}, {Sanders}, {Montgomery}, \& {Winget}}]{Kao2024}
{Kao}, M.~L., {Hawkins}, K., {Rogers}, L.~K., {et~al.} 2024, \apj, 970, 181

\bibitem[{Kohonen(1982)}]{Kohonen1982}
Kohonen, T. 1982, Biol. Cybern., 43, 59

\bibitem[{Kong {et~al.}(2018)Kong, Luo, Li, Wang, Li, \& Zhao}]{Kong2018}
Kong, X., Luo, A.-L., Li, X.-R., {et~al.} 2018, Publications of the Astronomical Society of the Pacific, 130, 084203

\bibitem[{{Levi} {et~al.}(2019){Levi}, {Allen}, {Raichoor}, {Baltay}, {BenZvi}, {Beutler}, {Bolton}, {Castander}, {Chuang}, {Cooper}, {Cuby}, {Dey}, {Eisenstein}, {Fan}, {Flaugher}, {Frenk}, {Gonzalez-Morales}, {Graur}, {Guy}, {Habib}, {Honscheid}, {Juneau}, {Kneib}, {Lahav}, {Lang}, {Leauthaud}, {Lusso}, {de la Macorra}, {Manera}, {Martini}, {Mao}, {Newman}, {Palanque-Delabrouille}, {Percival}, {Allende Prieto}, {Rockosi}, {Ruhlmann-Kleider}, {Schlegel}, {Seo}, {Song}, {Tarle}, {Wechsler}, {Weinberg}, {Yeche}, \& {Zu}}]{DESI2019}
{Levi}, M., {Allen}, L.~E., {Raichoor}, A., {et~al.} 2019, in Bulletin of the American Astronomical Society, Vol.~51, 57

\bibitem[{{Manser} {et~al.}(2024){Manser}, {Izquierdo}, {G{\"a}nsicke}, {Swan}, {Koester}, {Robert}, {Xu}, {Inight}, {Amroota}, {Fusillo}, {Koposov}, {Kim}, {Dey}, {Prieto}, {Aguilar}, {Ahlen}, {Blum}, {Brooks}, {Claybaugh}, {Cooper}, {Dawson}, {de la Macorra}, {Doel}, {Forero-Romero}, {Gazta{\~n}aga}, {Gontcho}, {Honscheid}, {Kisner}, {Kremin}, {Lambert}, {Landriau}, {Le Guillou}, {Levi}, {Li}, {Meisner}, {Miquel}, {Moustakas}, {Nie}, {Palanque-Delabrouille}, {Percival}, {Poppett}, {Prada}, {Rezaie}, {Rossi}, {Sanchez}, {Schlafly}, {Schlegel}, {Schubnell}, {Seo}, {Silber}, {Tarl{\'e}}, {Weaver}, {Zhou}, \& {Zou}}]{Manser2024}
{Manser}, C.~J., {Izquierdo}, P., {G{\"a}nsicke}, B.~T., {et~al.} 2024, \mnras, 535, 254

\bibitem[{{Marsh}(1989)}]{Marsh1989}
{Marsh}, T.~R. 1989, \pasp, 101, 1032

\bibitem[{{McCleery} {et~al.}(2020){McCleery}, {Tremblay}, {Gentile Fusillo}, {Hollands}, {G{\"a}nsicke}, {Izquierdo}, {Toonen}, {Cunningham}, \& {Rebassa-Mansergas}}]{McCleery2020}
{McCleery}, J., {Tremblay}, P.-E., {Gentile Fusillo}, N.~P., {et~al.} 2020, \mnras, 499, 1890

\bibitem[{{McInnes} {et~al.}(2018){McInnes}, {Healy}, \& {Melville}}]{McInnes2018}
{McInnes}, L., {Healy}, J., \& {Melville}, J. 2018, arXiv e-prints, arXiv:1802.03426

\bibitem[{{Montegriffo} {et~al.}(2023){Montegriffo}, {De Angeli}, {Andrae}, {Riello}, {Pancino}, {Sanna}, {Bellazzini}, {Evans}, {Carrasco}, {Sordo}, {Busso}, {Cacciari}, {Jordi}, {van Leeuwen}, {Vallenari}, {Altavilla}, {Barstow}, {Brown}, {Burgess}, {Castellani}, {Cowell}, {Davidson}, {De Luise}, {Delchambre}, {Diener}, {Fabricius}, {Fr{\'e}mat}, {Fouesneau}, {Gilmore}, {Giuffrida}, {Hambly}, {Harrison}, {Hidalgo}, {Hodgkin}, {Holland}, {Marinoni}, {Osborne}, {Pagani}, {Palaversa}, {Piersimoni}, {Pulone}, {Ragaini}, {Rainer}, {Richards}, {Rowell}, {Ruz-Mieres}, {Sarro}, {Walton}, \& {Yoldas}}]{Montegriffo2023}
{Montegriffo}, P., {De Angeli}, F., {Andrae}, R., {et~al.} 2023, \aap, 674, A3

\bibitem[{{Napiwotzki} {et~al.}(2020){Napiwotzki}, {Karl}, {Lisker}, {Catal{\'a}n}, {Drechsel}, {Heber}, {Homeier}, {Koester}, {Leibundgut}, {Marsh}, {Moehler}, {Nelemans}, {Reimers}, {Renzini}, {Str{\"o}er}, \& {Yungelson}}]{Napiwotzki2020}
{Napiwotzki}, R., {Karl}, C.~A., {Lisker}, T., {et~al.} 2020, \aap, 638, A131

\bibitem[{{O'Brien} {et~al.}(2023){O'Brien}, {Tremblay}, {Gentile Fusillo}, {Hollands}, {G{\"a}nsicke}, {Koester}, {Pelisoli}, {Cukanovaite}, {Cunningham}, {Doyle}, {Elms}, {Farihi}, {Hermes}, {Holberg}, {Jordan}, {Klein}, {Kleinman}, {Manser}, {De Martino}, {Marsh}, {McCleery}, {Melis}, {Nitta}, {Parsons}, {Raddi}, {Rebassa-Mansergas}, {Schreiber}, {Silvotti}, {Steeghs}, {Toloza}, {Toonen}, {Torres}, {Weinberger}, \& {Zuckerman}}]{Obrien2023}
{O'Brien}, M.~W., {Tremblay}, P.~E., {Gentile Fusillo}, N.~P., {et~al.} 2023, \mnras, 518, 3055

\bibitem[{{O'Brien} {et~al.}(2024){O'Brien}, {Tremblay}, {Klein}, {Koester}, {Melis}, {B{\'e}dard}, {Cukanovaite}, {Cunningham}, {Doyle}, {G{\"a}nsicke}, {Gentile Fusillo}, {Hollands}, {McCleery}, {Pelisoli}, {Toonen}, {Weinberger}, \& {Zuckerman}}]{Obrien2024}
{O'Brien}, M.~W., {Tremblay}, P.~E., {Klein}, B.~L., {et~al.} 2024, \mnras, 527, 8687

\bibitem[{Pedregosa {et~al.}(2011)Pedregosa, Varoquaux, Gramfort, Michel, Thirion, Grisel, Blondel, Prettenhofer, Weiss, Dubourg, Vanderplas, Passos, Cournapeau, Brucher, Perrot, \& Duchesnay}]{Pedregosa2011}
Pedregosa, F., Varoquaux, G., Gramfort, A., {et~al.} 2011, Journal of Machine Learning Research, 12, 2825

\bibitem[{{P{\'e}rez-Couto} {et~al.}(2024){P{\'e}rez-Couto}, {Pallas-Quintela}, {Manteiga}, {Villaver}, \& {Dafonte}}]{Perez-Couto2024}
{P{\'e}rez-Couto}, X., {Pallas-Quintela}, L., {Manteiga}, M., {Villaver}, E., \& {Dafonte}, C. 2024, \apj, 977, 31

\bibitem[{{Rebassa-Mansergas} {et~al.}(2013){Rebassa-Mansergas}, {Agurto-Gangas}, {Schreiber}, {G{\"a}nsicke}, \& {Koester}}]{Rebassa2013}
{Rebassa-Mansergas}, A., {Agurto-Gangas}, C., {Schreiber}, M.~R., {G{\"a}nsicke}, B.~T., \& {Koester}, D. 2013, \mnras, 433, 3398

\bibitem[{{Sion} {et~al.}(1983){Sion}, {Greenstein}, {Landstreet}, {Liebert}, {Shipman}, \& {Wegner}}]{Sion1983}
{Sion}, E.~M., {Greenstein}, J.~L., {Landstreet}, J.~D., {et~al.} 1983, \apj, 269, 253

\bibitem[{{Torres} {et~al.}(2019){Torres}, {Cantero}, {Rebassa-Mansergas}, {Skorobogatov}, {Jim{\'e}nez-Esteban}, \& {Solano}}]{Torres2019}
{Torres}, S., {Cantero}, C., {Rebassa-Mansergas}, A., {et~al.} 2019, \mnras, 485, 5573

\bibitem[{{Torres} {et~al.}(2023){Torres}, {Cruz}, {Murillo-Ojeda}, {Jim{\'e}nez-Esteban}, {Rebassa-Mansergas}, {Solano}, {Camisassa}, {Raddi}, \& {Doliguez Le Lourec}}]{Torres2023}
{Torres}, S., {Cruz}, P., {Murillo-Ojeda}, R., {et~al.} 2023, \aap, 677, A159

\bibitem[{{Torres} {et~al.}(1998){Torres}, {Garc{\'\i}a-Berro}, \& {Isern}}]{Torres1998}
{Torres}, S., {Garc{\'\i}a-Berro}, E., \& {Isern}, J. 1998, \apjl, 508, L71

\bibitem[{{Tremblay} {et~al.}(2019{\natexlab{a}}){Tremblay}, {Cukanovaite}, {Gentile Fusillo}, {Cunningham}, \& {Hollands}}]{Tremblay2019b}
{Tremblay}, P.~E., {Cukanovaite}, E., {Gentile Fusillo}, N.~P., {Cunningham}, T., \& {Hollands}, M.~A. 2019{\natexlab{a}}, \mnras, 482, 5222

\bibitem[{{Tremblay} {et~al.}(2019{\natexlab{b}}){Tremblay}, {Fontaine}, {Gentile Fusillo}, {Dunlap}, {G{\"a}nsicke}, {Hollands}, {Hermes}, {Marsh}, {Cukanovaite}, \& {Cunningham}}]{Tremblay2019}
{Tremblay}, P.-E., {Fontaine}, G., {Gentile Fusillo}, N.~P., {et~al.} 2019{\natexlab{b}}, \nat, 565, 202

\bibitem[{{Tremblay} {et~al.}(2020){Tremblay}, {Hollands}, {Gentile Fusillo}, {McCleery}, {Izquierdo}, {G{\"a}nsicke}, {Cukanovaite}, {Koester}, {Brown}, {Charpinet}, {Cunningham}, {Farihi}, {Giammichele}, {van Grootel}, {Hermes}, {Hoskin}, {Jordan}, {Kepler}, {Kleinman}, {Manser}, {Marsh}, {de Martino}, {Nitta}, {Parsons}, {Pelisoli}, {Raddi}, {Rebassa-Mansergas}, {Ren}, {Schreiber}, {Silvotti}, {Toloza}, {Toonen}, \& {Torres}}]{Tremblay2020}
{Tremblay}, P.~E., {Hollands}, M.~A., {Gentile Fusillo}, N.~P., {et~al.} 2020, \mnras, 497, 130

\bibitem[{{Vincent} {et~al.}(2024){Vincent}, {Barstow}, {Jordan}, {Mander}, {Bergeron}, \& {Dufour}}]{Vincent2024}
{Vincent}, O., {Barstow}, M.~A., {Jordan}, S., {et~al.} 2024, \aap, 682, A5

\bibitem[{{Vincent} {et~al.}(2023){Vincent}, {Bergeron}, \& {Dufour}}]{Olivier2023}
{Vincent}, O., {Bergeron}, P., \& {Dufour}, P. 2023, \mnras, 521, 760

\bibitem[{{Weiler} {et~al.}(2023){Weiler}, {Carrasco}, {Fabricius}, \& {Jordi}}]{Weiler2023}
{Weiler}, M., {Carrasco}, J.~M., {Fabricius}, C., \& {Jordi}, C. 2023, \aap, 671, A52

\bibitem[{{Zuckerman} {et~al.}(2007){Zuckerman}, {Koester}, {Melis}, {Hansen}, \& {Jura}}]{Zuckerman2007}
{Zuckerman}, B., {Koester}, D., {Melis}, C., {Hansen}, B.~M., \& {Jura}, M. 2007, \apj, 671, 872

\end{thebibliography}

\begin{appendix}

\begin{figure*}[h!]
\section{Color-magnitude diagrams of the \emph{Gaia} 500-pc classification sample}
\label{s:ann1}
\vspace{1.0cm}

\centering
    \includegraphics[width=1.0\columnwidth,trim=-20 0 0 0, clip]{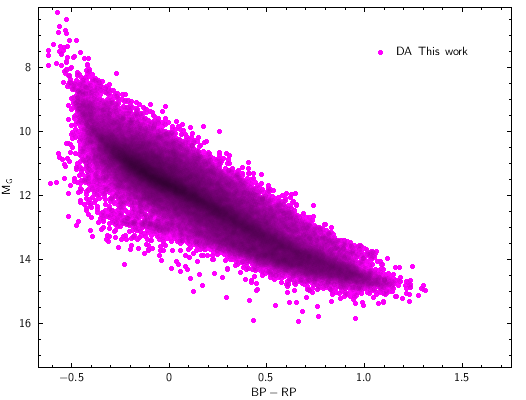}
    \includegraphics[width=1.0\columnwidth,trim=0 0 -20 0, clip]{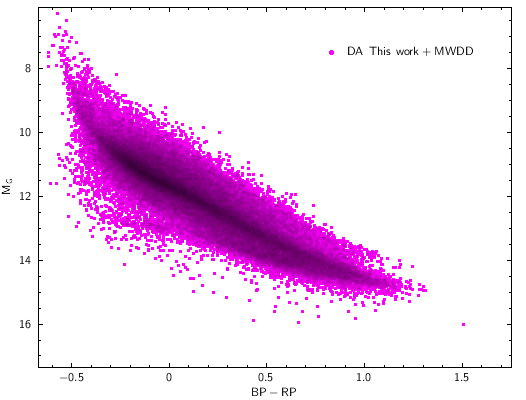}
        \vspace{0.5cm}
    \centering
    \includegraphics[width=1.0\columnwidth,trim=-20 0 0 0, clip]{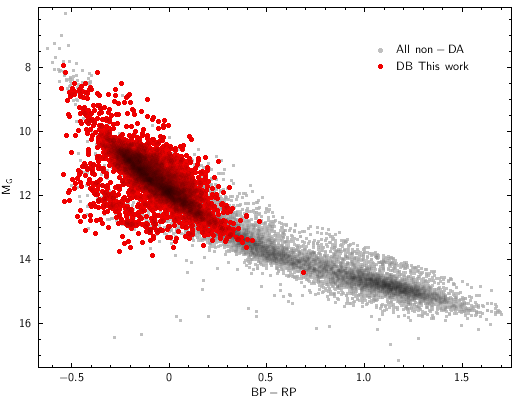}
    \includegraphics[width=1.0\columnwidth,trim=0 0 -20 0, clip]{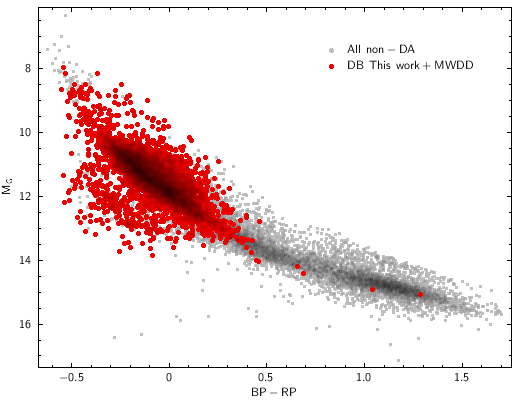}
        \vspace{0.5cm}
    \centering
    \includegraphics[width=1.0\columnwidth,trim=-20 0 0 0, clip]{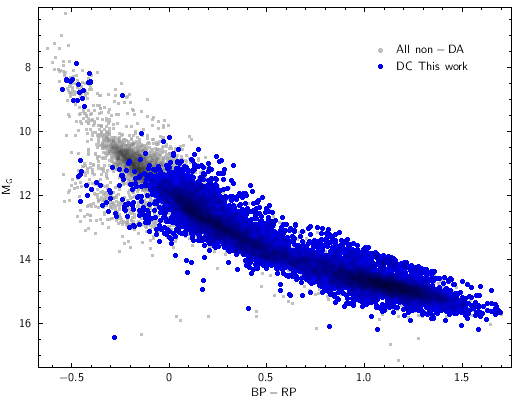}
    \includegraphics[width=1.0\columnwidth,trim=0 0 -20 0, clip]{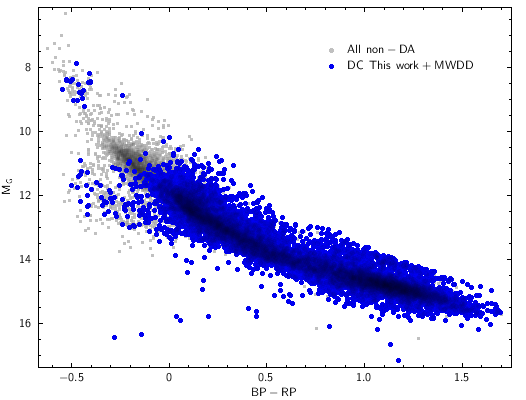}
    
    \caption{{\it Left panels}: White dwarfs classified as DA, DB and DC in this work. {\it Right panels}: Entire population (those classified in this work and those labeled in MWDD) of DA, DB and DC white dwarfs.}
    \label{f:500pc_HR}
        \vspace{0.5cm}
\end{figure*}

\begin{figure*}[h!]
\centering
    \includegraphics[width=1.0\columnwidth,trim=-20 0 0 0, clip]{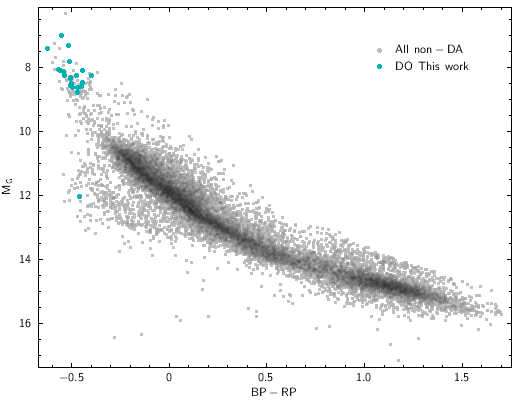}
    \includegraphics[width=1.0\columnwidth,trim=0 0 -20 0, clip]{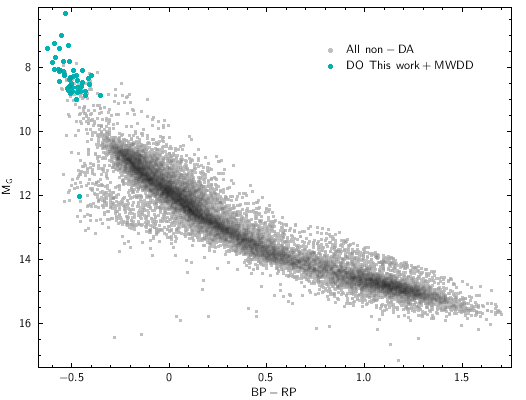}
        \vspace{0.5cm}
    \centering
    \includegraphics[width=1.0\columnwidth,trim=-20 0 0 0, clip]{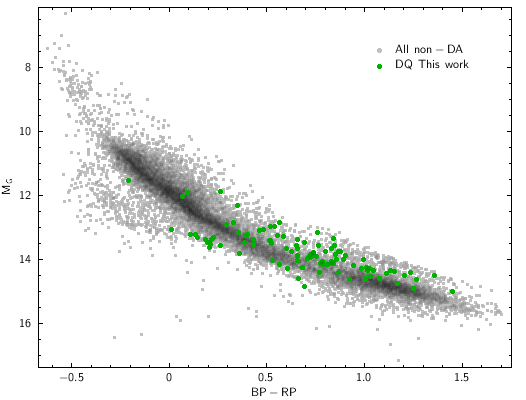}
    \includegraphics[width=1.0\columnwidth,trim=0 0 -20 0, clip]{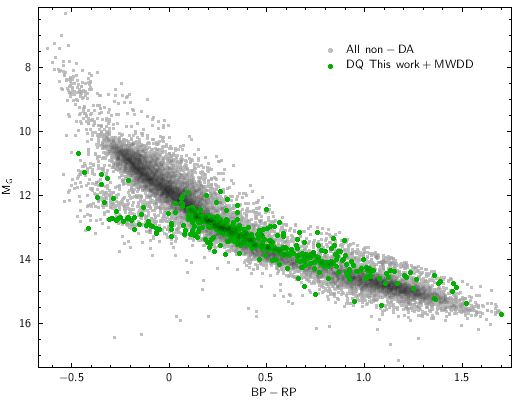}
        \vspace{0.5cm}
    \centering
    \includegraphics[width=1.0\columnwidth,trim=-20 0 0 0, clip]{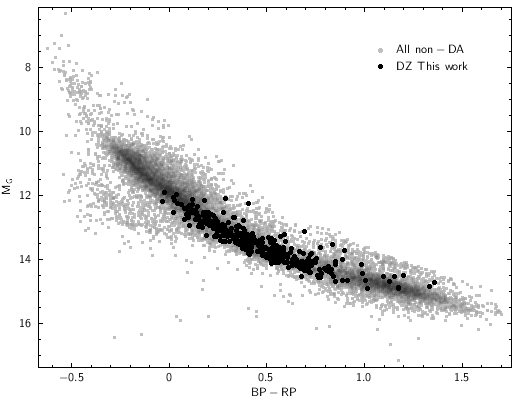}
    \includegraphics[width=1.0\columnwidth,trim=0 0 -20 0, clip]{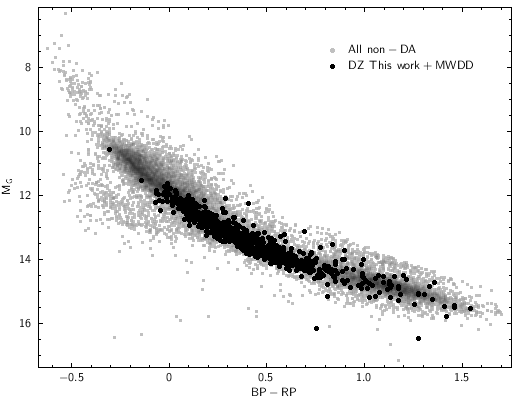}
    
    \caption{{\it Left panels}: White dwarfs classified as DO, DQ and DZ in this work. {\it Right panels}: Entire population (those classified in this work and those labeled in MWDD) of DO, DQ and DZ white dwarfs.}
    \label{f:500pc_HR_2}
        \vspace{0.5cm}
\end{figure*}


\begin{figure*}[h!]
\section{Golden sample color-magnitude diagrams}
\label{s:ann2}
\vspace{1.0cm}

\centering
    \includegraphics[width=1.0\columnwidth,trim=-20 0 0 0, clip]{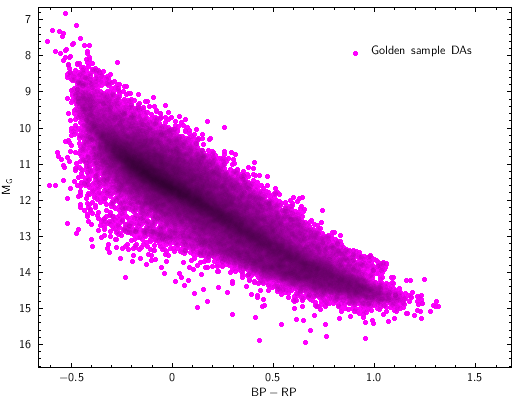}
    \includegraphics[width=1.0\columnwidth,trim=0 0 -20 0, clip]{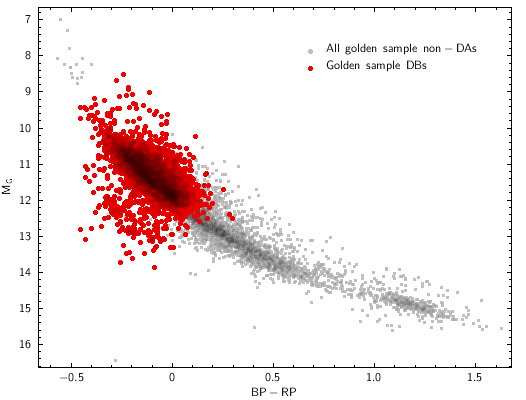}
        \vspace{0.5cm}
    \centering
    \includegraphics[width=1.0\columnwidth,trim=-20 0 0 0, clip]{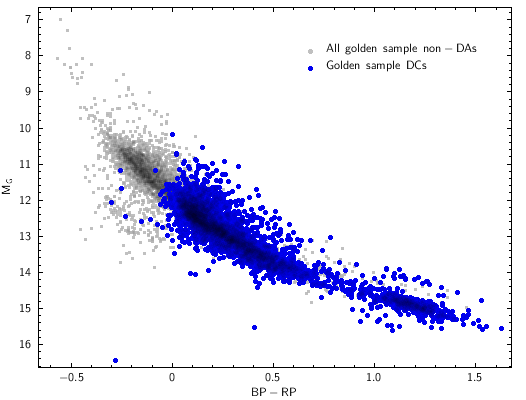}
    \includegraphics[width=1.0\columnwidth,trim=0 0 -20 0, clip]{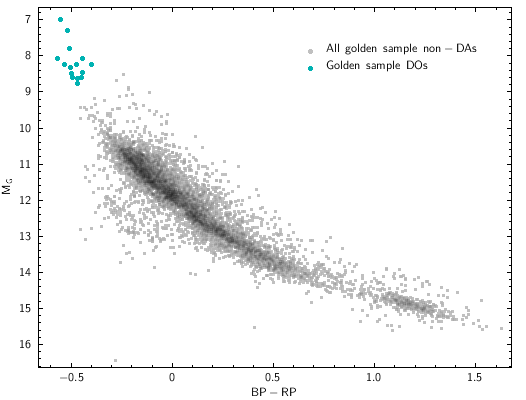}
        \vspace{0.5cm}
    \centering
    \includegraphics[width=1.0\columnwidth,trim=-20 0 0 0, clip]{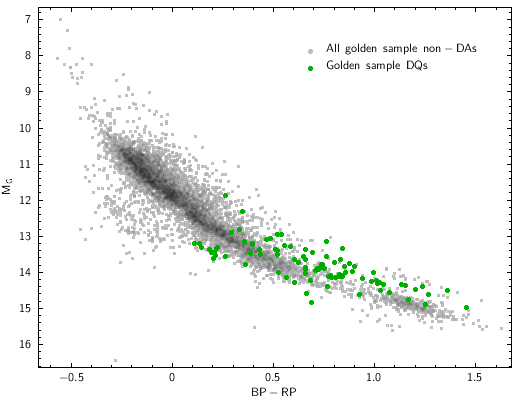}
    \includegraphics[width=1.0\columnwidth,trim=0 0 -20 0, clip]{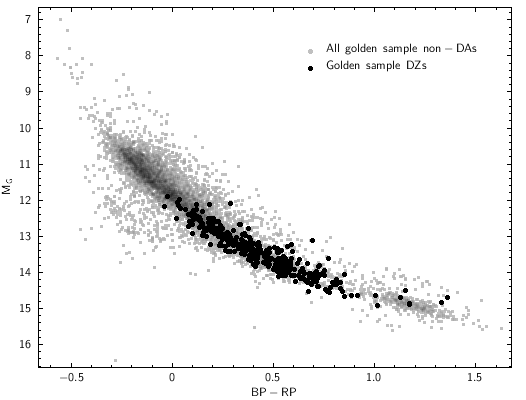}
    \caption{White dwarfs of the Golden Sample classified as DA, DB ,DC, DO, DQ, and DZ (left to right, top to bottom, respectively).}
    \label{f:GS_HR}
        \vspace{0.5cm}
\end{figure*}


\begin{table*}[p]
\section{Observational versus prediction comparison table}
\label{a:Table}
    \caption[]{Observational versus Random Forest spectral prediction of a selected sample of white dwarfs}
    \label{tab:obs_pred}
\begin{center}
    \begin{tabular}{cccccccc}
            \noalign{\smallskip}
            \hline\hline
            \noalign{\smallskip}
        \emph{Gaia} Source ID & RA & Dec & BP-RP & $M_G$ & Observational & Predicted \\
            & (deg) & (deg) & (mag) & (mag) & type & type \\
            \noalign{\smallskip}
            \hline
        508787598324392448 & 30.459 & 62.359 & -0.045 & 11.819 & DBA & DBA\\
        336814913062739456 & 41.53 & 41.623 & 0.026 & 11.976 & DBA & DBA\\
        467032338586262144 & 46.945 & 63.660 & -0.051 & 10.731 & DB & DB\\
        435123656452918784 & 47.864 & 47.653 & -0.107 & 10.346 & DB & DB\\
        547793872828407680 & 50.393 & 74.542 & 0.518 & 13.395 & DQ & DQ\\
        5161807767825277184 & 52.140 & -12.329 & 0.286 & 13.425 & DZ & DZ\\
        44403089893740288 & 54.486 & 17.321 & 0.423 & 13.401 & DZ & DZ\\
        120059179330248704 & 55.691 & 29.574 & 0.388 & 13.080 & DZ & DZ\\
        3190140908188585472 & 60.975 & -11.536 & 0.236 & 12.984 & DZ & DZ\\
        233493113910461952 & 61.818 & 45.275 & 0.378 & 12.786 & DZ & DZ\\
        3260058921320198016 & 62.021 & 2.718 & 0.465 & 13.682 & DZ & DZ\\
        3252501909182919424 & 62.599 & -2.397 & 0.312 & 13.109 & DZ & DZ\\
        3298587423664896640 & 62.745 & 8.796 & 0.521 & 13.337 & DZ & DZ\\
        3176664640844839424 & 62.964 & -13.900 & 0.332 & 13.348 & DZ & DZ\\
        3300518578399661952 & 63.053 & 10.160 & -0.090 & 11.475 & DBA & DB\\
        3191926927390852608 & 63.352 & -10.736 & 0.670 & 14.270 & DZ & DZ\\
        476005109382589184 & 65.937 & 63.200 & 0.007 & 10.466 & DB & DB\\
        3306772527522597632 & 66.009 & 10.905 & -0.034 & 11.957 & DAH & DAH\\
        252658602311901184 & 68.625 & 44.157 & -0.010 & 10.776 & DB & DB\\
        3231073419693322240 & 69.379 & 0.854 & 0.119 & 12.118 & DZ & DZ\\
        3185120018405133184 & 71.181 & -9.032 & 0.641 & 13.903 & DZ & DZ\\
        3281864642080410112 & 71.769 & 4.978 & -0.510 & 7.796 & DO & DO\\
        158445436654807680 & 72.542 & 30.263 & 0.553 & 13.243 & DQ & DQ\\
        3226269756470965120 & 73.148 & -2.246 & 0.024 & 12.053 & DZ & DZ\\
        3214603044747246208 & 77.858 & -2.781 & 0.258 & 12.982 & DZ & DZ\\
        3420572943072331008 & 78.266 & 25.230 & 0.357 & 12.934 & DZ & DZ\\
        3387989607472006912 & 78.270 & 11.788 & -0.495 & 8.611 & DO & DO\\
        187330225592045824 & 78.728 & 37.505 & 0.740 & 13.778 & DQ & DQ\\
        3242039746806764672 & 78.927 & 7.479 & -0.445 & 8.072 & DO & DO\\
        267738709226387328 & 79.666 & 55.896 & 0.139 & 12.531 & DZ & DZ\\
        3210273928166901632 & 79.955 & -4.717 & -0.400 & 8.249 & DO & DO\\
        262271632470042240 & 81.637 & 51.939 & 0.279 & 13.114 & DZ & DZ\\
        264990106251426432 & 84.118 & 54.815 & -0.475 & 8.238 & DO & DO\\
        2997076768116088576 & 86.785 & -12.845 & 0.011 & 12.710 & DAH & DAH\\
        214530012958427008 & 88.086 & 50.457 & 0.281 & 12.941 & DZ & DZ\\
        484896928635700992 & 88.902 & 69.421 & 0.464 & 13.088 & DQ & DQ\\
        3335944632592001280 & 88.950 & 10.187 & 0.503 & 13.949 & DZ & DZ\\
        3350410499060702976 & 89.425 & 17.882 & 0.098 & 12.681 & DZ & DZ\\
        2997192526074656640 & 90.687 & -13.851 & -0.517 & 7.295 & DO & DO\\
        1005918380521751552 & 91.181 & 61.404 & 0.377 & 13.658 & DZ & DZ\\
        3122591513688353792 & 91.826 & 0.579 & 0.009 & 12.038 & DBA & DB\\
        3346019045322167808 & 92.068 & 16.009 & 0.279 & 13.164 & DZ & DZ\\
        2994361966531540608 & 92.702 & -14.04 & 0.562 & 14.001 & DZ & DZ\\
        3433732624085019264 & 94.768 & 28.246 & 0.027 & 12.064 & DBA & DBA\\
        3433733208197186560 & 94.870 & 28.246 & 0.261 & 13.396 & DZ & DZ\\
        1104287803208980096 & 97.141 & 66.606 & 0.149 & 12.638 & DZ & DZ\\
        3351595630861258624 & 101.691 & 11.861 & -0.113 & 11.090 & DBA & DBA\\
        890846762030633984 & 106.264 & 33.126 & -0.038 & 11.766 & DB & DBA\\
        3044597415857442944 & 107.291 & -13.543 & 0.276 & 12.923 & DZ & DZ\\
        3368315320066689152 & 108.139 & 23.255 & 0.388 & 13.675 & DZ & DZ\\
        3156946991449113728 & 110.163 & 11.545 & 0.079 & 12.739 & DZ & DZ\\
        1085368678428090880 & 114.414 & 57.478 & -0.035 & 12.708 & DB & DBA\\
        1136073108116352128 & 114.627 & 74.409 & 0.721 & 14.404 & DZ & DZ\\
        3040162810580901376 & 115.577 & -10.737 & 0.775 & 14.550 & DZ & DZ\\
        985948088265005696 & 116.349 & 55.854 & -0.022 & 11.895 & DZ & DZ\\
        
            \noalign{\smallskip}
            \noalign{\smallskip}
            \hline
    \end{tabular}
\end{center}
\end{table*}

\begin{table*}[p]
    \caption[]{Random Forest algorithm classification of {\it Gaia} 500-pc white dwarfs (continuation)}
    \label{tab:obs_pred_2}
\begin{center}
    \begin{tabular}{cccccccc}
            \noalign{\smallskip}
            \hline\hline
            \noalign{\smallskip}
        \emph{Gaia} Source ID & RA & Dec & BP-RP & $M_G$ & Observational & Predicted \\
            & (deg) & (deg) & (mag) & (mag) & type & type \\
            \noalign{\smallskip}
            \hline
    3080844435869554176 & 117.175 & -3.393 & -0.032 & 11.783 & DBA & DBA\\
    3037191315987645312 & 117.659 & -11.549 & 0.250 & 13.083 & DZ & DZ\\
    1142428555905843072 & 118.280 & 81.331 & 0.308 & 13.411 & DZ & DZ\\
    3144625662286579968 & 119.277 & 8.0551 & 0.260 & 13.010 & DZ & DZ\\
    5727797557366321024 & 122.000 & -11.933 & 0.782 & 14.075 & DQ & DQ\\
    5750300407152168064 & 130.670 & -9.568 & 0.351 & 13.154 & DZ & DZ\\
    5736541423584135936 & 133.514 & -12.692 & 0.028 & 12.153 & DZ & DZ\\
    5735535370444130944 & 135.847 & -13.185 & -0.320 & 11.391 & DB & DB\\
    5740146241175408512 & 142.780 & -11.042 & 0.193 & 12.789 & DZ & DZ\\
    1121031651608385536 & 145.394 & 73.839 & 0.229 & 13.013 & DZ & DZ\\
            \noalign{\smallskip}
            \noalign{\smallskip}
            \hline
    \end{tabular}
\end{center}
\end{table*}

\begin{figure*}[h!]
\section{\emph{Gaia} and GTC-OSIRIS sample spectra}
\label{a:gaiaGTC}
\vspace{0.5cm}
\centering
        \includegraphics[width=1.0\columnwidth,trim=-20 0 0 0, clip]{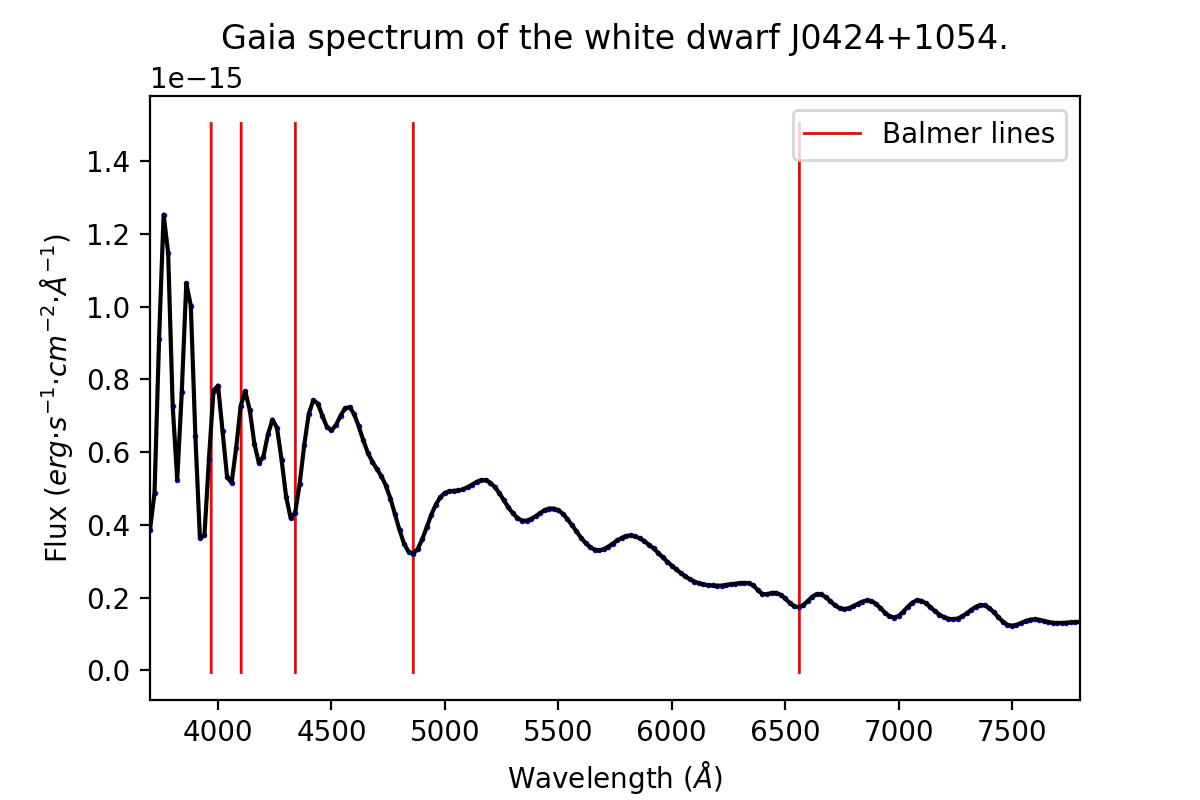}
            \includegraphics[width=1.0\columnwidth,trim=0 0 -20 0, clip]{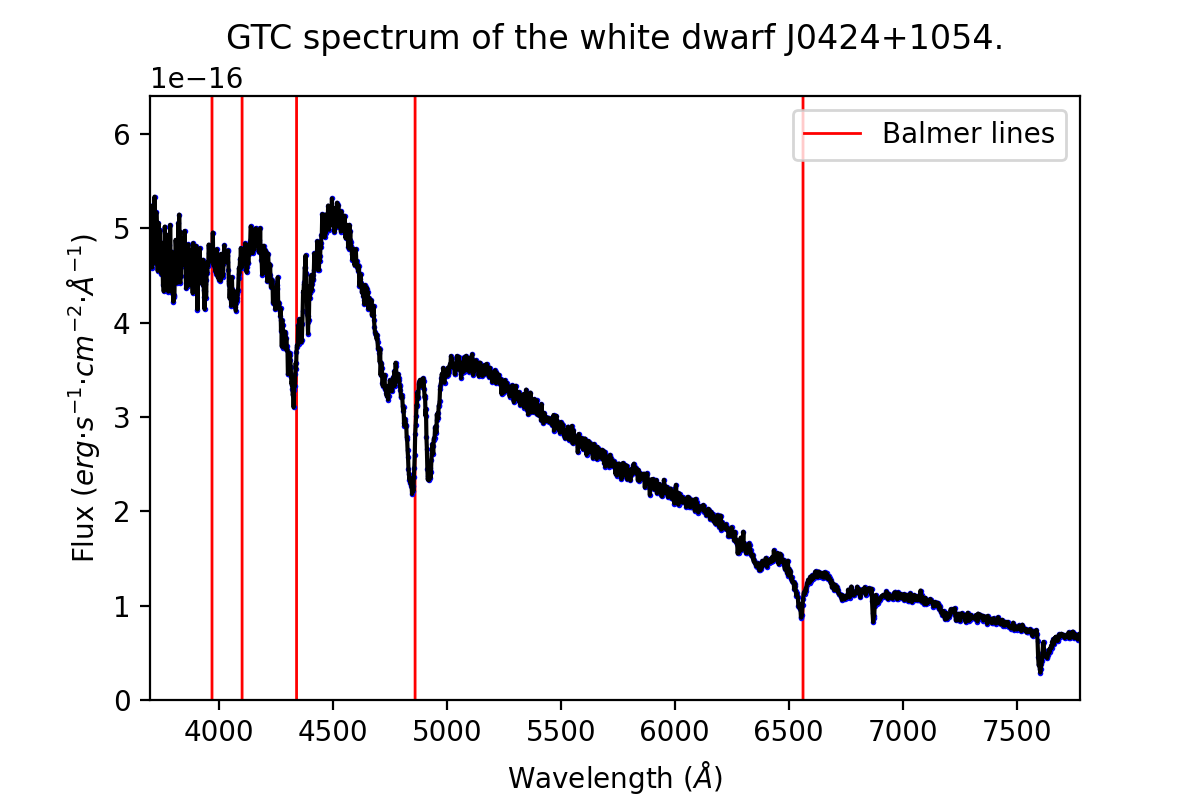}
            
        \includegraphics[width=1.0\columnwidth,trim=-20 0 0 0, clip]{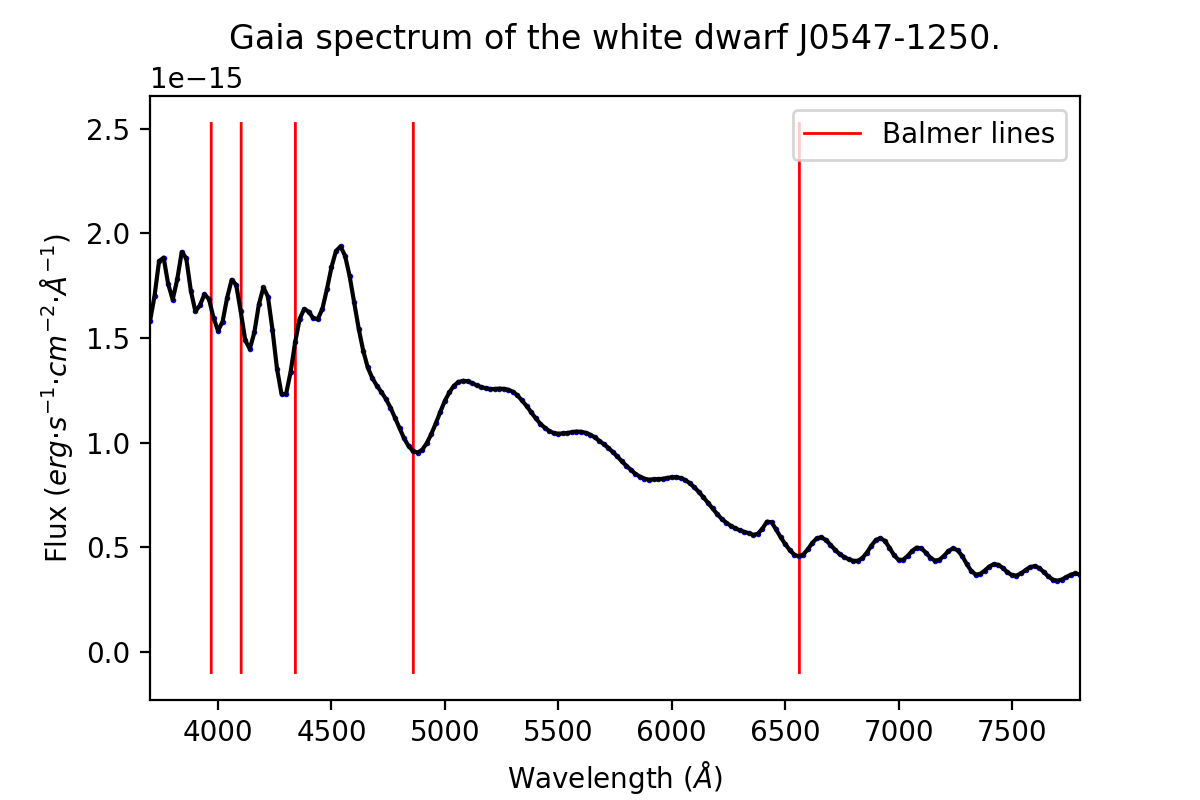}
            \includegraphics[width=1.0\columnwidth,trim=0 0 -20 0, clip]{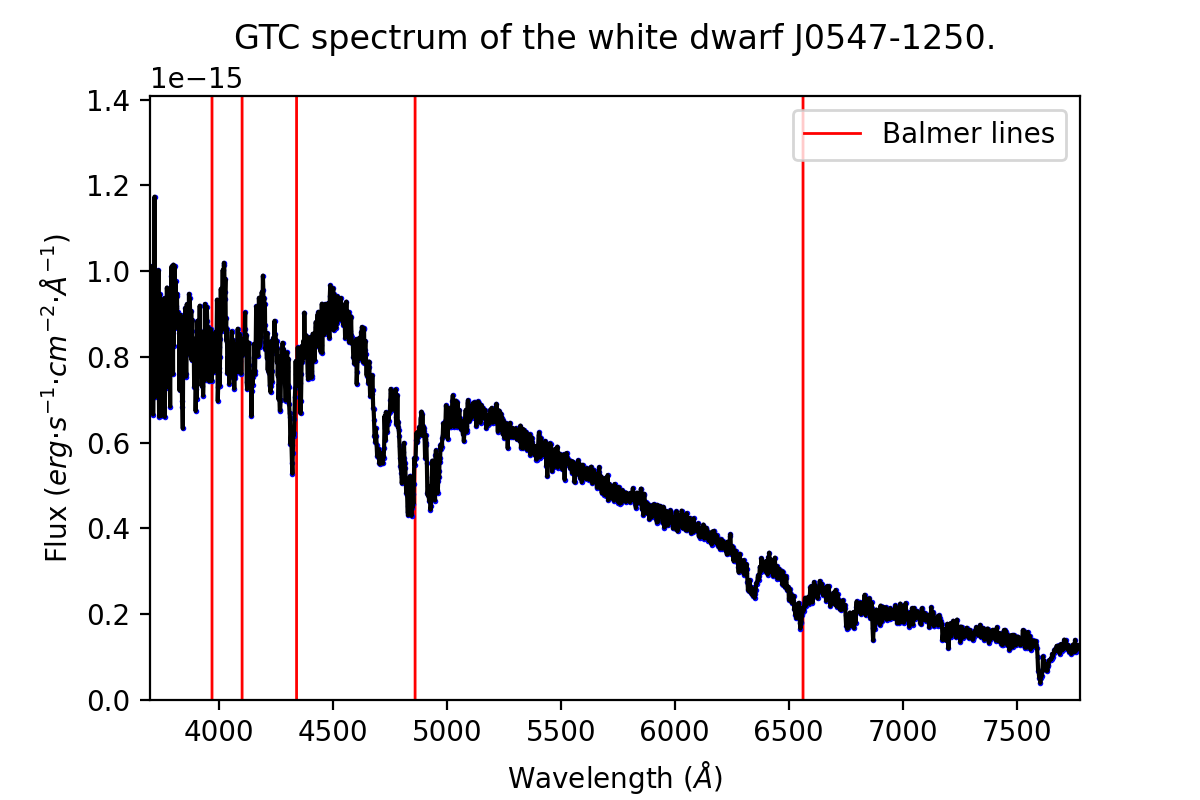}

        \includegraphics[width=1.0\columnwidth,trim=-20 0 0 0, clip]{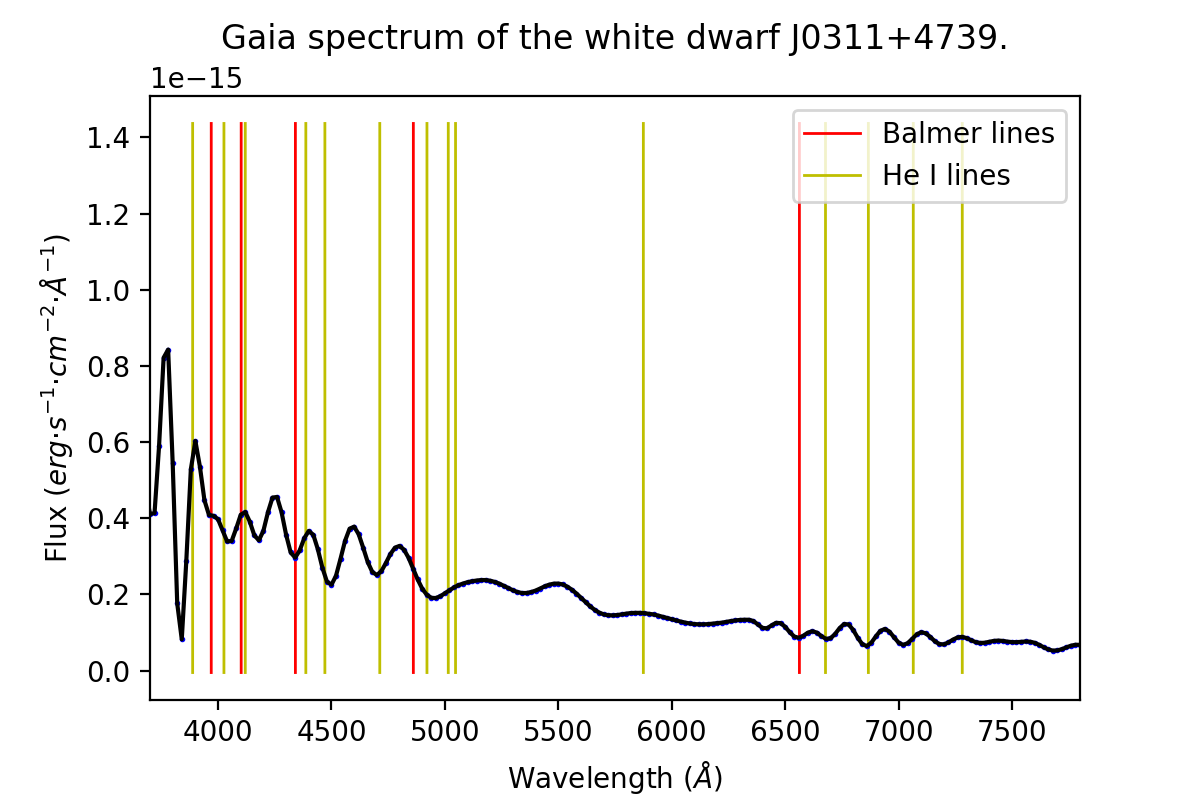}
            \includegraphics[width=1.0\columnwidth,trim=0 0 -20 0, clip]{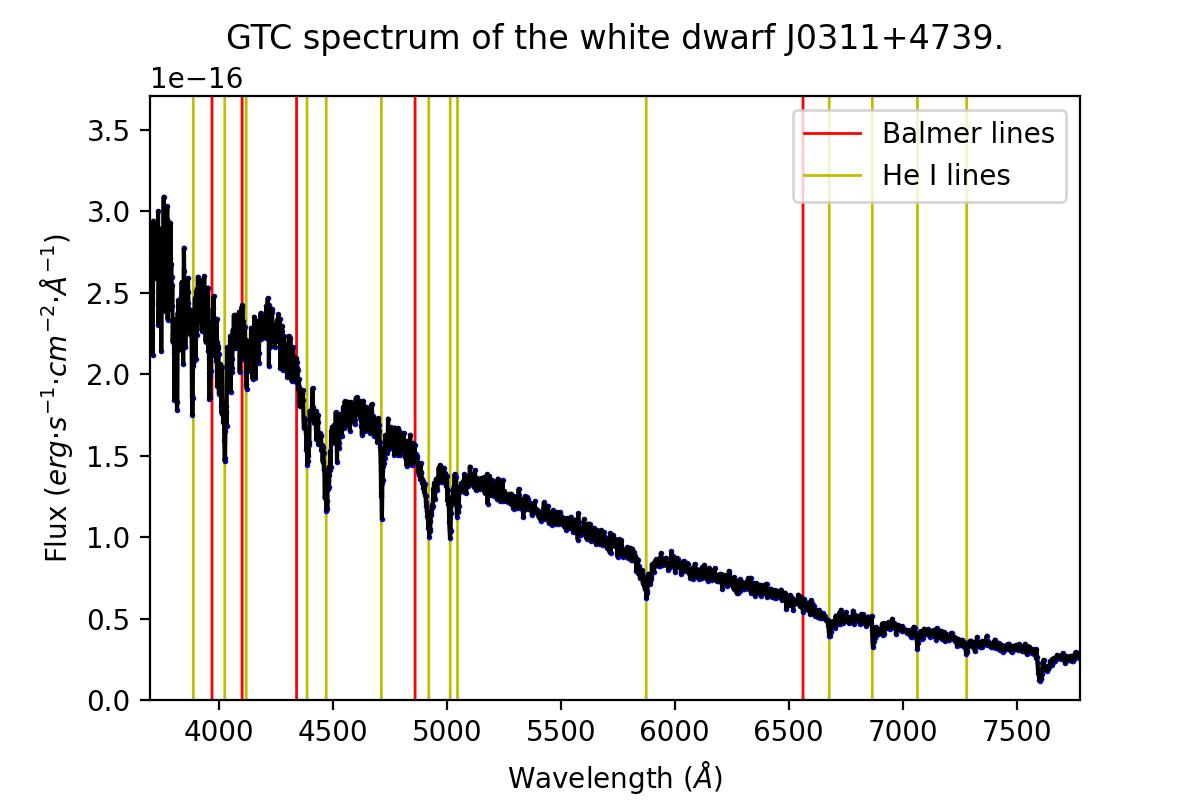}

       \includegraphics[width=1.0\columnwidth,trim=-20 0 0 0, clip]{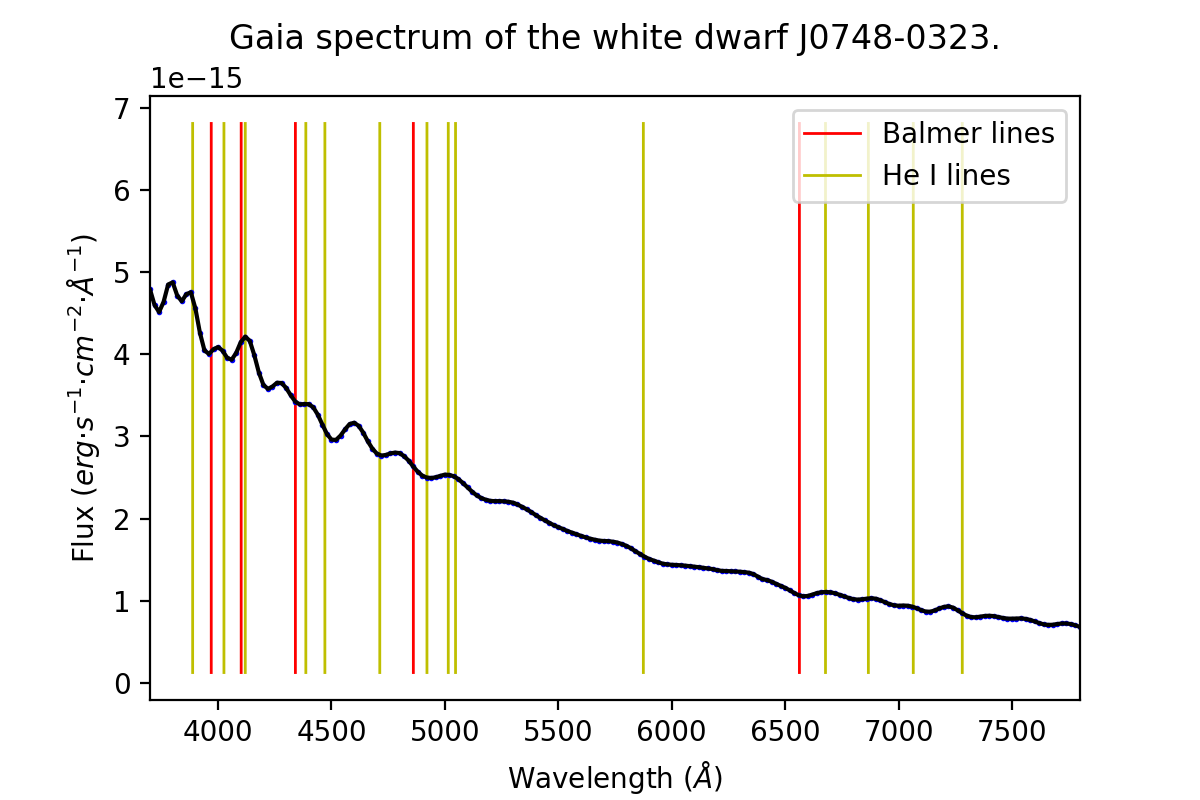}
            \includegraphics[width=1.0\columnwidth,trim=0 0 -20 0, clip]{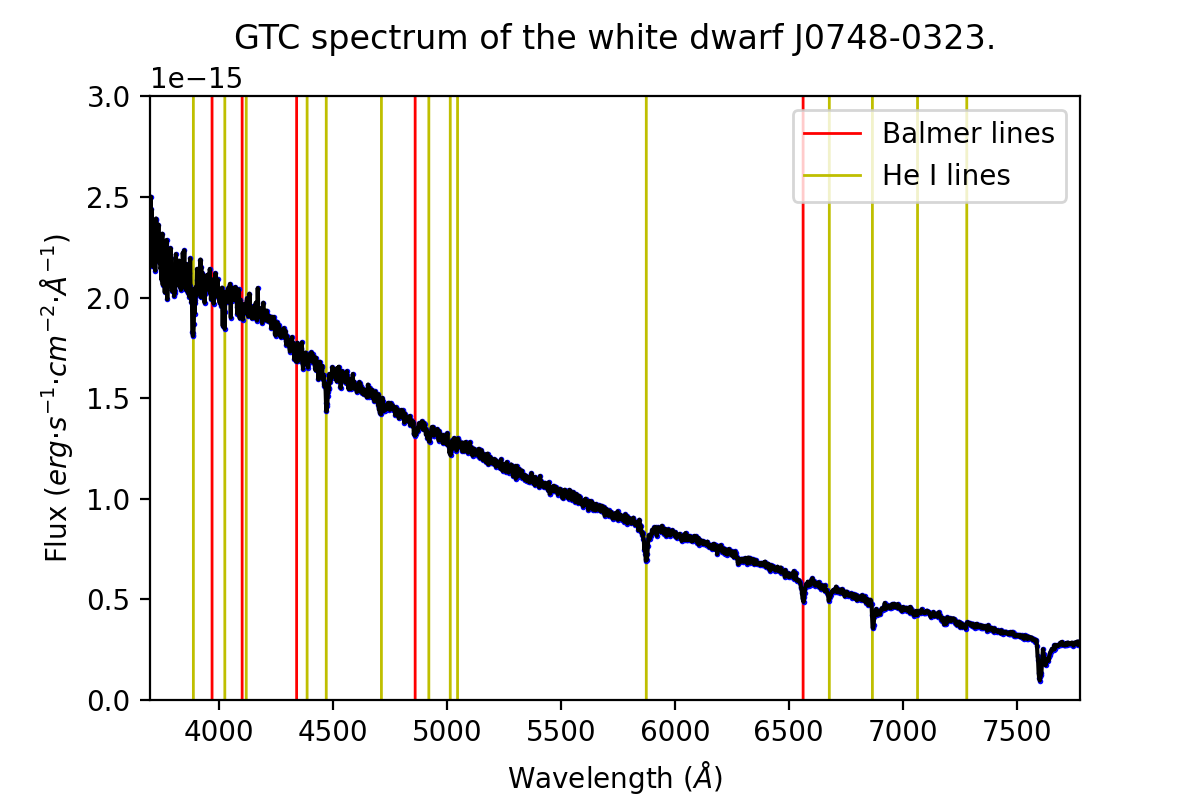}
            
        \caption{\emph{Gaia} spectra (left column) and GTC spectra (right column) of objects classified as DAH, DB and DBA.}
    \label{f:Gaia_GTC_1}
\end{figure*}

\begin{figure*}[h!]
         \includegraphics[width=1.0\columnwidth,trim=-20 0 0 0, clip]{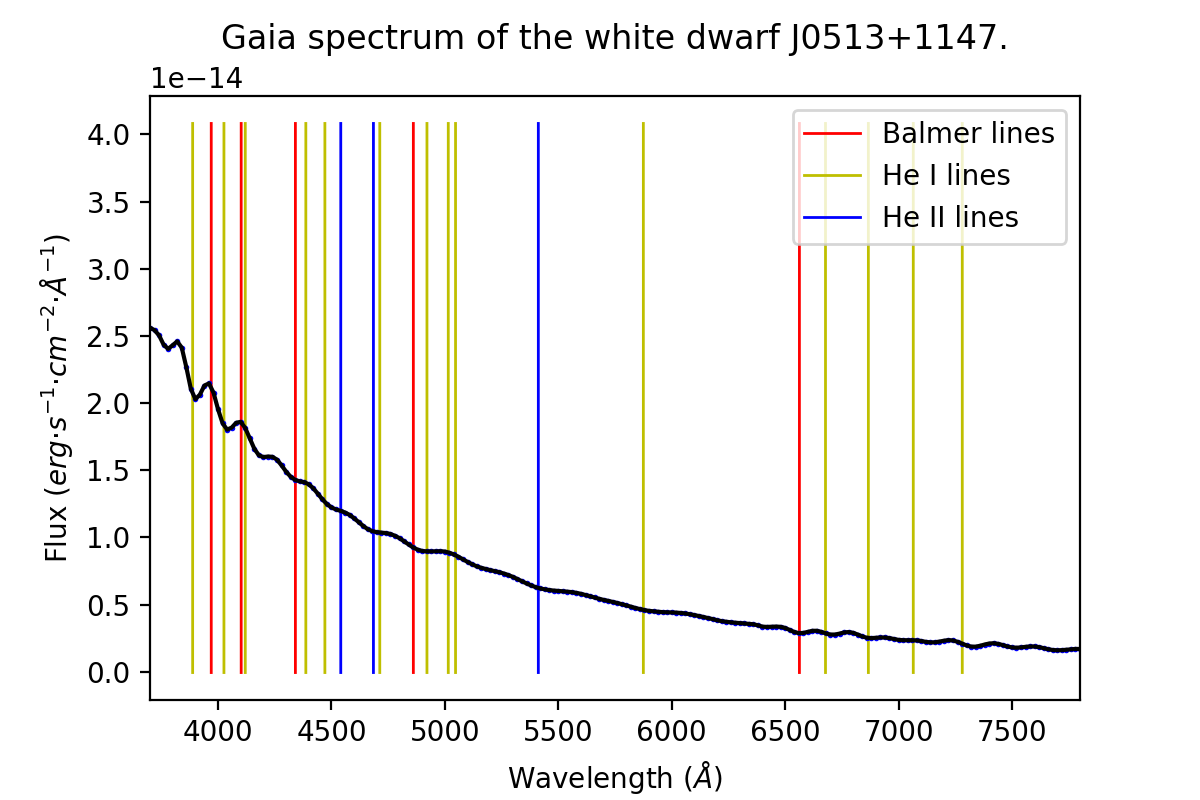}
            \includegraphics[width=1.0\columnwidth,trim=0 0 -20 0, clip]{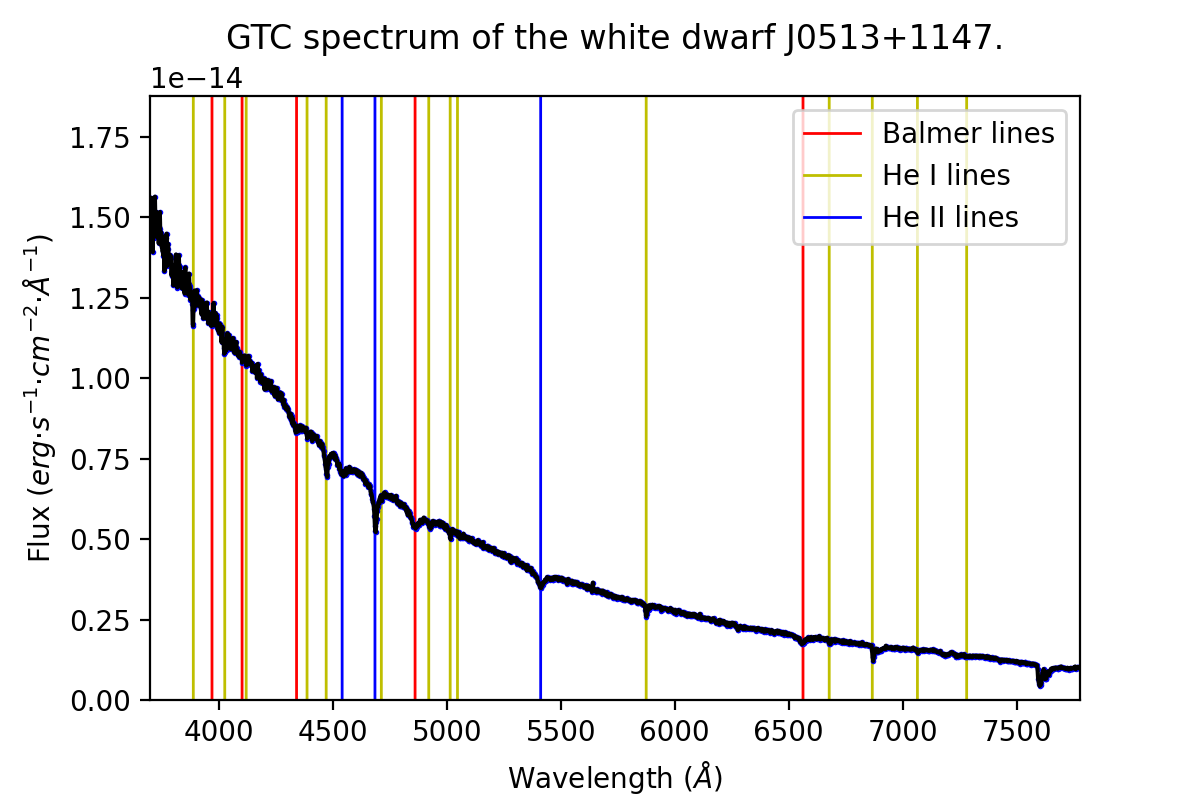}
            
        \includegraphics[width=1.0\columnwidth,trim=-20 0 0 0, clip]{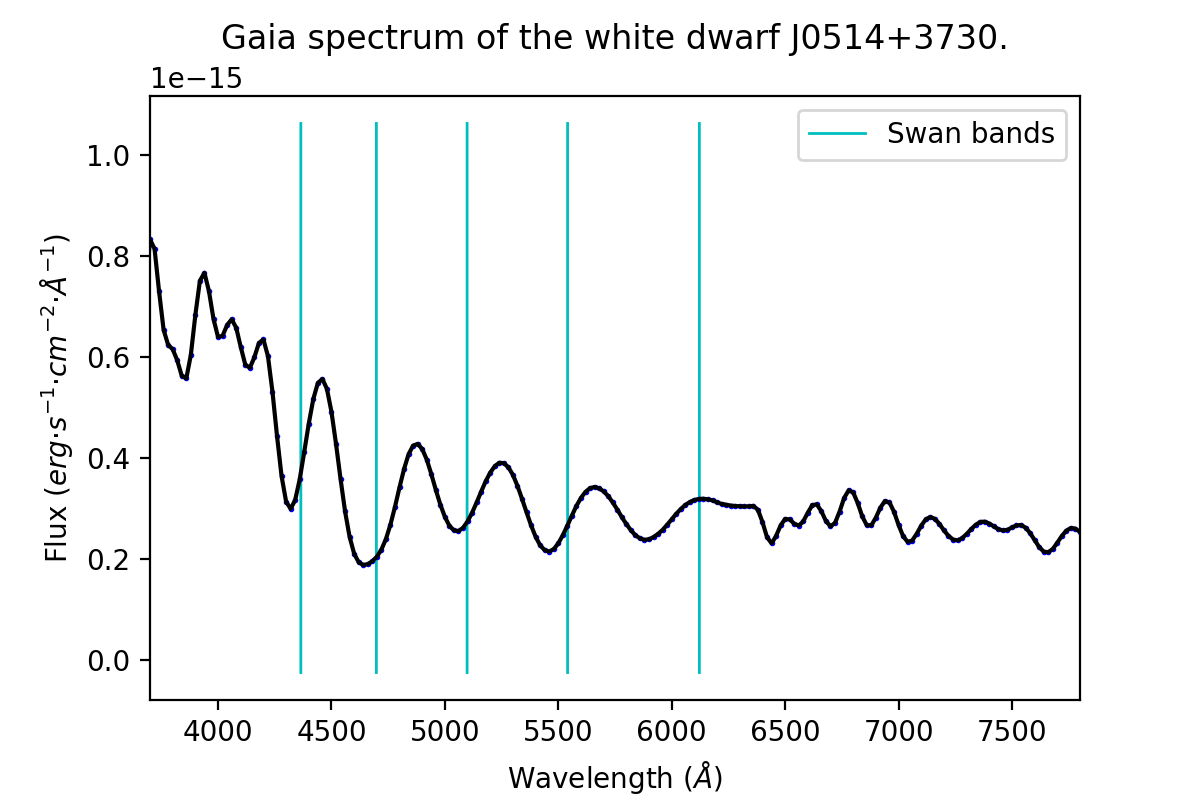}
            \includegraphics[width=1.0\columnwidth,trim=0 0 -20 0, clip]{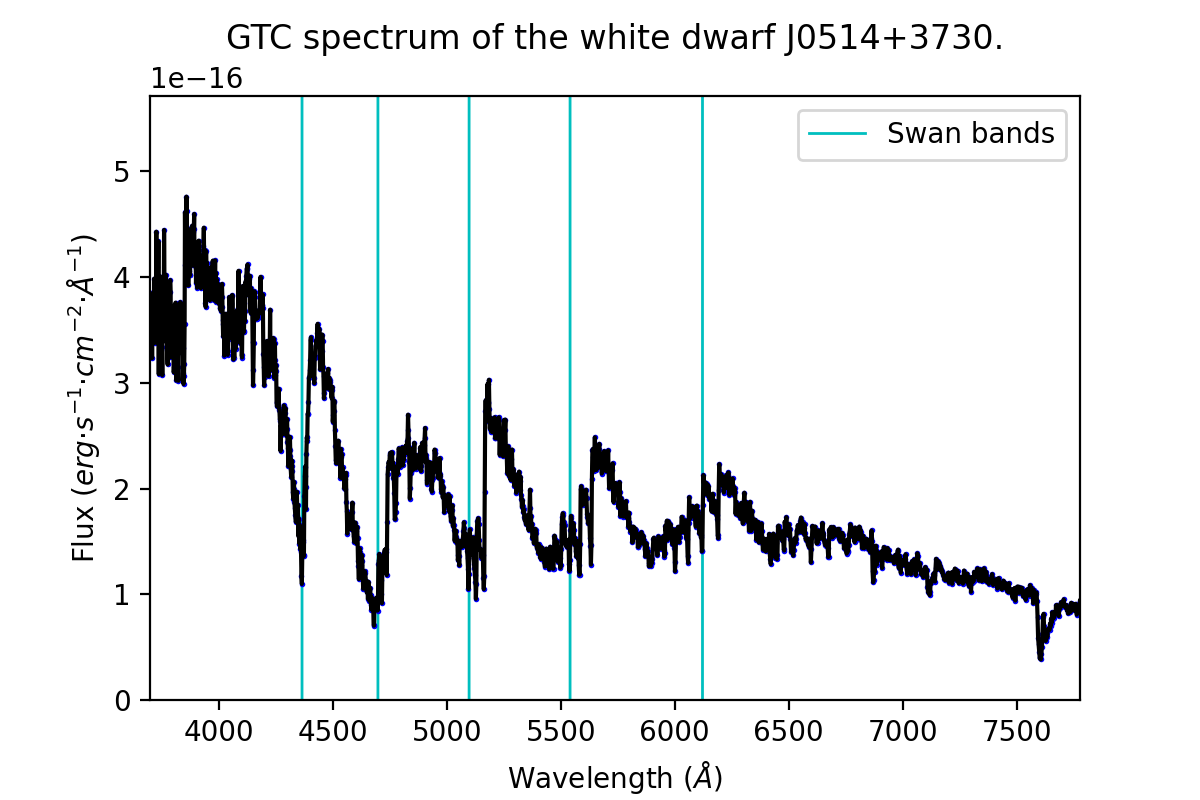}
            
        \includegraphics[width=1.0\columnwidth,trim=-20 0 0 0, clip]{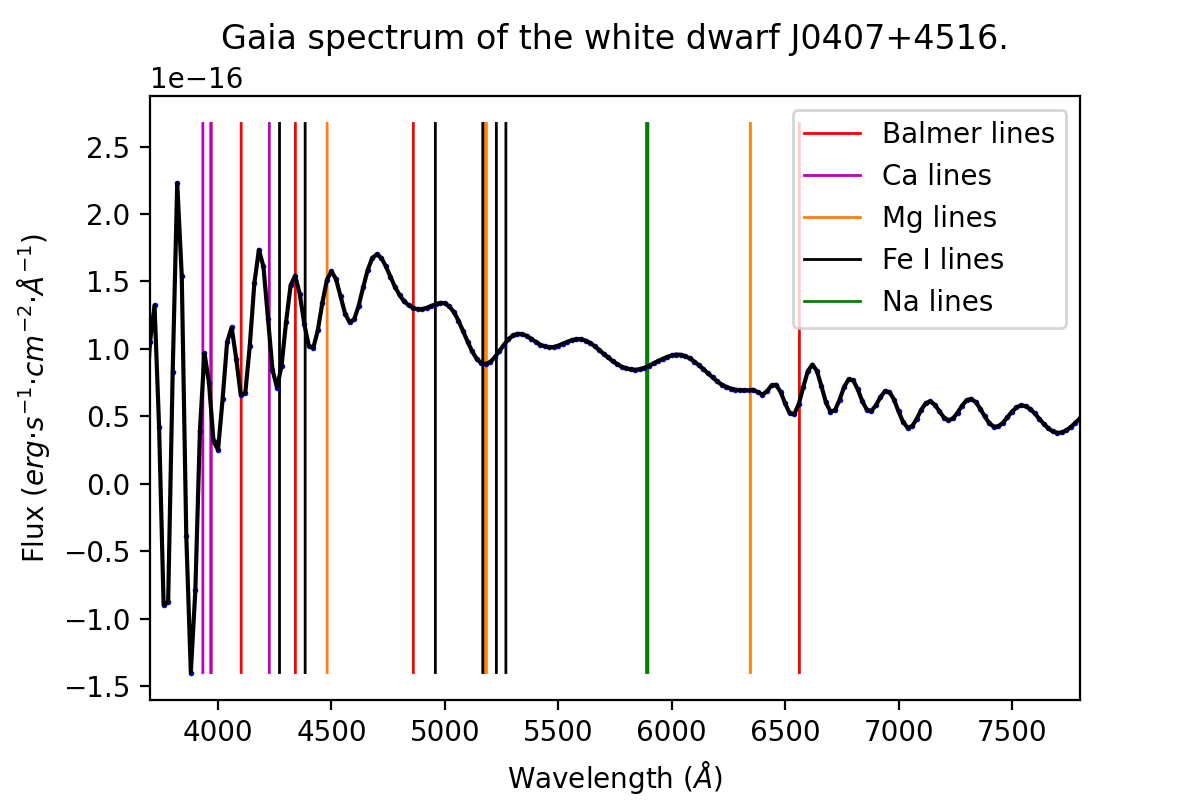}
            \includegraphics[width=1.0\columnwidth,trim=0 0 -20 0, clip]{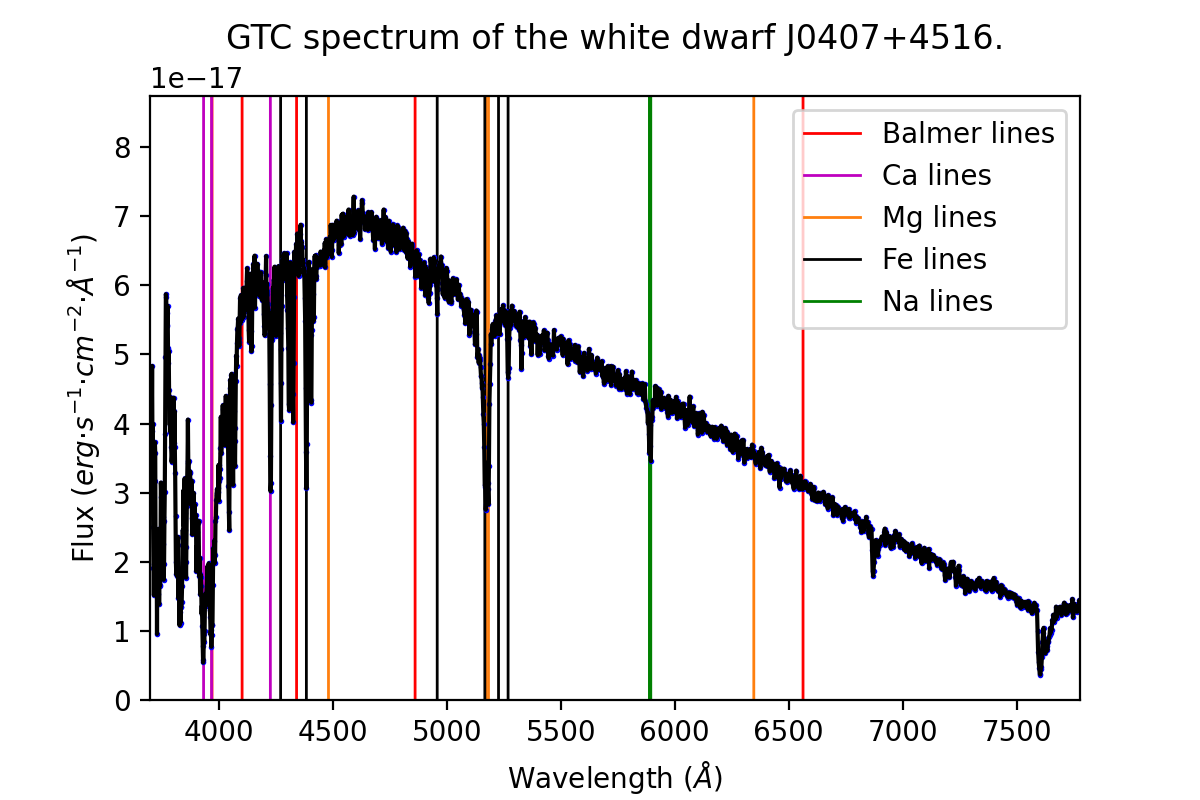}

        \includegraphics[width=1.0\columnwidth,trim=-20 0 0 0, clip]{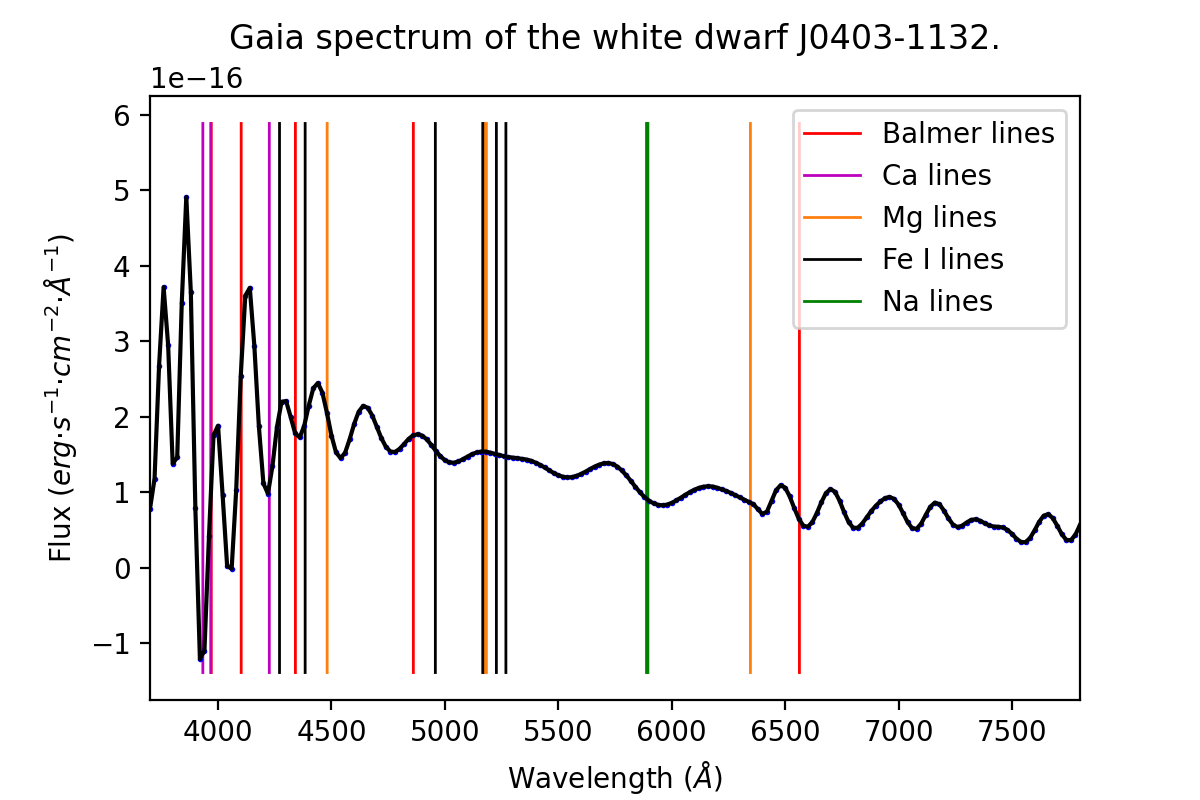}
            \includegraphics[width=1.0\columnwidth,trim=0 0 -10 0, clip]{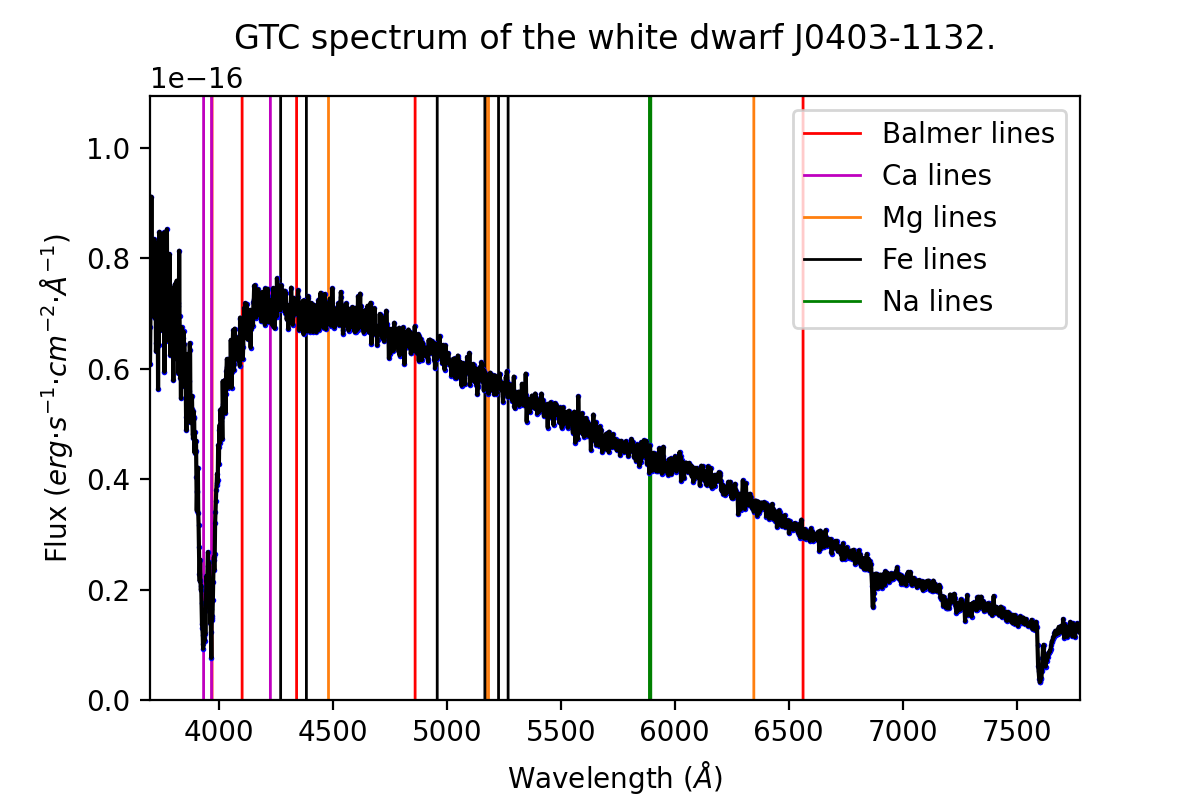}

        \caption{\emph{Gaia} spectra (left column) and GTC spectra (right column) of objects classified as DO, DQ and DZ.}
    \label{f:Gaia_GTC_2}
\end{figure*}

\end{appendix}
\end{document}